\documentclass[%
 reprint,
groupedaddress,
nofootinbib,
 amsmath,amssymb,
 aps,
prb,
]{revtex4-2}

\usepackage{graphicx}
\usepackage{dcolumn}
\usepackage{bm}
\usepackage{dsfont}
\usepackage{amsmath}
\usepackage{amssymb}
\usepackage{braket}
\usepackage{courier}
\usepackage{appendix}
\usepackage[normalem]{ulem}
\usepackage{graphicx}
\usepackage{cancel}
\usepackage{printlen}
\usepackage[caption=false]{subfig}
\usepackage[dvipsnames]{xcolor}
\usepackage[normalem]{ulem}
\usepackage[colorlinks=true,
    linkcolor=blue,
    citecolor=red,
    filecolor=magenta,      
    urlcolor=cyan,
    pdftitle={Gauge},]{hyperref}
\usepackage{pgfplots}
\usepackage{tikz}
\usepackage{float}
\usepackage{comment}
\usepackage{framed}

\definecolor{mygray}{gray}{0.85}
\colorlet{shadecolor}{mygray}

\DeclareMathOperator*{\foo}{\scalerel*{+}{\sum}}

\def\msquare{\mathord{\scalerel*{\Box}{gX}}}
\usepackage{scalerel}

\begin{document}

\title{The Physics of (good) LDPC Codes I. Gauging and dualities} 

\author{Tibor Rakovszky}
\author{Vedika Khemani}
\affiliation{%
 Department of Physics, Stanford University, Stanford, California 94305, USA
}%

\begin{abstract}

Low-depth parity check (LDPC) codes are a paradigm of error correction that allow for spatially non-local interactions between (qu)bits, while still enforcing that each (qu)bit interacts only with finitely many others. When defined on \emph{expander graphs}, they can give rise to so-called ``good codes'' that combine a finite encoding rate with an optimal scaling of the code distance, which governs the code's robustness against noise. Such codes have garnered much recent attention due to two breakthrough developments: the construction of good quantum LDPC codes and good \emph{locally testable} classical LDPC codes, using similar methods. Here we explore these developments from a physics lens, establishing connections between LDPC codes and ordered phases of matter defined for systems with non-local interactions and on non-Euclidean geometries. We generalize the physical notions of Kramers-Wannier dualities and gauge theories to this context, using the notion of \emph{chain complexes} as an organizing principle. We discuss gauge theories based on generic classical LDPC codes and make a distinction between two classes of codes, based on whether their excitations are point-like or extended. For the former, we describe Kramers-Wannier dualities, analogous to the 1D Ising model and describe the role played by ``boundary conditions''. For the latter we generalize Wegner's duality in a way that gives rise to generic \emph{quantum} LDPC codes, which appear in deconfined phases of a $\mathbb{Z}_2$ gauge theory defined on a chain complex. We show that all known examples of good quantum LDPC codes are obtained by gauging locally testable classical codes.  We also construct cluster Hamiltonians from arbitrary classical codes, related to the Higgs phase of the gauge theory, and relate these to the classical codes by formulating generalizations of the Kennedy-Tasaki duality transformation. We discuss the symmetry protected topology of the resulting models, using the chain complex language to define boundary conditions and order parameters, initiating the study of SPT phases in non-Euclidean geometries. 
\end{abstract}

\maketitle

\section{Introduction}

It has been long appreciated that there exists a close connection between the topics of error correcting codes in computer science, and ordered phases of matter in condensed matter physics. At the classical level, the Ising model, a paradigmatic example of spontaneous symmetry breaking (SSB), can also be recognized as a repetition code, the simplest error correcting scheme. On the quantum side, the toric code~\cite{kitaev2003fault} is both the most well studied example of a quantum error correcting code, and a toy model of topological order~\cite{wen1990topological} (TO). More generally, the Knill-Laflamme condition of error correction~\cite{knill1997theory} is also the condition of local indistinguishability of ground states, characteristic of topologically ordered phases~\cite{bravyi2006lieb,chen2010local,bravyi2010topological}. 

In the physical context, the simple commuting Hamiltonians that can be equated with stabilizer error correcting codes play a crucial role as stable fixed points that are associated to entire phases of matter~\cite{kitaev2003fault,chen2010local}. In particular, they are often robust to perturbations that move us away from the fixed point limit (as long as they respect the underlying symmetries, in the case of SSB). This robustness is closely related to the resilience of the corresponding code to noise: in both cases, the crucial feature is that distinct states in the relevant low-energy / code subspace cannot be connected by local (symmetric) operations~\cite{bravyi2010topological}. Thus, error correcting codes can serve as a basis for understanding gapped phases of matter in general, an approach that has already proven fruitful in the context of fracton phases~\cite{haah2011local,vijay2016fracton}.

In physics, one typically considers models defined by local interactions in $D$-dimensional Euclidean space. When used to encode information, such spatially local systems have some fundamental limitations. Notably, they exhibit a tradeoff between the amount of logical information one can encode (the \emph{code rate} $k$) and its robustness (the \emph{code distance} $d$, which is the size of the smallest undetectable error)~\cite{bravyi2010tradeoffs,flammia2017limits}. To overcome these limitations, one is lead to consider a more general paradigm of error correction, called \emph{low-depth parity check} (LDPC) codes. LDPC codes are defined by the fact that every (qu)bit interacts with a finite number of other (qu)bits. This includes spatially local models, but it also allows for much more general geometries.  Of particular interest are \emph{expander graphs}~\cite{sipser1996expander,breuckmann2021quantum}, which are defined (heuristically) by the property that the surface area of a region grows proportionally to its volume. Indeed, it has been known for decades that expander graphs allow for the construction of \emph{good} classical LDPC (cLDPC) codes~\cite{sipser1996expander}, where ``goodness" is defined by the property that both $k$ and $d$ are proportional to the number of physical bits $n$ (which is the best achievable scaling). In contrast, a construction for good quantum LDPC (qLDPC) codes had remained a major outstanding challenge until very recently. 

A series of recent breakthroughs have achieved long elusive results for both classical and quantum LDPC codes~\cite{hastings2021fiber,breuckmann2021balanced,panteleev2022asymptotically,leverrier2022quantum,dinur2023good,lin2022good, dinur2022locally,lin2022c}. On the quantum side, good qLDPC codes have now been constructed~\cite{breuckmann2021balanced,panteleev2022asymptotically,leverrier2022quantum,dinur2023good,lin2022good}. On the classical side, we now have good cLDPC codes that are also \emph{locally testable}~\cite{dinur2022locally,panteleev2022asymptotically,lin2022c}. A locally testable code (LTC) is one in which a local ``tester" can distinguish (with high probability) between a state in the code subspace and one far from it by looking at only a small --- typically constant --- number of (qu)bits. LTCs have important applications in complexity theory, for instance in the construction of probabilistically checkable proofs (PCPs)~\cite{goldreich2005short}\footnote{The construction of good \emph{quantum} LDPC codes which are also locally testable is still an outstanding challenge.}. Remarkably, the constructions that led to good qLDPC codes were closely related to (or, in come cases, the same as) the ones that led to good classical LTCs, suggesting a physical connection between these properties. 

Our goal in this paper is to understand these developments from a physicist's lens, and to use this to push the frontier for quantum many-body physics to incorporate more general (LDPC) interactions and non-Euclidean geometries. Such non-Euclidean geometries present a promising arena for many-body physics: the locality of interactions on the graph means that some crucial features of local physics still go through: energies are extensive, so one can meaningfully talk about thermodynamics, and a version of the Lieb-Robinson bound~\cite{hastings2006spectral}, which plays a fundamental role in our understanding of phases of matter and their stability~\cite{hastings2005quasiadiabatic,bravyi2010topological}, still applies. While there exist investigations of phases on non-Euclidean graphs~\cite{laumann2009absence,laumann2010aklt,manoj2023arboreal,gorantla2023gapped}, compared to their Euclidean counterparts they are relatively unexplored, and indeed the recent breakthroughs in LDPC codes suggest that much remains to be understood. A glimpse into intriguing novel phenomena that can appear in this context was provided by a recent result~\cite{anshu2023nlts} which showed that a certain family of good qLDPC codes exhibits a particularly strong form of topological order (known as the``no low-energy trivial states'', or NLTS, property~\cite{freedman2013quantum}). There also now exists an increasing number of quantum simulation platforms where non-Euclidean geometries can be realized experimentally~\cite{Kollar2019,Periwal2021,Bluvstein2022,LukinLDPC}, further motivating their study from a physical perspective.

\begin{figure*} 
    \centering
    \includegraphics[trim={1cm 0 0 0},width = 1.\linewidth]{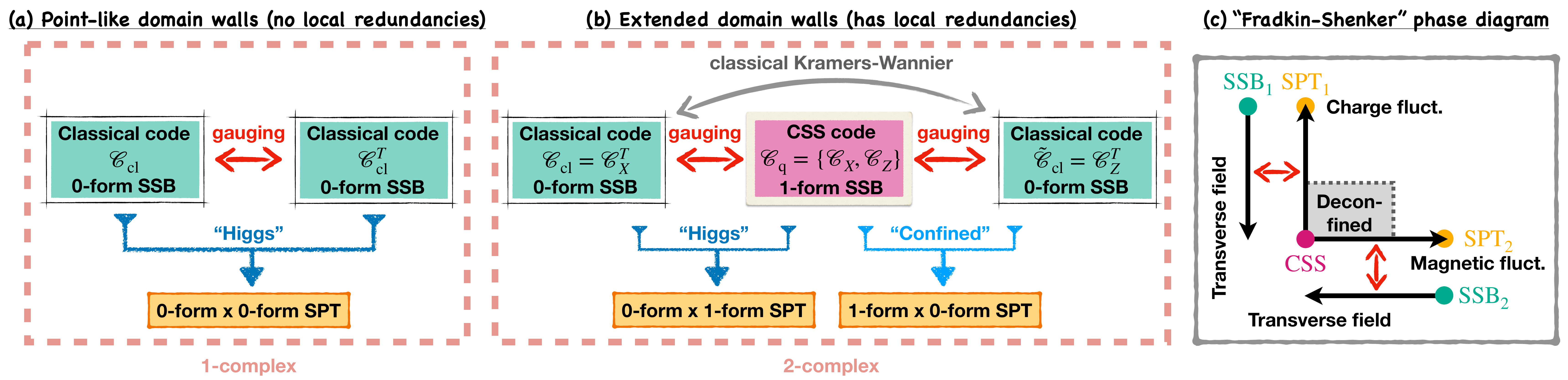}
    \caption{Summary of some of the dualities that appear in this paper. Our starting points are classical LDPC codes that fall into two categories: with and without local redundancies. Physically, this determines the dimensionality of domain wall excitations, which become new degrees of freedom upon gauging. (a) With point-like excitations, gauging gives rise to a duality between two classical codes, both exhibiting spontaneous symmetry breaking (SSB) order. By combining the code and its dual, one can construct a cluster Hamiltonian which exhibits features of symmetry protected topological (SPT) order. (b) When domain walls are extended, gauging gives rise to a CSS code appearing in the deconfined phase of the $\mathbb{Z}_2$ gauge theory. In this case there are two distinct classical codes dual to the same CSS code; the two classical codes are related to each other via a classical Kramers-Wannier duality. One can now construct two cluster Hamiltonians, using either of thw two classical codes as a starting point. (c) Shows a sketch of the expected phase diagram corresponding to the case depicted in (b). In this case, we end up with a two-dimensional phase diagram analogous to the seminal work of Fradkin and Shenker~\cite{fradkin1979phase}; the CSS code is situated inside a deconfined phase, while the two SPT Hamiltonians are related to the extreme limits along thw two axes. The parts of the phase diagram along the axes can be related, via gauging, to the phase diagram of the classical codes in the presence of a transverse field.}
    \label{fig:GaugeDualities}
\end{figure*}

We begin an initial foray into this territory, using classical and quantum LDPC codes as a guide. We do so by incorporating them into the language of \emph{gauge theories} familiar from high energy and condensed matter physics. In particular, we develop a very general \emph{gauging} procedure, extending those that have appeared in the context of usual $\mathbb{Z}_2$ gauge theory~\cite{wegner1971duality, fradkin1979phase} and fracton phases~\cite{vijay2016fracton,shirley2019foliated,williamson2016fractal,kubica2018ungauging}, which takes as its input an arbitrary classical LDPC code and outputs a generalized $\mathbb{Z}_2$ gauge theory. We argue that for a certain family of classical codes, the resulting gauge theory should host a stable \emph{deconfined} or topologically ordered phase, whose fixed point is associated to a quantum LDPC code. This situates both types of codes within larger phase diagrams, and gives rise to duality transformations between them, paving the way both towards understanding the stability of these phases, as well as properties of the critical points between phases (see Fig.~\ref{fig:GaugeDualities}). This approach also reveals a relationship between the two aforementioned developments: we find that all the recent constructions of good qLDPC codes arise from gauging good classical LTCs.

A central organizing principle in our framework is a \emph{chain complex} associated to a classical or quantum code~\cite{kitaev2003fault,bombin2007homological,bravyi2014homological}. These allow the properties of codes to be formulated in terms of homology theory, and allows one to extract code properties from the topological features of the complexes. Importantly for us, these will allow us to define geometrical features directly from the code itself without reference to an underlying Euclidean lattice. In particular, an important role is played in our discussion by local and global \emph{redundancies}\footnote{These are also variously called \emph{relations} and \emph{meta-checks} in the coding literature.} (i.e., constraints between the local interactions terms defining the code, which force certain products of these terms to be the identity, see Fig.~\ref{fig:IsingOnATorus}) of the classical codes. We will make a distinction between two classes of classical codes - those with and without local redundancies - which we interpret in terms of the dimensionality of their ``domain-wall" excitations, which can be either point-like or extended (see Fig.~\ref{fig:IsingOnATorus} for a graphical illustration of the basic idea). We will show that these two classes are distinguished by the effective dimensionality of the chain complexes defining the classical codes, which locally define a notion of geometry from the code. Broadly, codes with local redundancies can be associated with 2-dimensional chain complexes which have a notion of ``plaquettes", while codes without local redundancies are effectively one-dimensional. It is the former that give rise to robust deconfined phases, and thus to qLDPC codes, upon gauging. We also discuss the relationship between redundancies and boundary conditions, which plays an important role in gauging. 
While the idea of boundary conditions is again non-obvious when we have no underlying lattice, we show that we can define them directly from the classical code itself, by considering its global redundancies.   

In the recent breakthrough papers, the structures required to achieve good qLDPC codes or good classical LTCs were obtained by constructing two-dimensional chain complexes out of two ones-dimensional ones, similar to ideas in topology where higher dimensional manifolds can be constructed from lower dimensional ones. The one-dimensional chain complexes themselves can be interpreted as classical codes, so these constructions can equivalently be thought of as constructing one quantum code out of two classical codes. Various such \emph{product constructions} have been developed~\cite{tillich2013quantum,bravyi2014homological,hastings2021fiber,breuckmann2021balanced,panteleev2022asymptotically,breuckmann2021quantum} to achieve various coding aims.  We will explore these constructions from the perspective of condensed matter physics in upcoming work~\cite{LDPCProduct}.

We note that the relationship between codes, chain complexes and gauging has appeared previously in the literature~\cite{vijay2016fracton,williamson2016fractal}, in particular in Ref. \onlinecite{kubica2018ungauging}. Our work generalizes and extends these discussions, and also takes a different perspective which puts emphasis on the classical code as the starting point from which gauge theories can be obtained. In particular, we will discuss how different kinds of classical codes give rise to different physics upon gauging and interpret the difference in terms of the redundancy structure of the codes and the properties of the domain wall excitations of the classical codes. 

Apart from SSB and TO, the third major class of gapped phases is that of \emph{symmetry protected topological} (SPT) order~\cite{pollmann2010entanglement,chen2013symmetry}. To further underline the usefulness of the gauge theory perspective, we draw upon recent work~\cite{devakul2019fractal,verresen2022higgs} to show how SPT physics arises generically from gauging any classical code, thus bringing together all three class of phases under the same unifying framework. We show how various known SPTs~\cite{you2018subsystem,devakul2019fractal} arise very naturally in this way, and discuss a set of generalized Kennedy-Tasaki transformations~\cite{kennedy1992hidden1,kennedy1992hidden2,else2013hidden,li2023non} that can be used to map between SSB and SPT phases.  In our discussion of SPT features, we again rely on the idea of chain complexes, for example in defining edge modes, which are a fundamental feature of these phases. Not only does this allow us to discuss a large class of different SPTs in the same unified terms, it also brings up the possibility of extending the notion of SPT phases to expander graphs and other non-Euclidean geometries. 

Overall, our results provide a ``web of dualities'' between classical LDPC codes, quantum LDPC codes, and cluster states, each of which can serve as a fixed point within an appropriate (SSB, TO or SPT) phase, formulated in a way that applies directly in the non-Euclidean setting. Many of these ideas are summarized in Fig.~\ref{fig:GaugeDualities} (see also Fig.~\ref{fig:KTduality}). The following section situates our work within the broader context of gauge theories in physics, and provides a roadmap through the rest of the paper. 

\tableofcontents

\section{Background and summary of results}\label{Sec:overview}

Our work connects two main strands: gauge theories in physics, and (classical and quantum) LDPC codes in computer science. In particular, we develop a general gauging procedure, describing how to couple an arbitrary cLDPC code to gauge fields, giving rise to a wide array of different gauge theories. When the excitations (``domain walls'') of the classical code are extended (``loop-like'') objects, the resulting gauge theory hosts a stable \emph{deconfined} or topologically ordered phase, described at its fixed point by a \emph{quantum} error correcting code. On the other hand, the \emph{Higgs} phases of these gauge theories give rise to symmetry protected topological phases. 

The preceding paragraph raises a great many questions. What do we mean by gauging and gauge fields in this context? How to make sense of the notion of extended domain walls in the absence of a Euclidean lattice? How to define the deconfined and Higgs phases? In fact, some of these questions are fairly intricate even for conventional gauge theories. In Sec.~\ref{subsec:Manifesto} we overview these various notions (using electromagnetism as a way to ground our intuition). Along the way, we will stop and point out where the relevant generalizations appear later in the paper. In Sec.~\ref{subsec:Ising}, we outline how these various ideas play out in the context of the Ising model and its associated gauge theory in different dimensions, which will further help orient the discussion in the rest of the paper. 

\subsection{Gauge theories and their phases}\label{subsec:Manifesto}

Gauge theories are fundamental to our understanding of nature, thanks to the key part that they play in the Standard Model of particle physics. In condensed matter physics also, gauge theories appear regularly in describing various phases of matter, from the quantum Hall effect~\cite{zhang1989effective} to spin liquids~\cite{savary2016quantum} to superconductors~\cite{anderson1963plasmons,hansson2004superconductors}. Despite this success, defining what constitutes a gauge theory is notoriously difficult and remains a matter of debate to the present day~\cite{harlow2021symmetries}. 

On a surface level, a gauge theory as usually defined refers to a system with two types of degrees of freedom---matter and gauge fields---obeying certain local constraints---the familiar Gauss's law of electromagnetism and its generalizations. These constraints generate the gauge transformations under which the theory is invariant. While these are sometimes colloquially referred to as ``gauge symmetries'', they are not really symmetries in the usual sense; rather, the states related by such gauge transformations are treated as physically equivalent\footnote{When the theory is defined on a lattice, this means that the physical Hilbert space, made up by gauge equivalence classes, necessarily lacks a tensor product structure.}. 

Even though the gauge transformations are not symmetries, they correspond to some ``gauge group'' $G$, and one can speak of e.g. $\mathbb{Z}_2$ gauge theory or $U(1)$ gauge theory (the latter corresponding to usual electromagnetism). A standard way of arriving at such a gauge theory is through the procedure of \emph{``gauging''}, which starts from a theory of matter fields that has $G$ as a global symmetry, and constructs a gauge theory by coupling them to gauge fields in a systematic way, through a prescription called ``minimal coupling''~\cite{kogut1979introduction,levin2012braiding,haegeman2015gauging,trott2013gauge}; this is often referred to as ``making the global symmetry local''.

\begin{shaded}
    In Sec.~\ref{subsec:classicalLDPC} we discuss how every classical LDPC code is associated to a set of global (including subsystem) symmetries, defined by their \emph{logical operators}. 
\end{shaded}

The gauging  procedure can be understood as a two-step process~\cite{barkeshli2019symmetry,thorngren2018gauging,harlow2021symmetries}: (i) first, given the global symmetry $G$, one can introduce \emph{classical background gauge fields} corresponding to it. When $G$ is a continuous symmetry, these appear as source terms that couple to the currents of the conserved charge that generates $G$\footnote{For example, one component of the background gauge field is the chemical potential that can be used to tune the charge density.}. More generally, the background gauge fields are a way of introducing \emph{symmetry defects}, such as domain walls (for a discrete symmetry) or vortices (for a continuous one). (ii) The second step is to make these gauge fields into dynamical degrees of freedom, by adding additional terms to the Hamiltonian that allow them to fluctuate (or by integrating over them, in the path integral formulation). 

\begin{shaded}
         In Sec.~\ref{sec:Gauging} we describe how an arbitrary classical code might be coupled to background gauge fields, which correspond to holonomies around ``loops'' which correspond naturally to \emph{redundancies} of the code (see Fig.~\ref{fig:IsingOnATorus}). We also discuss how these gauge fields can be used to define a set of generalized Kramers-Wannier dualities.  
\end{shaded}

However, their defining property---the local gauge freedom---makes the notion of gauge theory inherently ambiguous. One can always eliminate the gauge freedom by choosing some appropriate set of variables. In some cases (e.g. 1D $\mathbb{Z}_2$ or 2D $U$(1) gauge theories), this can be done in a way such that the resulting ``gauge fixed'' theory becomes equivalent to the original theory involving matter fields alone. In this sense, ``gauge symmetry'' is ``\emph{not a property of Nature, but rather a property of how
we choose to describe Nature}''~\cite{tong2018gauge}. Despite this, their aforementioned ubiquity implies that there must be something physically meaningful about the notion of a gauge theory, and indeed there are phenomena that are characteristic to them, such as robustly gapless particles for continuous $G$ (e.g. photons in a $U(1)$ gauge theory) or topological order for discrete $G$. How to account for these in unambiguous terms?

We can gain insight into this question by considering the limit of \emph{pure gauge theory}, when no matter fields are present. Considering electromagnetism for concreteness, Gauss's law in this case has a clear physical meaning: it states that electric field lines can have no endpoints. Thus, the pure gauge theory is a theory of \emph{closed loops}~\cite{polyakov1987gauge,harlow2021symmetries}. The same is true of pure $\mathbb{Z}_2$ gauge theory (equivalent to the toric code). More generally, pure gauge theories are theories of various kinds of extended objects. This helps explain why gauge theories arise from the gauging procedure outlined above: symmetry defects of the matter fields, such as domain walls, often naturally form such extended objects. This also helps explain why the $\mathbb{Z}_2$ gauge theory hosts a robust topologically ordered phase in 2D but not in 1D: in the latter case, domain walls are point-like, non-extended objects. We will discuss this point in more detail in the next subsection.

\begin{shaded}
    How to make sense of the notion of extended domain walls without an underlying lattice? This is the topic of Sec.~\ref{sec:Redundancies} where we argue that they can be defined in terms of \emph{local redundancies}, which can be used to form a \emph{two-dimensional chain complex}. This also allows us to define a notion of ``boundary conditions'', which play an important role in gauging, using the \emph{global} redundancies of the classical code. 
\end{shaded}

The importance of the pure gauge theory arises from the fact that, in the non-trivial cases, it describes the low-energy behavior within an entire phase, the \emph{deconfined phase} of the gauge theory\footnote{The term ``deconfined'' refers to the fact that the electric charges can be arbitrarily separated from each other at a finite energy cost.}. Thus, even when we allow electric charges, which serve as the endpoints of electric field lines, their presence turns out to be irrelevant (in the RG sense) as long as their coupling to the gauge fields is sufficiently weak. On large enough length scales, the theory is still that of closed loops. 

\begin{figure} 
    \centering
    \includegraphics[trim={1cm 0 0 0},width = 0.8\linewidth]{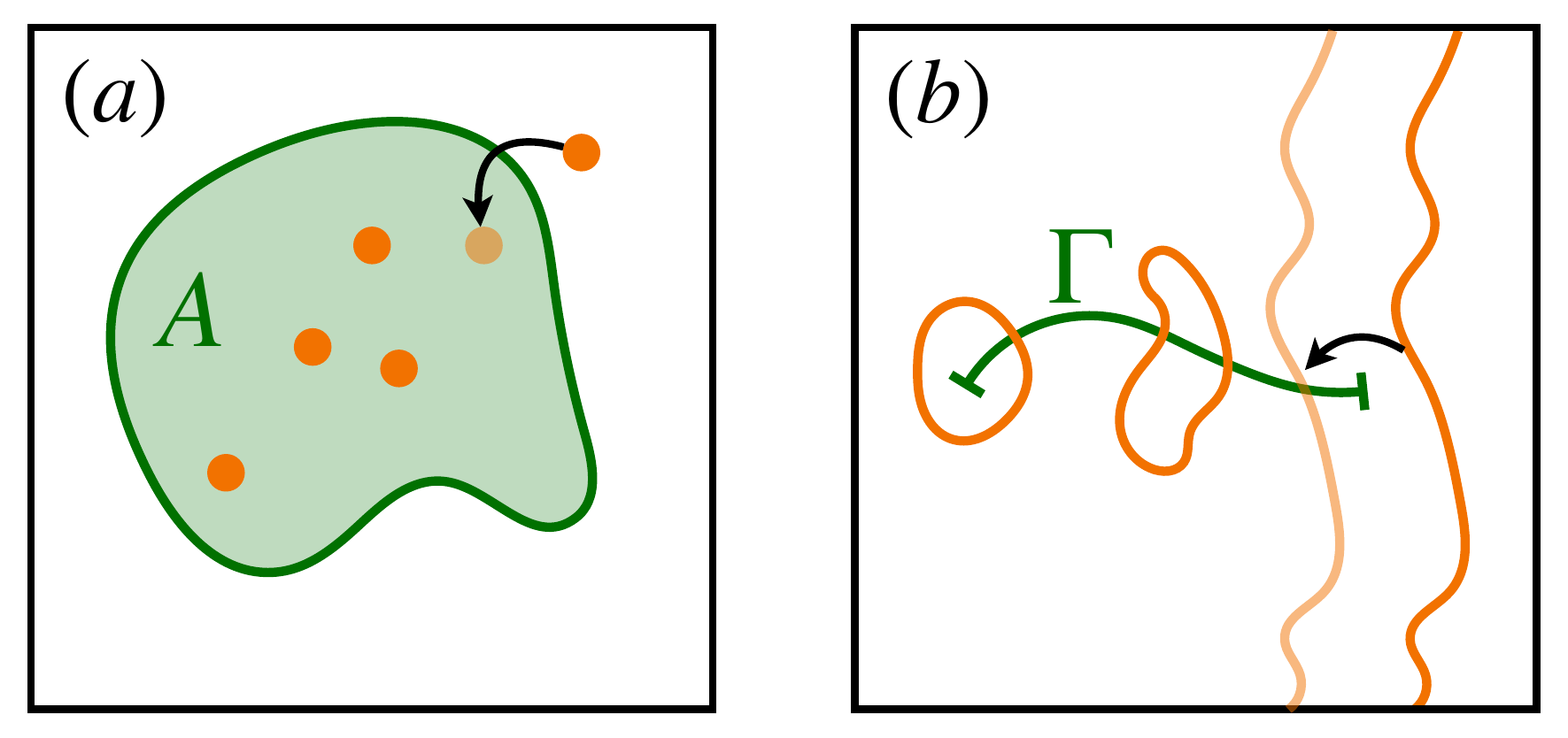}
    \caption{Higher form symmetries. (a) in a ``usual'' (0-form) symmetry, charge is associated to a region of space ($A$) and its value can only change by charges (which are point-like) moving through the boundary. (b) In a higher form symmetry, the charge is measured through a line/surface ($\Gamma$), counting line-like charged objects crossing it. It can only change when a line moves through the boundary of $\Gamma$, which implies that the charges must form closed loops. Deforming $\Gamma$, while keeping its boundary fixed, does not change the charge measured through it.}
    \label{fig:HigherForm}
\end{figure}

There is a modern understanding of all this in terms of the notion of \emph{higher form symmetry}~\cite{gaiotto2015generalized,mcgreevy2023generalized}. The fact that field lines must be closed in pure electromagnetism means that the electric flux through some surface (i.e., the number of field lines crossing it) can only change by pulling a field line through the boundary of the surface (see Fig.~\ref{fig:HigherForm}). This can be interpreted as a generalized continuity equation, where the role of charges is played by the field lines. When the surface is closed, the associated charge must remain constant, an example of a higher-form conservation law. More generally, for a \emph{$p$-form symmetry}, charged objects are $p$-dimensional; correspondingly, the symmetry operators are associated to objects (lines, surfaces, etc.) with co-dimension $p$, such as the surface through which we measure the flux in the example above. An important aspect of these higher-form symmetry operators is that they are \emph{topological}, in the sense that one can deform the surface continuously without changing the amount of charge associated to it, as long as its boundaries remain fixed---in the EM example this is a simple consequence of Gauss's law. Since the ``charge'' is only truly conserved for closed surfaces, and contractible surfaces can by definition be shrunk to zero, the non-trivial symmetry sectors correspond to values of the conserved charge associated to topologically non-trivial surfaces. 

From this symmetry perspective, the deconfined phase is interpreted as the \emph{spontaneous breaking} of the higher form symmetry: the expectation value of loops of electric field lines (which are the charges of the symmetry) remains finite even as the size of the loop diverges, which can be interpreted in analogy with conventional long-range order for $0$-form symmetries~\cite{gaiotto2015generalized,mcgreevy2023generalized,kapustin2017higher}. The stability of the deconfined phase then amounts to saying that, within this phase, there is an \emph{emergent} higher form symmetry, which, although broken at the microscopic level by coupling to charges, re-emerges at low energies. The intuition behind this stability is that, since the perturbations one is allowed to add to the theory need to be local, they cannot be charged under the higher form symmetry, which therefore cannot be broken in the same way as $0$-form symmetries are. There have been attempts at making this intuition more rigorous~\cite{iqbal2022mean,pace2023exact}.

\begin{shaded}
    In Sec.~\ref{sec:Deconfined} we describe how the deconfined phase is associated to a \emph{quantum} LDPC code. The stability of the phase is tied to the code's ability to correct all local errors. We also discuss how \emph{good} qLDPC codes arise from gauging classical \emph{locally testable codes}.     
\end{shaded}

As the coupling to matter fields gets stronger, the gauge theory eventually undergoes a ``Higgs transition'' which marks the end of the deconfined phase. Is there still anything distinct about a gauge theory in this regime? The conventional wisdom would say no: at the Higgs transition, the distinct features of gauge theory, such as gapless photons or topological order, disappear, and the resulting theory is trivial
~\cite{harlow2021symmetries}. This conclusion is reinforced by the classic study of Fradkin and Shenker~\cite{fradkin1979phase}. They considered a 2-parameter phase diagram, where one axis corresponds to the coupling to matter fields, controlling the amount of charge fluctuations, while the other axis is coupling to magnetic fluctuations (Fig.~\ref{fig:GaugeDualities}(c)). Tuning either coupling will eventually lead to a transition out of the deconfined phase; one is the aforementioned Higgs transition, driven by a proliferation of charges, and the other, conventionally referred to as ``confinement'', is driven by a proliferation of magnetic excitations (fluxes, or monopoles). As Fradkin and Shenker showed, while the two regimes, Higgs and confined, look physically quite different near the transition, they are in fact adiabatically connected to each other, making everything outside the confined phase into a single trivial phase. 

However, it has been pointed out recently, that this narrative needs to be refined when symmetries are properly taken into account~\cite{verresen2022higgs,thorngren2023higgs}. First, one should note that the pure gauge theory has not one but two higher-form symmetries; in pure electromagnetism, electric and magnetic field lines both need to form closed loops (the corresponding symmetry operators are generated by electric and magnetic flux through closed surfaces, respectively). While coupling to charges breaks the electric higher-form symmetry, it maintains the magnetic one, which is therefore present on both sides of the Higgs transition. Similarly, introducing magnetic mononpoles allows the magnetic field lines to have endpoints, but electric field lines remain closed. It is only when both couplings are non-zero (away from either axis of the phase diagram) that the higher-form symmetries are fully broken. 

As was pointed out in Refs. \onlinecite{verresen2022higgs,thorngren2023higgs}, in the regimes where one of the symmetries is present, the theory remains non-trivial, even outside the deconfined phase. In particular, these phases are examples of \emph{symmetry protected topological phases}~\cite{pollmann2010entanglement,chen2013symmetry,yoshida2016topological}; indeed, there is no way to adiabatically connect the Higgs and confined phases without breaking both higher-form symmetries along the way. Along with the higher-form symmetry of the gauge fields, these SPT phases also require a global $0$-form symmetry; in the Higgs phase, this is nothing but the original symmetry $G$ of the ungauged model, which acts on the matter fields. Thus, the Fradkin-Shenker phase diagram hosts three non-trivial phases: the deconfined phase, whose fixed point is the pure gauge theory, and two distinct SPTs, whose fixed points are in the limit of infinite coupling to either charge or magnetic fluctuations (with the other coupling kept at zero). 

\begin{shaded}
    In Sec.~\ref{sec:SPT} we discuss how to construct SPT phases from an arbitrary classical LDPC code, generalizing the ``decorated domain wall'' construction of SPTs~\cite{chen2014symmetry}.     
\end{shaded}


\subsection{Gauging the Ising model}\label{subsec:Ising}

\begin{figure} 
    \centering
    \includegraphics[trim={1cm 0 0 0},width = 1.\linewidth]{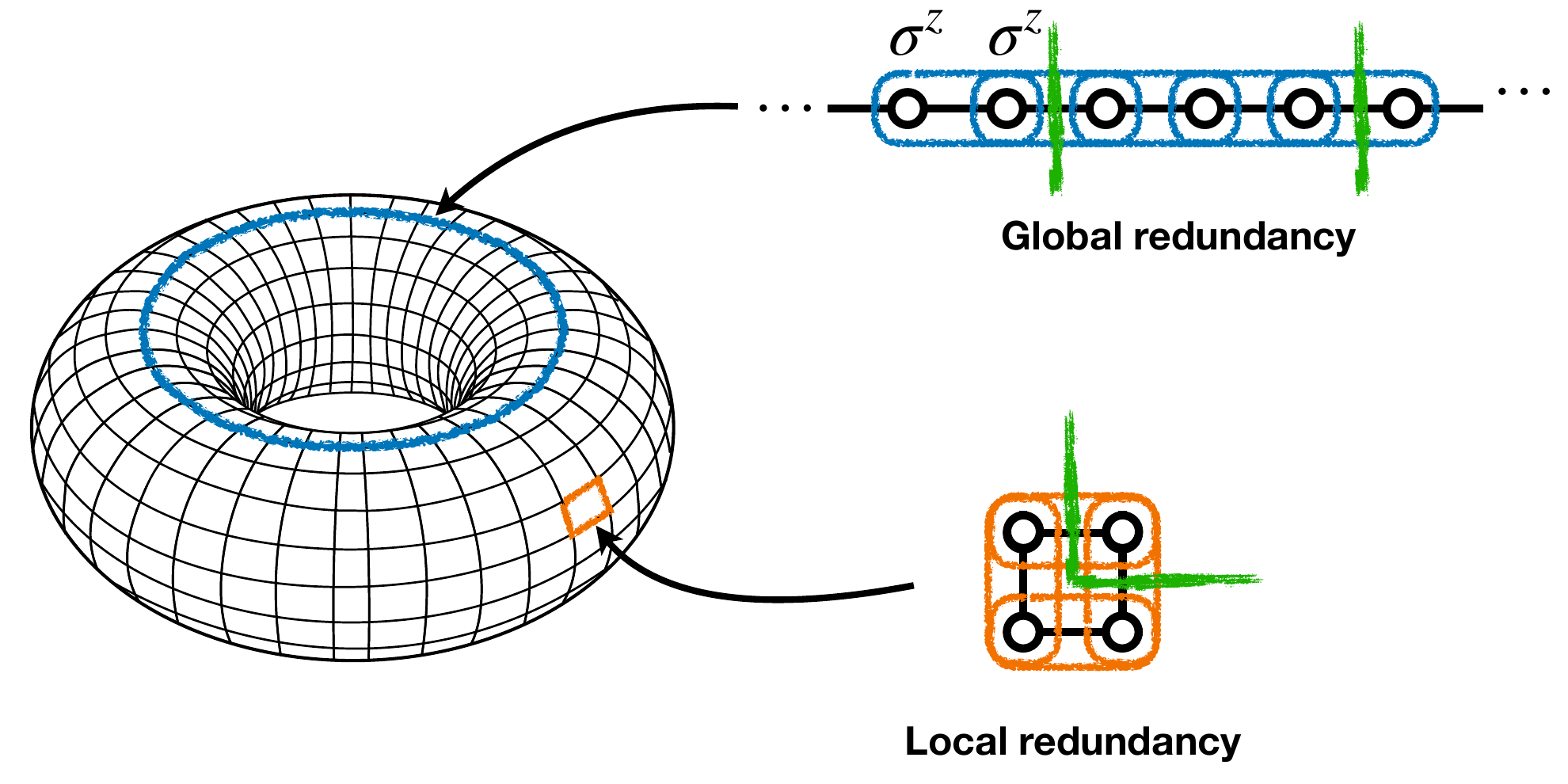}
    \caption{Illustration of the notion of local and global redundancies of a classical code in the examples of the Ising model on a torus.  Each edge hosts a two-body $\sigma_i^z\sigma_j^z$ interaction. Redundacies are products of the interactions (checks) which multiply to the identity.
    Global redundancies wrap around the torus, while local ones correspond to individual plaquettes. Domain walls (green lines) have to intersect redundancies an even number of times. For local redundancies, these enforce that the domain walls form closed loops.}
    \label{fig:IsingOnATorus}
\end{figure}

We now review how the ideas of the previous section are realized in what is arguably their simplest version, the $\mathbb{Z}_2$ lattice gauge theory obtained from gauging the Ising model. Here, we give a broad overview, while a more detailed discussion, illustrating the machinery developed in later sections, is presented in App.~\ref{app:Ising}. While most of this is fairly standard material~\cite{wegner1971duality,kogut1979introduction,levin2012braiding}, we will highlight various subtleties that will become important when we come to gauging arbitrary classical codes in Sec.~\ref{sec:Gauging}. 

To obtain the $\mathbb{Z}_2$ gauge theory, we start with the $D$-dimensional Ising model, defined by spins (qubits) on a $D$ dimensional hypercubic lattice. Nearest neighbors interact via the ferromagnetic Ising interaction $-J\sigma_i^z\sigma_j^z$, defined in terms of Pauli matrices $\sigma^z$ on sites $i$ and $j$; they are also subject to a transverse field $-h \sigma_i^x$. Both of these terms are invariant under a global Ising symmetry $\prod_i \sigma_i^x$, which flips all spins, and depending on the relative strength of the two, the system is either in the symmetric ($h \gg J$) or the symmetry broken ($h \ll J$) phase. 

The minimal coupling procedure for gauging the Ising symmetry consists of introducing a new variable, the gauge field $\tau_{ij}$, for every nearest neighbor bond of the lattice, and modifying the Ising interaction to $-J\sigma_i^z \tau_{ij}^z \sigma_j^z$. This is supplemented by a local Gauss law constraint, $G_i = \sigma_i^x \prod_{j \in N(i)} \tau_{ij}^x$, where the product runs over all bonds meeting at $i$. These commute with the Hamiltonian, ``making the global symmetry local" and allowing the spin on a site to be flipped at the cost of flipping the gauge fields adjacent on that site. This gauge freedom gives rise an equivalence relation on the combined configurations of $\sigma$ and $\tau$ variables. Two configurations related to each other by applications of $G_i$ are treated as physically equivalent. Their equivalence classes define the physical states, satisfying $G_i = +1$, which form the physical subspace.

The value of $\tau_{ij}^z$ changes the sign of the Ising terms in the Hamiltonian. These can be used to induce domain walls, which is seen most easily in one dimension: if we introduce a single bond with $\tau_{ij}^z=-1$, the set of ground states at $h=0$ will involve configurations with a pair of Ising domain walls, one located at the bond that was flipped, and the other at an arbitrary position along the chain. Note that the introduction of $\tau$ in this case can also be seen as a change of boundary condition, from periodic to anti-periodic. 

A similar picture holds in two dimensions when the bonds with $\tau^z=-1$ form closed loops on the dual lattice (these are called ``flat'' gauge field configurations): at their locations, one can freely create domain walls with no energy cost. One can think of this as the system being divided into a number of distinct patches, with the configuration of $\tau^z$ encoding the boundary conditions that determine how these various patches should be glued together~\cite{harlow2021symmetries,thorngren2018gauging,cheng2023lieb,barkeshli2019symmetry}. The gauge freedom comes about because one can introduce a new patch at the cost of flipping all the spins inside it. 

One can use this gauge freedom to get rid of the original variables and rewrite the theory entirely in terms of $\tau$ spins. This can be done by enforcing $\sigma_i^z = +1$ everywhere, in which case the Ising term turns into $-J\tau_{ij}^z$ while the transverse field is equivalent to $-h\prod_{j}\tau_{ij}^x$, due to Gauss's law. This procedure (coupling to background gauge fields and then eliminating $\sigma$ variables) maps the original Hamiltonian into a new Hamiltonian in terms of $\tau$ operators. In 1D, this is nothing but the celebrated Kramers-Wannier duality~\cite{kramers1941statistics}. This is usually described as a mapping from the Ising chain to itself, with the two coupling constants ($J$ and $h$) interchanged. However, as the above discussion shows, one should be careful: the dual model is really the original Ising chain in the presence of background gauge fields. This necessitates a careful treatment of boundary conditions~\cite{radicevic2018spin,li2023non}. 

So far, the $\tau$ variables we introduced are static background gauge fields, with no dynamics on their own, as can be seen from the fact that the gauged Hamiltonian commutes with $\tau_{ij}^z$. We can make them dynamical by adding additional terms, the simplest of which is $-\Gamma \tau_{ij}^x$. More generally, we can also consider other gauge-invariant terms and study the resulting phase diagram. Of particular importance is the \emph{deconfined phase}, which appears in dimensions $D \geq 2$ but not in $D=1$. In 2D, it is obtained by adding the ``plaquette term'' $-K \prod_{\msquare}\tau_{ij}^z$ to the Hamiltonian. This has the effect of energetically favoring flat gauge configurations. When all other terms commute with the plaquette term, it provides an example of the aforementioned 1-form symmetries. It ensures that at low energies the gauge fields must form closed loops, which originates from the domain walls of the original Ising model. The deconfined phase is characterized by a spontaneous breaking of this 1-form symmetry~\cite{gaiotto2015generalized,mcgreevy2023generalized}. 

Despite this, the deconfined phase is in fact stable even to perturbations that violate the plaquette terms and create open strings. When such perturbations are sufficiently weak, the symmetry is still present at low energies in an emergent way, making the deconfined phase absolutely stable~\cite{hastings2005quasiadiabatic,gaiotto2015generalized,mcgreevy2023generalized,pace2023exact}; the fixed-point limit of this phase is described by the toric code Hamiltonian. This should be contrasted with the one-dimensional case, where domain walls are point-like. As a consequence, the gauge theory only has conventional (0-form) symmetries and no absolutely stable phase. 

We can identify the source of this difference between 1D and 2D in the fact the in the former case, domain walls of the Ising model are point-like, while in 2D they form extended, loop-like objects. This, in turn, follows from the fact that the 2D Ising model exhibits \emph{local redundancies}: the product of Ising terms around a square plaquette gives the identity operator (see Fig.~\ref{fig:IsingOnATorus}), which forces the domain walls to form loops. This should be contrasted with the 1D Ising model, which only has a single, global redundancy (the product of \emph{all} Ising terms). In the gauge theory, this distinction shows up in that in 1D, only flat gauge field configurations exist.

Another regime where the 1-form symmetry plays an important role is the \emph{Higgs phase} of the gauge theory. This is obtained when the coupling constant $J$ becomes much larger than other scales. It has been realized recently that this regime can be understood as a symmetry protected topological (SPT) phase~\cite{verresen2022higgs}. That is, while the 1-form symmetry is unbroken in this regime, it is realized in a non-trivial way which leads to the appearance of \emph{edge modes} when the system is put on a lattice with open boundaries. The same regime on the 1D gauge theory corresponds to the cluster chain, a canonical example of an SPT (in that case, protected by 0-form symmetries)~\cite{raussendorf2001one}. 

\section{Classical and quantum stabilizer codes}\label{sec:CodeReview}

In this section, we review various definitions relevant for classical and quantum codes. We do so from a physical perspective, associating a  Hamiltonian to each code and emphasizing the physical interpretation of various code properties. 

\subsection{Classical LDPC codes}\label{subsec:classicalLDPC}

To begin, let us consider a classical code, 
defined in terms of $n$ binary variables\footnote{To make contact with statistical physics, we use `spin' variables $\sigma=\pm 1$, rather than $\sigma = 0,1$ as would be more natural from the coding perspective.} $\sigma_i = \pm 1, \,\, i = 1,\ldots, n$. We will refer to these as bits or spins, interchangeably. The code is defined by a set of \emph{parity checks}, $C_a \equiv \prod_{i \in \delta(a)} \sigma_i$, each one specified by some subset of bits $\delta(a)$, labeled by $a = 1,\ldots,m$. 
The \emph{codewords} are those spin configurations that satisfy the conditions $C_a = + 1 \, \forall a$. These are the ground states of the classical Hamiltonian
\begin{equation}\label{eq:H_classical}
    H_\text{cl} = - \sum_a C_a = - \sum_a \prod_{i\in \delta(a)} \sigma_i.
\end{equation}

One useful language that we will make use of throughout the paper is to represent spin configurations as vectors in an $n$-dimensional linear space over binary field $\mathbb{Z}_2$. That is, we assign to each $i$ a basis vector $\ket{i}$ and consider all linear combinations of these, $\sum_i \alpha_i \ket{i}$, with $\alpha_i = 0,1$.  We can think of each such vector as corresponding to the subset of spins with $\alpha_i = 1$, with vector addition corresponding to the symmetric difference of subsets. Every check corresponds to a vector, defined by the subset $\delta(a)$, and the \emph{code subspace}, formed by all the codewords, is the orthogonal complement of the subspace spanned by all the checks. The dimension of this subspace is denoted by $k$, which means that the Hamiltonian of Eq.~\eqref{eq:H_classical} has $2^k$ ground states. Such a code encodes $k$ bits of classical information. 

By construction, $\sigma_i = +1 \, \forall i$ (the ``all up'' state) is always a codeword. We can define the \emph{code distance} $d$ as the smallest number of spins that need to be flipped in order to get to a different codeword\footnote{By linearity, this is also the same as the minimal number of spin flips -- known as the Hamming distance -- to go between \emph{any} two codewords.}. Each codeword forms a $\mathbb{Z}_2$ vector $\ket{\Sigma} = \sum_{i\in\Sigma} \ket{i}$. We can associate to this a \emph{logical operator} $\mathcal{X}_\Sigma$ that flips the sign of all spins in $\Sigma$, thus mapping from one ground state / codeword to another. By construction, these logical operators (or logicals, for short) leave all checks $C_a$ unchanged and are therefore symmetries of $H_\text{cl}$, which are spontaneously broken at zero temperature. 

Let us pick a basis of codewords, $\ket{\Sigma_\lambda}$, labeled by $\lambda = 1,\ldots,k$. We can also consider a dual set of vectors, $\ket{\Theta_\lambda}$, such that $\braket{\Theta_\lambda|\Sigma_{\lambda'}} = \delta_{\lambda\lambda'}$. To these, we can associate the combination $\mathcal{Z}_\lambda = \prod_{i \in \Theta_\lambda} \sigma_i$; this has the property that it is flipped by $\mathcal{X}_\lambda$ and none of the other logicals. As such, its expectation value serves as an order parameter for the breaking of this particular symmetry. We can always choose the basis such that $\Theta_\lambda$ corresponds to a single site\footnote{To see this, start with some arbitrary basis $\ket{\Sigma_\lambda}$ and choose $i_1 \in \Sigma_1$. If any other $\Sigma_{\lambda \geq 2}$ contains $i_1$ in its support, redefine them by adding $\ket{\Sigma_1}$ (which corresponds to the multiplication of the logicals). This gives a new basis, where $\ket{i_1}$ is orthogonal to all the basis elements other than $\ket{\Sigma_1}$. This can be repeated iteratively until we construct a set $\{\ket{i_\lambda}\}$.} $i_\lambda$, making these local order parameters. The values of $\{\sigma_{i_\lambda}\}_{\lambda=1}^{k}$ can be used to label the $2^k$ distinct ground states; in the coding language, we can think of them as the $k$ bits of information that we are encoding into our $n$-bit code.

We thus have a code characterized by a triple of integers, $[n,k,d]$. We will be interested in families of codes where the number of bits $n$ can be increasing without bound, in order to take a ``thermodynamic limit''. In that case, one can ask how $k$ and $d$ scale as $n$ is increased. In particular, in order for the information to be well-protected from noise, one would generically want $d$ to scale with $n$ in some non-trivial way. In a physics context, different scalings of $d$ with $n$ can distinguish e.g. global symmetries and various kinds (planar, fractal, etc.) of subsystem symmetries. 

While the definition so far is completely general, one often wants to impose certain restrictions on it in terms of locality. This is certainly reasonable if one thinks of codes as physical systems, but also often useful from a coding perspective. For the particular and much-studied class of LDPC codes, the LDPC condition can be informally phrased as ``every bit talks to finitely many bits''. More precisely, we want both that a) the support of each check, $\delta(a)$ is finite, independent of $n$, and b) that each bit $i$ appears in only finitely many of the sets $\delta(a)$. Clearly, this definition is satisfied by local interactions on any finite-dimensional Euclidean lattice, but, as we will see, is more general, allowing for more general codes. In what follows, we will focus entirely on LDPC codes.

\subsubsection{Examples}

Before moving on, we list some examples of known models of classical statistical mechanics that fit into the definition of an LDPC code given above. 

\begin{figure} 
    \centering
    \includegraphics[trim={1cm 0 0 0},width = 1.\linewidth]{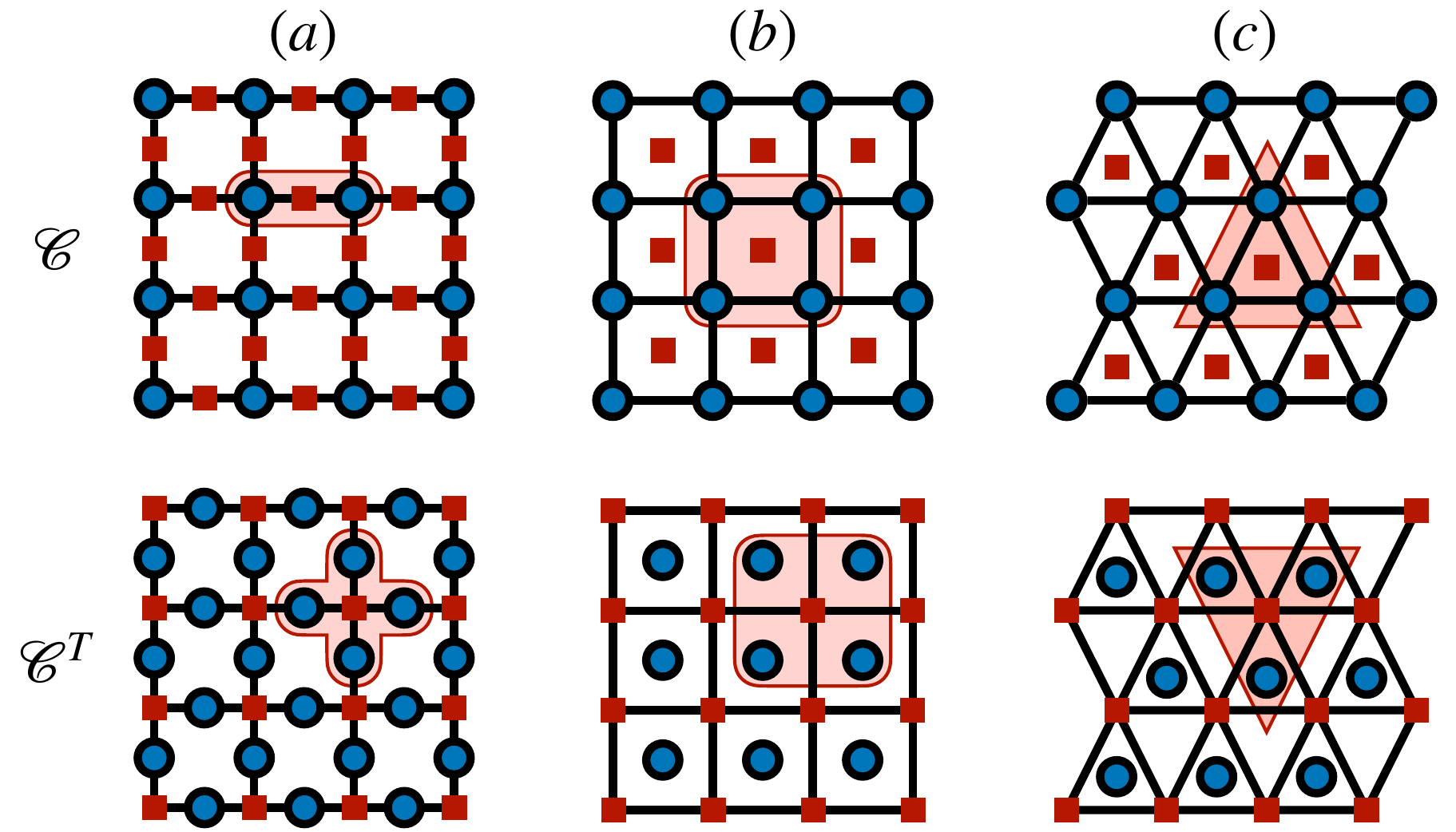}
    \caption{Three examples of classical codes living on two-dimensional lattices. Blue circles represent bits/spins and red squares represent checks. (a) Ising model 2-body checks assign to each edge of the square lattice. (b) Plaquette Ising model, with 4-body checks assigned to every square plaquette. (c) Newman-Moore model with 3-body checks assigned to every upward triangle. The lower row shows the corresponding transpose codes which interchanges the roles of bits and checks; in the latter two cases they are isomorphic to the original, while in (a), the checks of the transpose code are ``star terms'' acting on four edges meeting at a site.}
    \label{fig:ClassicalExamples}
\end{figure}

\paragraph{Ising model.} We consider the classical Ising model in $D$ dimensions (Fig.~\ref{fig:ClassicalExamples}(a)). The spin labels $i$ correspond to sites of a $D$-dimensional lattice of linear size $L$ in each direction, and the subsets $\delta(a)$ are given by all nearest neighbor (nn) pairs of $(i,j)$ i.e. defined by the edges of the lattice. There are two codewords (``all up" and ``all down") related by a single global symmetry corresponding to flipping all spins. Hence, these are $[n,k,d] = [L^D,1,L^D]$ codes. 

\paragraph{Plaquette Ising model} Another example is the plaquette Ising model~\cite{vijay2016fracton} (Fig.~\ref{fig:ClassicalExamples}(b)), on a $D\geq 2$ dimensional cubic lattice of linear size $L$. Here $i$ still labels sites, but $\delta(a)$ now corresponds to all the square plaquettes, formed by $4$ sites each. These have subsystem symmetries corresponding to flipping all spins along a $D-1$ dimensional plane, making them $[L^D,D(L-1)+1,L^{D-1}]$ codes. 

\paragraph{Newman-Moore model} Originally considered as an example of a classical spin glass without disorder~\cite{newman1999glassy}, the Newman-Moore model (Fig.~\ref{fig:ClassicalExamples}(c)), has spins assigned to sites of a two-dimensional triangular lattice and checks correspond to the product of three spins that form an upward pointing triangle. The logicals take the shape of Sierpinski triangles and we have $[n,k,d] = [L^2, O(L), O(L^{\eta})]$, where $\eta = \log{3}/\log{2}$ is the fractal dimension of the Sierpinski triangle and the precise values of $k$ and $d$ depend sensitively on the choice of $L$\footnote{For example, while the scaling $k = O(L)$ is true for generic choice of $L$, for $L= 2^p$, we end up with $k=0$.}. 

\subsubsection{Tanner graph representation}

\begin{figure} 
    \centering
    \includegraphics[trim={1cm 0 0 0},width = 1.\linewidth]{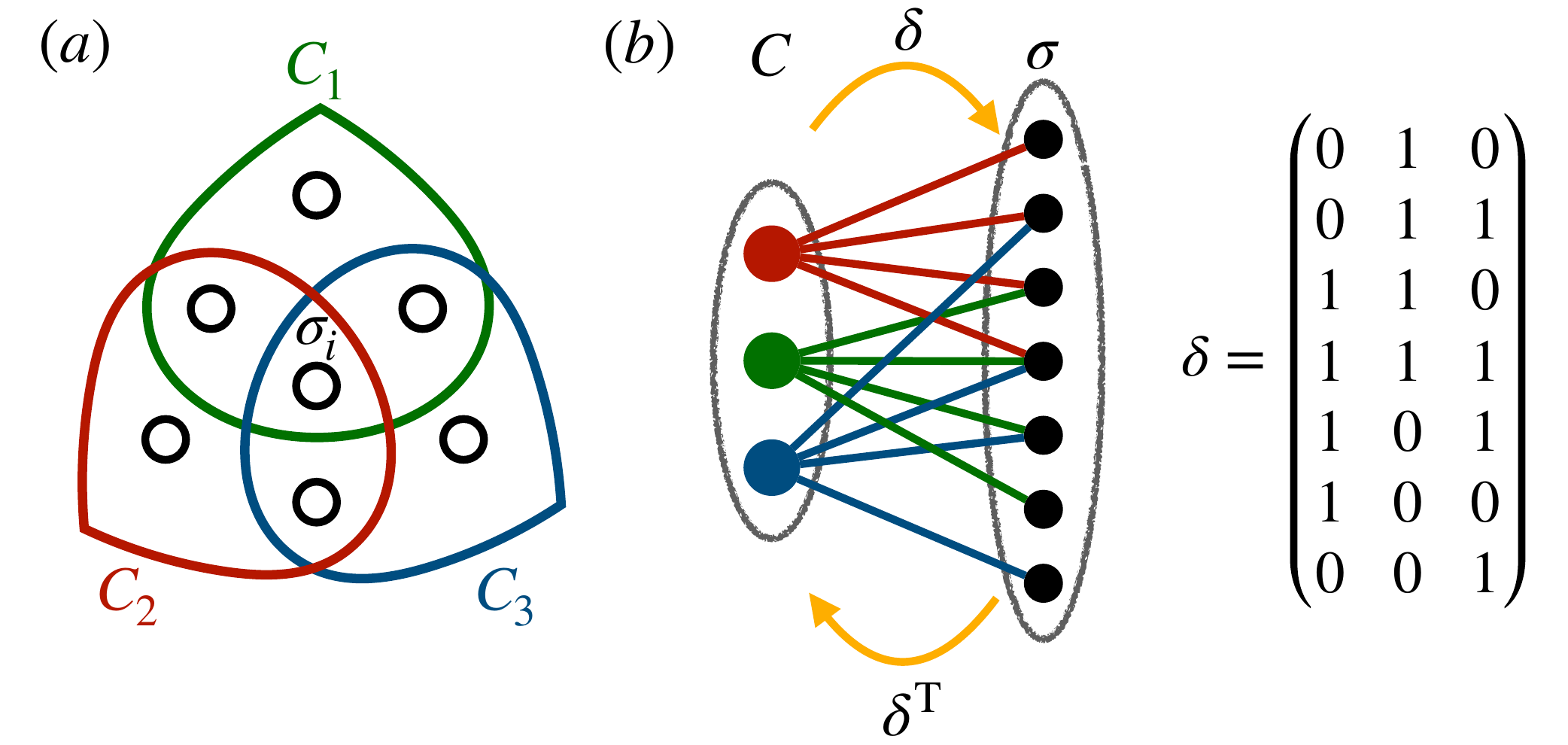}
    \caption{Representations of classical codes, illustrated on the example of the [7,4,3] Hamming code. (a) as a hypergraph, where each check $C_a$ corresponds to a collection of bits involved in it. (b) As a bipartite Tanner graph where one set of nodes corresponds to bits and the other to checks. $\delta$, the adjacency matrix of the tanner code, defines a map between two linear spaces.
    }
    \label{fig:Hypergraph}
\end{figure}

There are a number of useful `geometric' representations of classical codes which we will use below. One is to think of it as a \emph{hypergraph}. A hypergraph is defined by a set of vertices $V$ and a set of hyperedges $E$, where each hyperedge $e \in E$ is a collection of some number of vertices $e\equiv (v_1,v_2,\ldots,v_m)$ with $v_i \in V$. This generalizes the notion of a graph where every edge is a pair of two vertices and in the following we will abuse notation and simply refer to $e$ as an edge. Now, given a classical code $\mathcal{C}$, we take each variable to define a vertex labeled by $i$ and each check $C_a$ a hyperedge, corresponding to the subset $\delta(a)$, as illustrated by Fig.~\ref{fig:Hypergraph}(a). Another, equivalent, representation the so-called \emph{Tanner graph}, which is a bi-partite graph associated to the code. One defines two sets of nodes, $V_0$ and $V_1$, labeled by $i$ (bits) and $a$ (checks), respectively, and draws an edge between the pair $(i,a)$ iff $i \in \delta(a)$, i.e. if spin $i$ is part of the check $C_a$. See Fig.~\ref{fig:Hypergraph}(b).

A more algebraic perspective is given by considering both bits and checks as forming a linear space over the binary field $\mathbb{Z}_2$. We already described this for the bits, which form the space $\mathbb{Z}_2^n$. Similarly, we define a vector space of checks, $\sum_a \beta_a \ket{a}$, $\beta_a = 0,1$, forming a linear space $\mathbb{Z}_2^m$. We can then think of $\delta$ as a linear map between these spaces, mapping a basis vector $\ket{a}$ to the vector $\sum_{i\in\delta(a)} \ket{i}$ and extending linearly to all linear combinations of $\ket{a}$; i.e., to any set of checks it associates the set of spins acted upon by their product. This linear map has a representation by a binary matrix, $\delta_{ia} = 1$ iff $i \in \delta(a)$. This is nothing but the (bi)adjacency matrix of the Tanner graph.

There is also a transpose map, $\delta^T$, which maps from subsets of bits to subsets of checks. Physically, $\delta^T(N)$ is the set of checks that change sign when the subset of spins in $N$ are flipped. This perspective allows for a very simple description of the logical operators: $\ket{\Sigma}$ are precisely those vectors that are in the kernel of the transpose matrix, $\text{Ker}(\delta^T)$, and $k$ is the dimension of the subspace defined by this condition. 

What about $\text{Ker}(\delta)$? An element of this kernel correspond to some set of checks that, when multiplied together, act on none of the bits. In other words, they are \emph{redundancies}: $\prod_{a \in R} C_a = +1$ for $R \in \text{Ker}(\delta)$. Just like the logicals, the redundancies also form a linear space and one could pick a basis set $R_r$ labeled by some integers $r$. 

A natural object to consider from this perspective is the \emph{transpose code}, $\mathcal{C}^T$, defined by interchanging $\delta$ and $\delta^T$. In terms of the Tanner graph, this simply corresponds to switching the labels of the two sets of vertices: bits become checks and vice versa. This is illustrated in the second row of Fig.~\ref{fig:ClassicalExamples}. Moreover, taking the transpose also exchanges logicals and redundancies.

We will make use of transpose codes when we discuss Kramers-Wannier dualities in Sec~\ref{subsec:KramersWannier}. We will then combine it with the notion of redundancies to introduce background gauge fields in Sec.~\ref{subsec:BackgroundGauge}. Later, in Sec.~\ref{sec:Redundancies}, we will distinguish between \emph{local} and \emph{global redundancies} (see also Fig.~\ref{fig:IsingOnATorus}) and discuss the different roles they play in determining properties of the code. 

\subsection{Quantum CSS codes}\label{subsec:CSS}

We now come to consider quantum stabilizer codes\footnote{We focus on stabilizer codes throughout this paper. We briefly discuss how some of the ideas here generalize to subsystem codes in App.~\ref{app:subsystem}.}, defined on a set of $n$ qubits, labeled by $a=1,\ldots,n$, with Pauli operators $\tau_a^{x,y,z}$ acting on them\footnote{The choice of notation in this section---e.g. using $a$ to label qubits---is made to harmonize with the discussion in Sec.~\ref{sec:Gauging} below.}. In particular, we focus on so-called \emph{CSS codes}\footnote{We note that this restriction is unimportant in the sense that any $[[n,k,d]]$ quantum stabilizer code can be mapped onto a $[[4n,2k,2d]]$ CSS code without changing its locality (i.e., LDPC) properties~\cite{bravyi2010majorana}.}, named after Calderbank, Steane and Shor~\cite{calderbank1996good,steane1996multiple}. These are defined by a series of checks, which now come in two flavors: $X$- and $Z$-checks, which we will label by $i=1\ldots,m_X$ and $p=1,\ldots,m_Z$ respectively. Consequently, we now need two functions, $\delta_1$ and $\delta_2$, that define the set of spins on which a given check is acting on. E.g., $X$-checks take the form $A_i \equiv \prod_{a \in \delta_1^T(i)} \tau_a^x$. Similarly, a $Z$-check takes the form $B_p \equiv \prod_{a \in \delta_2(p)} \tau_a^z$. Note that we have defined $\delta_1$ through its transpose; i.e., $\delta_1$ as a map goes from bits to checks. The reason for this will become clear in a moment.

For a CSS stabilizer code, we require that all checks commute, $[A_i,B_p] = 0$ for any pair $(i,p)$. With this condition, we can consider common eigenstates of all checks, $A_i\ket{\psi} = B_p \ket{\psi} = \ket{\psi} \, \forall i,p$. These states form a $2^k$ dimensional \emph{code subspace} $\mathcal{H}_\text{code}$. Just like in the classical case, it is possible to combine all the checks into a Hamiltonian
\begin{equation}\label{eq:H_quantum}
    H_\text{q} = -\sum_i A_i - \sum_p B_p = -\sum_{i} \prod_{a \in \delta_1^T(i)} \tau_a^x - \sum_{p} \prod_{a\in \delta_2(p)} \tau_a^z,
\end{equation}
which has $\mathcal{H}_\text{code}$ as its ground state subspace. 

One useful perspective is to work in the $\tau_a^z$ basis and consider the $B_p$ as checks of a classical code $\mathcal{C}_Z$. The $A_i$ act as logical operators of this code and diagonalizing them selects out particular superpositions of the codewords of $\mathcal{C}_Z$. In particular, a set of basis states in $\mathcal{H}_\text{code}$ is formed by taking equal weight superpositions of all codewords of the classical code that are connected to each other by applications of $X$-checks. Of course, we could just as well work in the $\tau_a^x$ basis, in which we would consider $A_i$ as checks of a different classical code $\mathcal{C}_X$ which has $B_p$ among its logicals. 

The quantum CSS code has its own set of logical operators, which perform non-trivial operations within $\mathcal{H}_\text{code}$; these once again come in two flavors. A logical $Z$ operator, $\mathcal{Z}_\Gamma = \prod_{a \in \Gamma} \tau_a^z$ leaves the code subspace invariant, but acts on it non-trivially. In order to do so, it needs to satisfy two conditions: it should commute with all the $A_i$ (otherwise it would take us out of the code subspace), but it should not be a product of $B_p$ checks (otherwise it would act simply as the identity on $\mathcal{H}_\text{code}$). Logical $X$ operators, $\mathcal{X}_{\tilde{\Gamma}} = \prod_{a\in \tilde{\Gamma}} \tau_a^x$, are defined analogously. There are $k$ independent logicals of each type and we can choose bases $\Gamma_\lambda$, $\tilde{\Gamma}_\lambda$, labeled by $\lambda=1,\ldots,k$, such that $\mathcal{X}_\lambda$ anticommutes with $\mathcal{Z}_\lambda$ and commutes with all other logicals; in terms of $\mathbb{Z}_2$ vectors, this amounts to $\braket{\tilde{\Gamma}_\lambda|\Gamma_{\lambda'}} = \delta_{\lambda\lambda'}$.

We can define the $X$- and $Z$-distances of the code, $d_X$ and $d_Z$, as the smallest \emph{Pauli weight} (i.e., number of qubits being acted upon) that a logical operator of each type can have. The overall code distance is the smaller of the two: $d \equiv \min(d_X,d_Z)$. To distinguish them from classical codes, the parameters of a quantum code are written in double brackets: $[[n,k,d]]$.

By construction, the logical operators all commute with $H_\text{q}$ defined in Eq.~\eqref{eq:H_quantum} and are therefore symmetry operators. However, the physical interpretation of these symmetries is somewhat different from the classical one. A code distance that diverges in the $n\to\infty$ limit means that the different states within $\mathcal{H}_\text{code}$ are indistinguishable by any local observable. As such, they do not manifest spontaneous symmetry breaking in the usual sense, but rather 
some form of \emph{topological order}. This distinction is tied to a difference in the nature of the symmetries. In the classical case, these are ``rigid'', acting on a specified set of spins, i.e. some particular subsystem. The symmetries of $H_\text{q}$, on the other hand, are ``deformable'': we can multiply $\mathcal{X}_\Gamma$ by some check $A_i$ to get a different operator that acts the same way on the code subspace. In the physics context, such deformable symmetry operations naturally arise in the context of \emph{higher form} symmetries\footnote{These are most naturally defined in continuum quantum field theory. In that context, such freely deformable operators are themselves called ``topological''~\cite{gaiotto2015generalized}}~\cite{gaiotto2015generalized,mcgreevy2023generalized}. Indeed, as discussed earlier, it has been realized that (Abelian) topological order can be reformulated as spontaneous symmetry breaking of higher form symmetries~\cite{mcgreevy2023generalized}. We can thus think of the classical and quantum codes as both exhibiting SSB, but for different kinds of symmetries. 

\begin{figure} 
    \centering
    \includegraphics[trim={1cm 0 0 0},width = 0.95\linewidth]{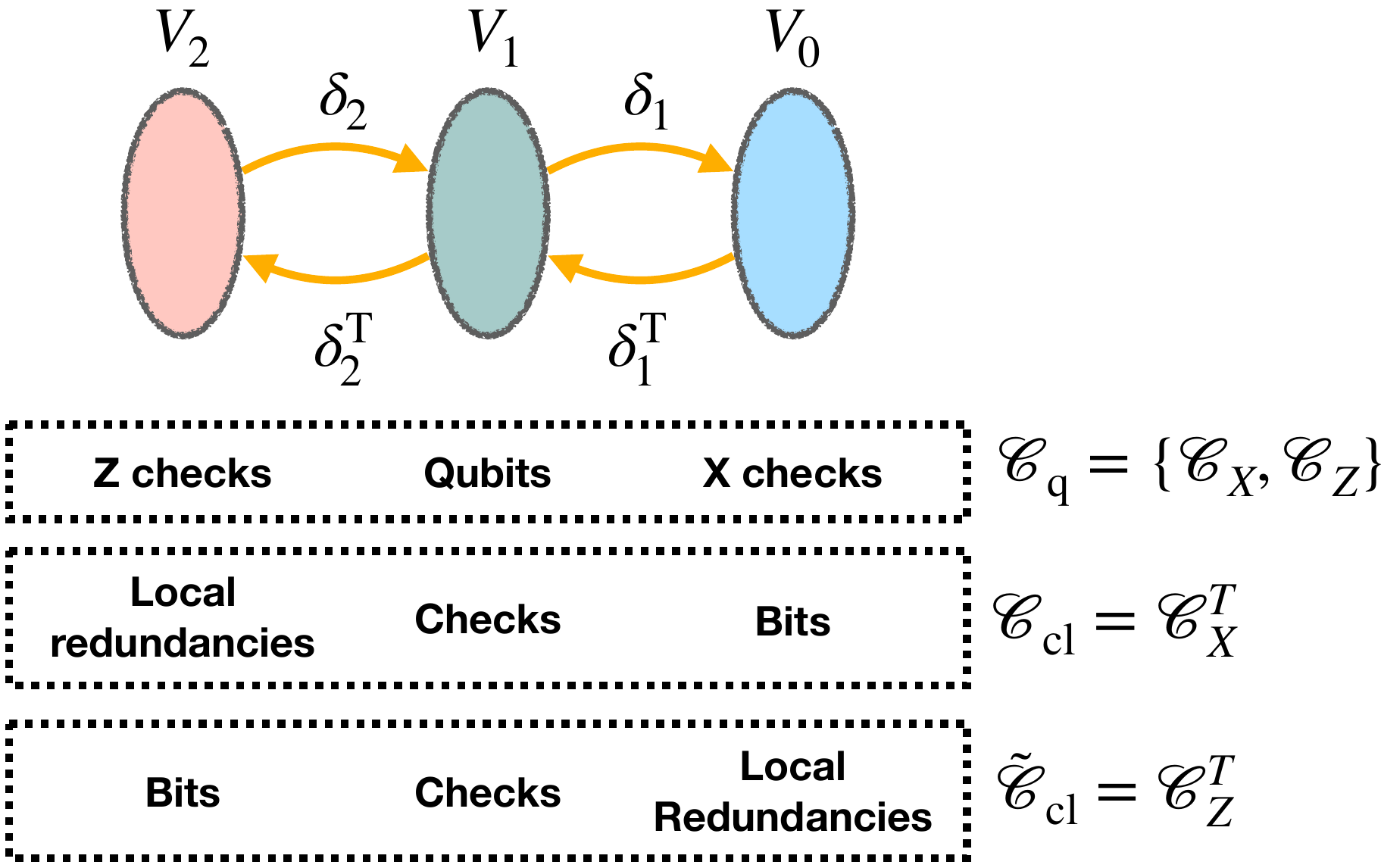}
    \caption{A 2-dimensional chain complex and its three associated codes: a quantum CSS code and two classical codes with local redundancies.
    }
    \label{fig:2complex}
\end{figure}

\subsubsection{Examples}

\paragraph{Toric code.} A simple example is given by the toric code on a $D\geq 2$ dimensional cubic lattice with periodic boundary conditions, where qubits are assigned to edges, $X$-checks to sites and $Z$-checks to plaquettes, such that $A_i$ acts on all edges meeting at site $i$ and $B_p$ on all edges forming the plaquette $p$. These form a code with $[[n,k,d]] = [[DL^D,D,L]]$, with the logical operators taking the shape of loops wrapping around the torus in one of the two directions.

\paragraph{$X$-cube model.} Another example is the $X$-cube model on a $D\geq 3$ dimensional cubic lattice\footnote{The definition of the $X$-cube model in Ref. \onlinecite{vijay2016fracton} can be obtained from the one we employ here by going to the dual lattice.}, where qubits are placed on plaquettes, $X$-checks on sites and $Z$-checks on cubes. $A_i$ acts on all plaquettes that touch site $i$. The definition of $B_p$ is more complicated: there are three of them for every cube, each one acting on $4$ qubits, involving the faces of the cube whose normal vectors are orthogonal to a particular lattice direction (e.g. $B^x$ involves plaquettes in the $x-y$ and $x-z$ directions etc). This is a $[[L^D D(D-1)/2,D(2L-1),L]]$ code. Logicals are again line-like, stretching through the lattice in one of the three directions.

\paragraph{Haah's code.} Haah's cubic code~\cite{haah2011local} is also defined on a 3D cubic lattice. There are two qubits on each site and $X$ and $Z$ checks are both associated to cubes, each involving 8 Pauli operators along the corners of the cube. There are $k = O(L)$ logicals (with the precise number being a complicated function of $L$) and the while the precise scaling of the code distance is unknown, it satisfies $d > L$ and $d \leq O(L^2)$.

\subsubsection{Chain complex representation}

As in the classical case, quantum CSS codes have a very useful geometric interpretation in terms of a \emph{chain complex}~\cite{kitaev2003fault,bombin2007homological,bravyi2014homological}. To draw an analogy with the Tanner graph representation of classical codes, we now represent the elements of the code as a tri-partite graph, formed by the three sets of vertices associated to $X$-checks ($V_0$), qubits ($V_1$) and $Z$-checks ($V_2$), respectively. Just like in the classical case, we can think of each of these as forming a linear space over the binary field. $\delta_{1,2}$ that specify the support of checks can then be thought of as maps between these spaces: $\delta_2: V_2 \to V_1$ and $\delta_1: V_1 \to V_0$ (see Fig.~\ref{fig:2complex}). The commutativity of checks, which implies  that the two subsets $\delta_1^T(i)$ and $\delta_2(p)$ should always overlap on an even number of qubits, has a particularly simple form in this language:
\begin{equation}\label{eq:ChainComplex}
    \delta_1 \delta_2 = 0,
\end{equation}
where matrix multiplication should be understood modulo $2$. What Eq.~\eqref{eq:ChainComplex} implies is that the objects $V_{q=0,1,2}$ and the maps $\delta_{q=1,2}$ together form a \emph{two-term chain complex} (see App.~\ref{app:ChainComplex}), or $2$-complex for short. Elements of $V_q$ can be thought of as $q$-dimensional objects, while $\delta_q$ is a \emph{boundary map} that map an $q$-dimensional object onto its $(q-1)$-dimensional boundary. Eq.~\eqref{eq:ChainComplex} then reads as ``the boundary of a boundary is zero'', which is the defining property of a chain complex.

Just as in the classical case, logical operators of the CSS code have simple descriptions in terms of the maps $\delta_{1,2}$, which can be given a geometrical interpretation in terms of the chain complex. For example, $\Gamma$ corresponding to the support of a $Z$-logical satisfies $\Gamma \in \text{Ker}(\delta_1) \setminus \text{Im}(\delta_2)$. In the chain complex language, elements of $\text{Ker}(\delta_1)$ are \emph{cycles}, i.e. closed loops with no boundaries, while $\Gamma \notin \text{Im}(\delta_2)$ means that these cycles do not arise as boundaries of some higher dimensional object, making them \emph{non-contractible}. The equivalence classes of $Z$-logicals, which have the same effect on $\mathcal{H}_\text{code}$ (i.e., they are related to each other by multiplication with $Z$-checks) correspond to the \emph{homology classes} of the chain complex, forming the homology group $\text{Ker}(\partial_1) / \text{Im}(\partial_2)$. Similarly, $X$-logicals correspond to on-contractible \emph{cocycles}, $\tilde{\Gamma}\in\text{Ker}(\delta_2^\text{T}) \setminus \text{Im}(\delta_1^\text{T})$, which form \emph{cohomology classes}, $\text{Ker}(\delta_2^\text{T}) / \text{Im}(\delta_1^\text{T})$. 

\subsection{Expander graphs and good codes}\label{subsec:expander}

While the definition of an LDPC code encompasses spatially local interactions, it goes beyond it, allowing for models in more general geometries that cannot be fit into Euclidean space in any finite dimension. A particularly important example is given by so-called \emph{expander graphs}~\cite{hoory2006expander}, which feature prominently in the construction of LDPC codes, both quantum and classical~\cite{sipser1996expander,breuckmann2021quantum}.

Expanders have a number of (ultimately equivalent) definitions; the conceptually simplest is that of a \emph{vertex expander}: we say that a graph with vertices $V$ is a $(\gamma,\alpha)$ vertex expander if for any set of vertices $A \subset V$, such that $|A| \leq \gamma |V|$, we have $|N(A)| \geq \alpha |A|$, where $N(A)$ is the set of vertices neighboring $A$. An \emph{edge expander} is defined similarly, with the number of neighbors replaced by the number of edges connecting $A$ to its complement. A third, related concept is that of \emph{spectral expansion}, which is measured by the gap between the two largest eigenvalues of the adjacency matrix. This is connected to the other notions by Cheeger's inequality~\cite{hoory2006expander}, which upper bounds the constant $\alpha$ in terms of the spectral gap. 

\begin{figure} 
    \centering
    \includegraphics[trim={1cm 0 0 0},width = 1.\linewidth]{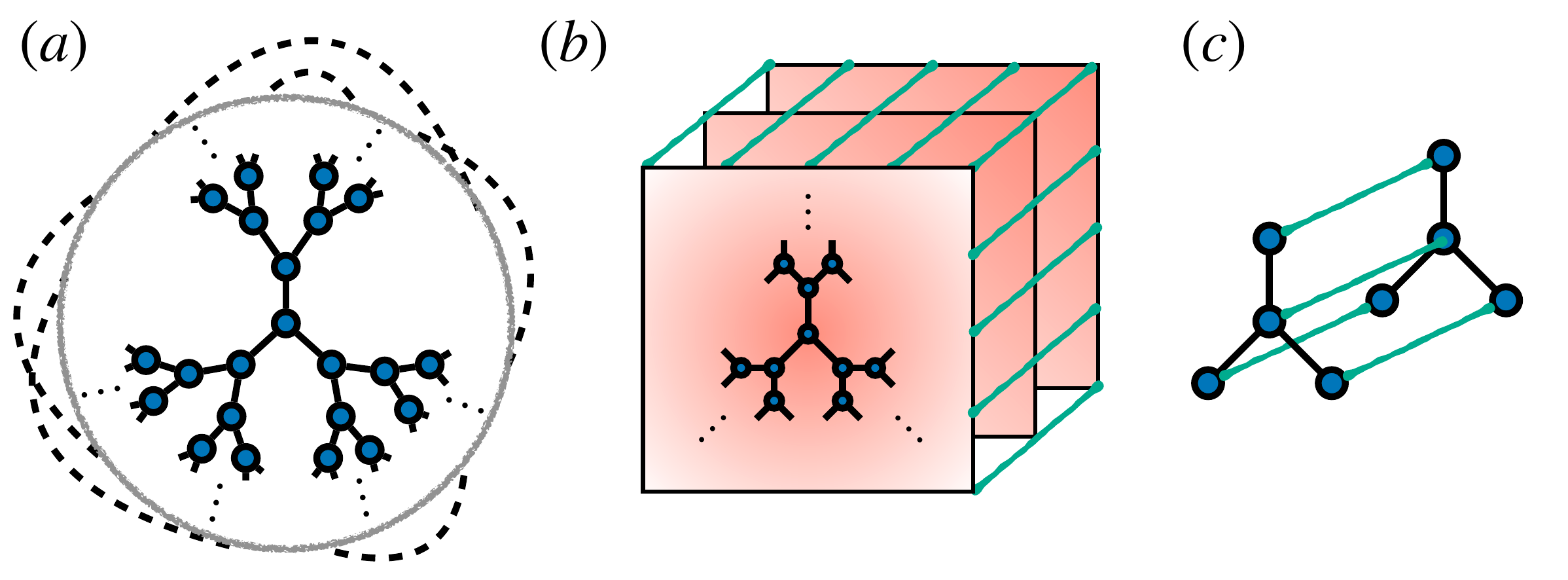}
    \caption{Codes on non-Euclidean graphs. (a) Sketch of an expander graph construction: locally, the graph resembles a tree (Bethe lattice), but has additional edges connecting the endpoints of the tree (dashed lines), making it expanding everywhere. (b) One can introduce short loops into this graph by taking multiple copies of it and adding ``vertical'' edges between the copies. This provides a simple example of a product construction~\cite{tillich2013quantum}. (c) Shows a local view of two subsequent copies, which have square-like plaquettes.}
\label{fig:ExpanderExamples}
\end{figure}

There are various constructions of expanders with a bounded degree. One can show that certain classes of random regular graphs satisfy the definition with high probability~\cite{hoory2006expander}. A more structured example is given by the Cayley graphs of certain discrete groups. Locally, these look like a tree graph (i.e., the Bethe lattice), as in the example shown in Fig.~\ref{fig:ExpanderExamples}(a). However, there are additional edges connecting the endpoints of the three in such a way as to make all vertices equivalent (see Fig. 1 in Ref. \onlinecite{lubetzky2016cutoff} for an illustration). These  provide examples of so-called \emph{Ramanujan graphs}, which are optimal spectral expanders~\cite{lubotzky1988ramanujan}.    

Why consider codes on such exotic geometries? One reason is given by the Bravyi-Terhal-Poulin (BTP) bounds~\cite{bravyi2010tradeoffs}. These concern geometrically local codes in $D$ dimensional Euclidean lattices and put upper bounds on the properties $k$ and $d$ that can be achieved in this setting. There is a separate bound for classical and quantum stabilizer codes, which read 
\begin{align}
    k d^\frac{1}{D} \leq O(n) & & \text{classical}, \label{eq:BravyiTerhal1} \\
    k d^{\frac{2}{D-1}} \leq O(N) & & \text{quantum}. \label{eq:BravyiTerhal2}
\end{align}
The former is saturated e.g. by the 1D Ising model, while the latter is by the toric code in two dimensions, both of which achieve an optimal scaling of $d$ at the cost of keeping $k$ at a finite, $n$-independent value. Thus Eqs.~\eqref{eq:BravyiTerhal1}-~\eqref{eq:BravyiTerhal2} put restriction on what code properties are obtainable in finite dimensions. 

Can codes on more general graphs, such as expanders, beat the Bravyi-Terhal-Poulin bounds? The answer is yes! In fact, it is possible to achieve the optimal scaling of \emph{both} code rate and code distance simultaneously, such that $k,d = O(n)$. Codes that achieve this are called \emph{good codes} in the literature. In the classical case, constructions of such good LDPC codes based on expander graphs have existed for a long time~\cite{sipser1996expander}. In the quantum case, on the other had, whether it is possible to achieve a scaling much better than $d = O(n^{1/2})$ remained an open problem for a long time, until a series of breakthrough results~\cite{hastings2021fiber,breuckmann2021balanced,panteleev2022asymptotically} in the last couple of years have proven that good quantum LDPC codes also exist. 

In the classical case, one way of constructing good codes is by making the Tanner graph of the code itself a (sufficiently good) expander. For example, consider a code where every bit is involved in $z$ distinct checks ($|\delta^T(i)| = z$, $\forall i$), whose Tanner graph is a $(\gamma,\alpha)$ vertex expander. It is easy to show that, if $\alpha > z/2$, then for any subset $A$ of bits, the number of checks they violate must grow linearly with $|A|$ and thus $d \geq \gamma n$. One way of constructing such Tanner graphs is by taking a random permutation on $c_1 n$ elements (for some integer $c_1$), represented as a bi-partite graph, and then fusing vertices together into groups of $c_1$ on the left, and groups of $m/(nc_1)$ on the right~\cite{gallager1962low}.

Another approach is the so-called \emph{Tanner code} construction (not to be confused with the Tanner graph). In this case, one takes as the starting point some graph $G$, which provides an underlying geometry. On every edge of this graph, one places a spin. Checks are associated to the vertices, and every check is acting on some subset of the edges adjacent to it. The same vertex can host multiple different checks which together form a ``small code'' on the adjacent bits. This way, one can construct an entire family of codes associated to the same graph, corresponding to different choices of the small codes. One can then prove a lower bound on the resulting code distance, which depends on both the spectral expansion of the graph, and properties ($k$ and $d$) of the small codes~\cite{sipser1996expander}. In particular, one can construct good classical codes on Ramanujan graphs~\cite{lubotzky1988ramanujan} of sufficiently high degree. 

In the quantum case, obtaining good codes has been an unsolved problem until the recent series of  breakthrough papers~\cite{hastings2021fiber,breuckmann2021balanced,panteleev2022asymptotically}. The main difficulty lies in making the $X$ and $Z$ checks commute with each other, which necessitates more structure and rules out the kind of simple random constructions that work in the classical case. To solve this issue, one can make use of the connection to chain complexes and import ideas from topology, where one can construct higher dimensional manifolds from lower dimensional ones\footnote{For illustration, one can imagine a torus being constructed out of a pair of circles, one repeated along every point of the other.}. An illustration is given in Figs.~\ref{fig:ExpanderExamples}(b) where we show a graph that consists of multiple copies of an expander graph connected by nearest-neighbor edges between copies; the resulting graph has small plaquettes by construction (Fig.~\ref{fig:ExpanderExamples}(c)). 

In these constructions, the one-dimensional chain complexes themselves can be interpreted as \emph{classical} codes, so these constructions can equivalently be thought of as constructing one quantum code out of two classical codes, which themselves can be taken to be good. Such \emph{product constructions} were initially explored in Refs. \onlinecite{tillich2013quantum,bravyi2014homological}, obtaining quantum codes with $k = O(n)$ and $d=O(n^{1/2})$. While this is a cast improvement over the BPT bound in terms of $k$, it is no better than the 2D toric code when it comes to $d$. This limitation was overcome by considering more general product constructions~\cite{hastings2021fiber,breuckmann2021balanced,panteleev2022asymptotically,breuckmann2021quantum}. We will explore these constructions from the perspective of condensed matter physics in upcoming work~\cite{LDPCProduct}.

\section{Gauging generic LDPC codes}\label{sec:Gauging}

We now turn to the description of our general gauging procedure for cLDPC codes. The reader might find is useful to consult App.~\ref{app:Ising} where we apply the procedure for the Ising model in one and two dimensions, which will help ground our discussion here. 

We develop the gauging procedure in three steps. First, in Sec.~\ref{subsec:KramersWannier}, we show how to define a generalized Kramers-Wannier duality for an arbitrary classical code, which relates it to its transpose code by providing a map between their symmetric subspaces. In Sec.~\ref{subsec:BackgroundGauge} we then show how to lift the symmetry restriction by introducing additional degrees of freedom which can be interpreted as background gauge fields. Finally, in Sec.~\ref{subsec:gauging}, we show how to achieve the same by introducing gauge fields in a local way, which can then be made dynamical. 

\subsection{Generalized Kramers-Wannier dualities}\label{subsec:KramersWannier}

In this section, we define a very general notion of Kramers-Wannier duality. Our starting point will be an arbitrary classical LDPC code $\mathcal{C}$, which we will embed into a quantum model (replacing classical spin variables with Pauli matrices). We will then show that there is a duality transformation between the model defined by $\mathcal{C}$ and its transpose code $\mathcal{C}^T$, defined in Sec.~\ref{sec:CodeReview}. In particular, we define a unitary map between the \emph{symmetric subspaces} of the two models, where the relevant symmetries are the logical operators of the two codes. In defining this mapping and interpreting it physically, we will find that the Tanner graph representation of classical codes proves to be a very useful tool.

We will begin by embedding the classical code $\mathcal{C}$ into a quantum Hilbert space $\mathcal{H}_\sigma = \mathbb{C}_2^{\otimes n}$ which has a qubit assigned to each label $i$, with Pauli matrices $\sigma_i^{x,y,z}$. We can then equate the checks in~\eqref{eq:H_classical} with $C_a = \prod_{i\in\delta(a)} \sigma_i^z$ and the logicals with $\mathcal{X}_\lambda = \prod_{i \in \Sigma_\lambda} \sigma_i^x$. Similarly for  $\mathcal{C}^T$, we can introduce a Hilbert space $\mathcal{H}_\tau = \mathbb{C}_2^{\otimes m}$ with a set of quantum spins $\tau_a^{x,y,z}$ and checks $A_i = \prod_{a\in\delta^T(i)} \tau_a^x$. Note that we picked a convention where the classical spin configurations of $\mathcal{C}^T$ are represented by basis states in the $\tau^x$, rather than the $\tau^z$, basis. Consequently, logicals of the transpose code correspond to $\mathcal{Z}_r = \prod_{a\in R_r} \tau_a^z$, where $R_r$ is the support of a redundancy in $\mathcal{C}$. 

Consider now the \emph{symmetric subspaces} of both codes, defined by requiring all logical operators to have eigenvalue $+1$. These subspaces have the same dimension,
\begin{equation}
    \text{dim}(\mathcal{H}_{\sigma}^\text{symm}) = 2^{n - k} = 2^{m - k^T} = \text{dim}(\mathcal{H}_{\tau}^\text{symm}),
\end{equation}
since $m-k^T$ is the number of linearly independent checks in $\mathcal{C}$. Therefore, these Hilbert spaces are isomorphic and there is some unitary, $U_\text{KW}$ mapping them to each other. It is easiest to define in terms of operators. The algebra of operators on $\mathcal{H}_\sigma^\text{symm}$ is generated by $\langle C_a, \sigma_i^x \rangle$, while $\mathcal{H}_\tau^\text{symm}$ is generated by $\langle A_i, \tau_a^z \rangle$. We thus define the \emph{Kramers-Wannier map} as follows:
\begin{align}\label{eq:KWmap}
    U_\text{KW} C_a U_\text{KW}^\dagger = \tau_a^z, & &     U_\text{KW} \sigma_i^x U_\text{KW}^\dagger = A_i.
\end{align}

In terms of states, $\mathcal{H}_\sigma^\text{symm}$ is spanned by states of the form  $\ket{N} \equiv \prod_{i \in N} \sigma_i^x \ket{\phi}$, where $\ket{\phi} \propto \sum_{\lambda} \mathcal{X}_\lambda \ket{0}^{\otimes n}$ is a state symmetrized over all logicals of $\mathcal{C}$. We take $U_\text{KW}\ket{\phi} = \ket{0}^{\otimes m}$, i.e. the state with $\tau_a^z = + 1\,\,\forall a$, which is clearly in the symmetric subspace $\mathcal{H}_\tau^\text{symm}$. We then have 
\begin{equation}\label{eq:KWdual}
    U_\text{KW} \ket{N} = \prod_{i \in N} A_i \ket{0}^{\otimes m} = \prod_{a \in \delta^T(N)} \tau_a^x \ket{0}^{\otimes m}.
\end{equation}
Physically, one should think of this as follows. After symmetrizing over all logicals / classical ground states, states are specified completely by the pattern of excitations in them. Each excitation corresponds to some term $C_a$ being violated and these excitations (i.e. the \emph{domain walls} of the classical model\footnote{Throughout this paper, we use ``domain wall'' to refer to a pattern of excitations as defined by the checks of some classical code. Note that the ``domains'' in question need not be fully polarized, and can instead be e.g. fractal shaped, as they are in the Newman-Moore model (Fig.~\ref{fig:ClassicalExamples}(c)) for example.}) become the new degrees of freedom, labeled by $\tau_a^z=\pm 1$. 

We can now ask how Hamiltonians map under the duality. Since the duality is restricted to the symmetric subspaces, we should consider Hamiltonians that commute with all logical operators. A particularly simple example is the Hamiltonian of $\mathcal{C}$ with an additional transverse field:
\begin{equation}\label{eq:C_trans_field}
    H = -J \sum_a C_a - g \sum_i \sigma_i^x.
\end{equation}
Under~\eqref{eq:KWdual}, this maps onto $\tilde{H} = -J \sum_a \tau_a^z - g \sum_i A_i$, i.e. a transverse field version of $\mathcal{C}^T$, but with the roles of the two terms reversed. Relatedly, $\ket{\phi}$ is the (symmetric) ground state of $H$ at $g=0$, which  $U_\text{KW}$ maps to the ground state of $\tilde{H}$ at the same choice of parameters. We thus see that the KW duality exchanges a symmetry broken phase with a trivial ``paramagnetic one''. One might wonder what order parameters to assign to these phases and how this map under the duality: we shall return to this question below in Sec.~\ref{sec:OrderParams}. 

The Kramers-Wannier duality introduced in this section is closely related to the ``ungauging map'' defined in Ref. \onlinecite{kubica2018ungauging}. From our perspective, the ungauging map is a special case of the KW duality in the case when the code $\mathcal{C}$ has local redundancies, in which case the duality naturally gives rise to a quantum CSS code---this will be discussed in more detail below in Sec.~\ref{subsec:LocalRed} and~\ref{sec:Deconfined}. An advantage of our approach is that it offers a natural way of extending the KW duality in such a way that it can also incorporate states beyond the symmetric subspaces, as we now describe.

\subsection{Coupling to background gauge fields}\label{subsec:BackgroundGauge}

\begin{figure} 
    \centering
    \includegraphics[trim={1cm 0 0 0},width = 0.9\linewidth]{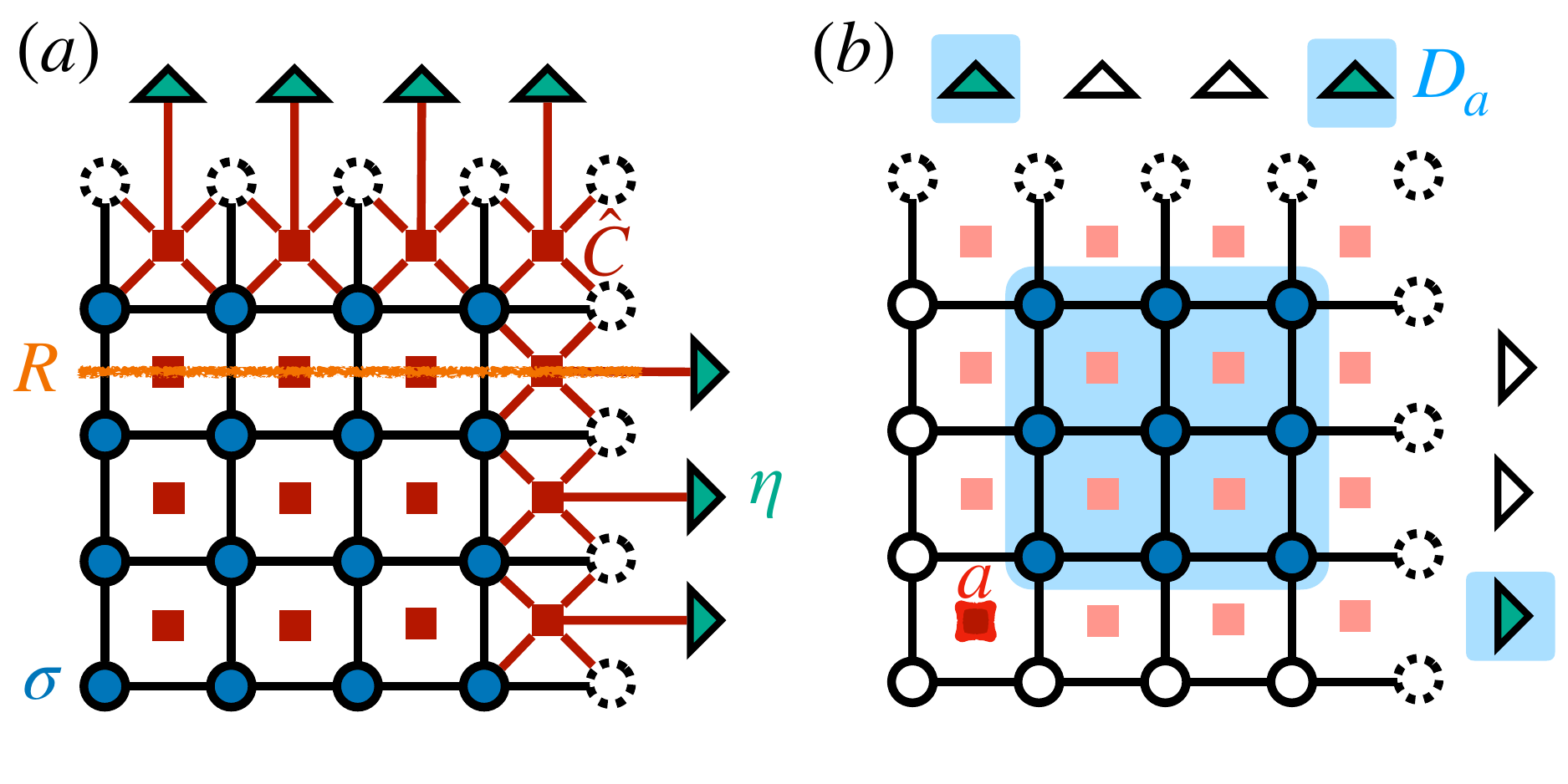}
    \caption{Coupling cLDPC codes to background gauge fields, illustrated on the example of the 2D plaquette Ising model (see also Fig.~\ref{fig:ClassicalExamples}(b)). (a) The checks involve four bits along a plaquette. We take periodic boundaries, such that the sites at the boundary (denoted with dotted lines) are identified with those at the opposite end. In this case, there is a redundancy $R$ for every column and row (horizontal orange line). These can be removed by adding one $\eta$ variable to each row/column, and modifying the checks along the boundary to 5-spin interactions involving one of the $\eta$ spins as shown in the figure. (b) The disorder operator $D_a$ flips $\sigma$ spins in the shaded region, along with a set of $3$ $\eta$ variables, creating a single excitation at the location of the check labeled by $a$.}
    \label{fig:BackgroundGauge}
\end{figure}

The KW duality we defined is a map between symmetric subspaces, a fact that was forced on us due to the fact that the symmetries on one side of the duality turn into redundancies on the other. Thus, for example, the ground state degeneracy of the classical code is not reflected as we only have access to the fully symmetrized state within the ground state manifold. We now describe how this restriction can be lifted. In order to do so, we will need to remove the redundancies from the classical codes (while maintaining all their logical operators). We will see that this procedure can be interpreted as coupling to (static) background gauge fields. 

To remove redundancies, we will introduce additional ancilla spins and modify the checks appropriately. In particular, let us fix some basis of redundancies of $\mathcal{C}$, labeled by integers $r=1,\ldots,k^T$, each corresponding to some redundancy $\prod_{a \in R_r} C_a = 1$. For each of these, we will introduce an ancilla spin $\eta_r$. We now want to modify the checks, $C_a \to \hat{C}_a$ such that $\prod_{R_r} \hat{C}_a = \eta_r$. 

How to define $\hat{C}_a$? Let us consider the redundancies as vectors $\ket{R_r}$ in the $\mathbb{Z}_2$ vector space of checks. We can find a set of dual vectors, $\ket{G_r}$ such that $\braket{G_r|R_{r'}} = \delta_{rr'}$ (see also our discussion of symmetries and order parameters in Sec.~\ref{subsec:classicalLDPC}). $G_r$ thus corresponds to a set of checks that overlap with $R_r$ an odd number of times and with all other redundancies an even number of times. Thus, if we modify the checks as $C_a \to \eta_r C_a$ iff $a \in G_r$, then this lifts the redundancy $R_r$ as required. This defines a different classical code, with the same set of codewords as $\mathcal{C}$ but no redundancies. This construction is illustrated for the 2D plaquette Ising model in Fig.~\ref{fig:BackgroundGauge}(a). 

We can apply the KW duality to this modified problem. The $\tau$ variables now have no symmetry constraint, since we removed all the redundancies of the original code. We thus obtain a map between the full Hilbert space $\mathcal{H}_\tau$ and a Hilbert space $\mathcal{H}_{\sigma,\eta}^\text{symm}$, which is the space of the combined $\sigma$ and $\eta$ variables with the symmetries (logicals) of $\mathcal{C}$ still enforced. These have dimensions $2^m = 2^{n-k+k^T}$. 

What do the operators $\tau_{a}^{x}$ correspond to under the duality? To preserve the commutation relations, they should correspond to operators that anticommute with the check $C_a$ but no other checks in $\mathcal{C}$. We will refer to these as \emph{disorder operators} and denote them by $D_a$; they correspond to a set of spin flips, $\sigma_i^x$, which creates a single domain wall at the location specified by $a$; an example, for the 2D plaquette Ising model, is shown in Fig.~\ref{fig:BackgroundGauge}(b) In the linear algebra language, they have the property $\braket{D_a|C_{a'}} = \delta_{aa'}$. We thus have $U_\text{KW} D_a U_\text{KW}^\dagger = \tau_a^x$ in this extended duality transformation. 

We note that for any code with a macroscopic code distance, the disorder operators need to be non-local, in the sense that there exist $a$ for which the support of $D_a$ must involve $O(d)$ spins. We can see this by adapting an argument from Ref. \onlinecite{ben2010locally} as follows. Since arbitrary spin flips are generated by a combination of logicals and disorder operators, we can write a single-site flip as
\begin{equation}\label{eq:Disorder}
    \sigma_i^x = \mathcal{X}_{\lambda_i} \prod_{a \in M_i} D_a,
\end{equation}
where $\lambda_i$ labels a logical of $\mathcal{C}$ and $M_i$ is some subset of the checks, both depending on the choice of $i$. Since each disorder operator violates one check, and they are all independent, we have that the energy cost of flipping spin $i$ is equal to $|M_i|$, the number of disorder operators that appear in Eq.~\eqref{eq:Disorder}. By assumption of the code being LDPC, this energy cost is some finite, $O(1)$ value for any $i$. On the other hand, since there are $n$ different spins, but only $n-k$ independent disorder operators, there has to exist some choice of $i$ for which $\lambda_i \neq 0$, i.e. the logical $\mathcal{X}_{\lambda_i}$ is different from the identity. Rearranging Eq.~\eqref{eq:Disorder}, we have $\mathcal{X}_{\lambda_i} = \sigma_i^x \prod_{a\in M_i} D_a$. Since the left hand side involves flipping $\geq d$ spins, it follows that that at least one of the $D_a$ also needs to have support of $O(d)$. Thus, if $d$ is macroscopic (scales non-trivially with $n$), then at least some disorder operators have to be non-local as well, showing that the KW duality is necessarily a highly non-local transformation. However, the disorder operators become effectively local when we restrict to $\mathcal{H}_{\sigma,\eta}^\text{symm}$, which is why they can be mapped to local excitations in the dual theory. 

\begin{figure} 
    \centering
    \includegraphics[trim={1cm 0 0 0},width = 0.9\linewidth]{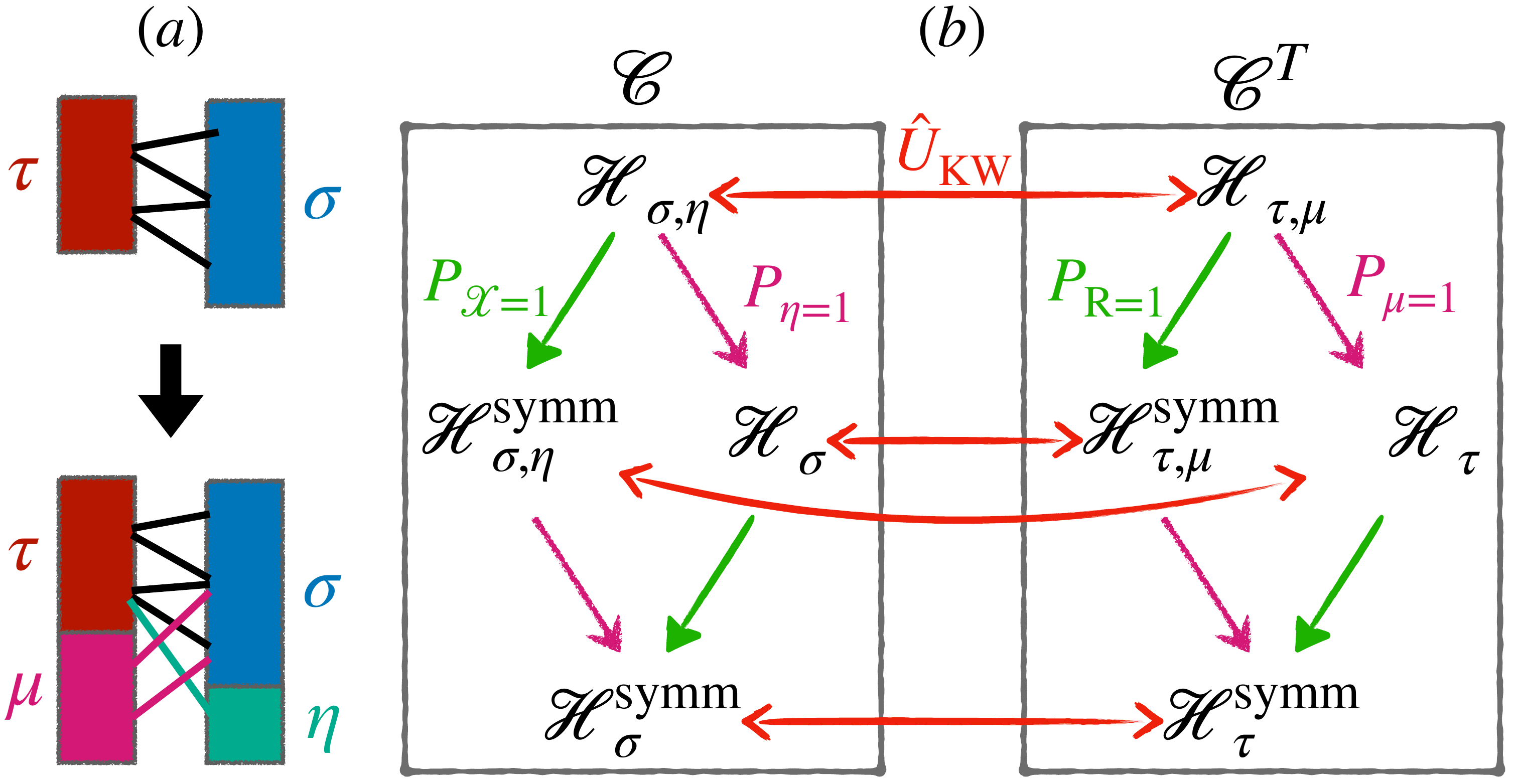}
    \caption{Kramers-Wannier and background gauge fields. (a) Background fields in terms of the Tanner graph: we can add new nodes, $\eta$ and $\mu$, to the two sides, which couple to the original degrees of freedom on the other side, but not to each other. The combined Hilbert space $\mathcal{H}_{\sigma,\eta}$ has the same size as $\mathcal{H}_{\tau,\mu}$ (b) Three levels of Kramers-Wannier duality. On the highest level, $U_\text{KW}$ connects the expanded Hilbert spaces, $\mathcal{H}_{\sigma,\eta}$ and $\mathcal{H}_{\tau,\mu}$, obtained from coupling the original variables ($\sigma$ and $\tau$) to background gauge field ($\eta$ and $\mu$). From these, we can obtain smaller Hilbert spaces, either by projecting to a trivial gauge field configuration ($P_{\eta=1}$ and $P_{\mu=1}$) or by by going to the symmetric subspaces ($P_{\mathcal{X}=1}$ and $P_{R=1}$). These are related to each other by restricted versions of $\hat{U}_\text{KW}$. The Hilbert spaces relate to each other by KW duality are denoted by the red arrows and have the same dimensions on the two sides. Notably, the full original Hilbert space on one side maps to the symmetric Hilbert space in the presence of gauge fields and vice versa.
    }
    \label{fig:KWDuality}
\end{figure}

We can also do the same procedure on the other side, introducing ancillary variables $\mu_\lambda$ corresponding to the logicals $\mathcal{X}_\lambda$ of $\mathcal{C}$ (the redundancies of $\mathcal{C}^T$) in a similar manner, which has the effect of removing the symmetry constraint from the $\sigma$ variables (see Fig.~\ref{fig:KWDuality}(a) for an illustration in terms of the code's Tanner graph). We thus have two extended Hilbert spaces $\mathcal{H}_{\sigma,\eta}$ and $\mathcal{H}_{\tau,\mu}$, which have dimension $2^{n+k^T} = 2^{m+k}$. This also allows us to define the analogues of the disorder operator, denoted by $E_i$, which correspond to a set of checks that multiply together to give $\sigma_i^z$. We can then write the fully extended KW duality map as 
\begin{align}
    \hat{U}_\text{KW} \sigma_i^x \hat{U}_\text{KW}^\dagger = A_i, & &  \hat{U}_\text{KW}^\dagger \tau_a^z \hat{U}_\text{KW} = C_a, \label{eq:KWfull1}\\
    \hat{U}_\text{KW} \sigma_i^z \hat{U}_\text{KW}^\dagger = E_i, & &  \hat{U}_\text{KW}^\dagger \tau_a^x \hat{U}_\text{KW} = D_a, \label{eq:KWfull2}\\
    \hat{U}_\text{KW} \eta_r^z \hat{U}_\text{KW}^\dagger = R_r, & &  \hat{U}_\text{KW}^\dagger \mu_\lambda^x \hat{U}_\text{KW} = \mathcal{X}_\lambda, \label{eq:KWfull3}\\
    \hat{U}_\text{KW} \eta_r^x \hat{U}_\text{KW}^\dagger = G_r, & &  \hat{U}_\text{KW}^\dagger \mu_\lambda^z \hat{U}_\text{KW} = \mathcal{Z}_\lambda. \label{eq:KWfull4}    
\end{align}
where we now also treated $\eta$ and $\mu$ as quantum variables. Eqs.~\eqref{eq:KWfull3}-\eqref{eq:KWfull4} follow from the way we introduced the ancilla variables: their eigenvalues correspond to the symmetry sectors upon dualizing. 

From this ``mother'' duality, the others follow by projecting down to appropriate subspaces, either by fixing the ancilla variables or the symmetry eigenvalues to be $+1$. This is illustrated in Fig.~\ref{fig:KWDuality}. Note that, while the transformations are unitary at every level, if one would restrict to the original Hilbert space, without adding any ancillary degrees of freedom, they would become non-invertible~\cite{li2023non,cao2023subsystem} (except for the initial mapping between symmetric subspaces introduced in the previous section). 

Let us finally comment on the physical interpretation of the ancilla variables $\eta$ and $\mu$. Recalling the hypergraph interpretation of classical codes from Sec.~\ref{subsec:classicalLDPC}, we observe that redundancies correspond to ``closed loops'' (a set of edges that go through every site an even number of times). Thus $\eta_r$ is encoding a \emph{holonomy} through the corresponding loop, a sign picked up by a $\mathbb{Z}_2$ charge as it moves around the loop. Assigning such holonomies is exactly the role usually played by gauge fields~\cite{tong2018gauge}; we can thus interpret $\eta$ as a configuration of background gauge fields. In the next section, we will see that they are indeed equivalent to a more conventional way of introducing gauge fields, which will also allow us to make them dynamical. 

\subsection{Dynamical gauge fields}\label{subsec:gauging}

We referred to the ancilla variables $\eta$ as background gauge fields. By ``background'' we mean that these degrees of freedom are, so far, non-dynamical. In particular, while we modified the checks $C_a \to \hat{C}_a$ in Eq.~\eqref{eq:C_trans_field}, we did not include additional transverse fields $\eta^x_r$. Indeed, the way we defined them, the $\eta$ variables might be very non-local in terms of the original code, coupling to many different checks (this happens e.g. in the 2D Ising model, see App.~\ref{app:Ising}). We now discuss an alternative formulation, where the gauge fields are explicitly local.

In our new gauging prescription, instead of adding one ancillary variable per redundancy, we introduce one \emph{gauge field} $\tau_a$ for every check $C_a$ of the classical code $\mathcal{C}$, forming a Hilbert space $\mathcal{H}_\tau$. To emphasize the analogy with $\mathbb{Z}_2$ gauge theory, we will also refer to the $\sigma_i$ variables as ``matter fields''; these form the Hilbert space $\mathcal{H}_\sigma$. The gauging procedure turns a Hamiltonian $H_0$ on $\mathcal{H}_\sigma$ into a Hamiltonian $H_\text{gauged}$ acting on an appropriate subspace $\mathcal{H}_\text{phys}$ of the combined Hilbert space $\mathcal{H}_\sigma \otimes \mathcal{H}_\tau$. The physical subspace, $\mathcal{H}_\text{phys}$, will be defined by a set of ``generalized Gauss law'' constraints. After enforcing these, the remaining degrees of freedom are equivalent to the set of original spins $\sigma_i$ along with the ancillas $\eta_r$ defined in the previous section. 

We assume that $H$, defined in terms of $\sigma$ variables, is symmetric under all the logicals of $\mathcal{C}$. This implies that it is generated some combination of $\langle C_a, \sigma_i^x \rangle$. Thus, to obtaine $H_\text{gauged}$, it is sufficient to map these to new operators acting on $\mathcal{H}_\sigma \otimes \mathcal{H}_\tau$. We achieve this by an appropriate ``minimal coupling'' prescription, given by
\begin{equation}\label{eq:MinCoupling}
    C_a \to \tau_a^z C_a,
\end{equation}
and leaving $\sigma_i^x$ unchanged. The $\tau_a^z$ therefore again act as background gauge fields, toggling the sign of various terms in the Hamiltonian, similar to the $\eta$ before. 

To make the two descriptions equivalent, this minimal coupling procedure is supplemented by a local ``Gauss law'' constraint, which projects the Hamiltonian down to the physical, gauge-invariant subspace. It is defined by the condition
\begin{equation}\label{eq:GaussLaw}
    G_i \equiv \sigma_i^x A_i = \sigma_i^x \prod_{a\in\delta_1^T(i)} \tau_a^x = +1 \quad \forall i.
\end{equation}
Note that this condition also implies that states are symmetric under the logical operators of the classical code, since $\prod_{i\in \Sigma_\lambda} G_i = \prod_{i\in \Sigma_\lambda} \sigma_i^x$ (where we used that $\delta^T(\Sigma_\lambda) = 0$). 

States satisfying Eq.~\eqref{eq:GaussLaw} define $\mathcal{H}_\text{phys}$. Within this subspace, the gauge fields $\tau_a^z$ become equivalent to the $\eta_r$ defined in the previous section. To see this, note that we can always multiply by the Gauss law $G_i$ to change the configuration of $\tau_a^z$ locally. The only invariant quantity is the product of $\tau_a^z$ on subsets that overlap with every check $A_i$ on an even number of locations; these are precisely the subsets that correspond to redundancies of $\mathcal{C}$. Thus, for some such redundancy, labeled by $r$, with support $R_r$, we have that $\prod_{a\in R_r} \tau_a^z$ is gauge invariant. Thus we see that the gauge fields encode exactly the same information as the ancilla qubits $\eta_r$ did in the previous section. Geometrically, one can think of the $\tau_a^z$ as encoding ``transition functions'' that define how the physical variables on different sites should be glued together, in analogy with the definition of a fibre bundle~\cite{harlow2021symmetries,thorngren2018gauging,cheng2023lieb}. 

We can also see how the Kramers-Wannier duality arises from this representation. What the above discussion shows is that $\mathcal{H}_\text{phys}$ is equivalent to the Hilbert space $\mathcal{H}_{\sigma,\eta}^\text{symm}$, defined by symmetric states of the $\sigma$ variables coupled to the gauge fields $\eta$. Another representation of $\mathcal{H}_\text{phys}$ is achieved by noting that there is exactly one Gauss law constraint per matter qubit. We can thus use these constraints to ``gauge fix'' $\sigma_i^z = +1$, $\forall i$, (in the gauge theory literature, this is called the \emph{unitary gauge}). Since we have gotten rid of all $\sigma$ spins this way, the resulting Hilbert space is equivalent to $\mathcal{H}_\tau$, the Hilbert space of $\tau$ spins. Thus, $\mathcal{H}_{\sigma,\eta}^\text{symm}$ and $\mathcal{H}_\tau$ are both equivalent to $\mathcal{H}_\text{phys}$, and therefore to each other. 

While so far the two prescriptions (in terms of $\eta_r$ and $\tau_a^z$) are equivalent, the present one, in terms of the gauge fields $\tau_a$, has the advantage of being more explicitly local. Therefore, in this language it becomes natural to make the gauge fields dynamical by including additional terms in the Hamiltonian. The simplest such term is of the form $-\Gamma \sum_a \tau_a^x$. We can also consider other terms acting solely on the gauge fields that could be added and study the phase diagram of the resulting Hamiltonian. We will return to this question in Sec.~\ref{sec:Deconfined} below. However, we first need to make a detour and discuss different classes of classical codes that will lead to qualitatively different behavior in their corresponding gauge theories. 


\section{Point-like vs extended domain walls}\label{sec:Redundancies}

In the previous section, we introduced a very general procedure that allows us to turn an arbitrary classical LDPC code into a gauge theory. However, the properties of these gauge theories will be different, depending on features of the underlying classical code. It will be instructive to begin by returning to the Ising model and contrasting the examples of 1D and 2D Ising models. 

As codes, the 1D and 2D Ising model share many properties: they both encode a single bit of information ($k=1$) and have maximal code distance ($d=n$); indeed, if we defined codes in terms of their codewords, rather than their checks, both of these were equivalent to the classical repetition code. Nevertheless, they are physically very much distinct. For example, in 1D, the symmetry breaking order is only present at zero temperature, while in 2D it is stable below some finite critical temperature. Another distinction appears when we gauge the two models (see App.~\ref{app:Ising} for details): in 1D, the resulting gauge theory turns out to be equivalent to the original Ising model, while in 2D, gauging the Ising model gives rise to a $\mathbb{Z}_2$ gauge theory that includes the toric code in its phase diagram. 

The key difference between the two cases is between \emph{the structure of their redundancies}. In the 1D Ising model with periodic boundary conditions, there is a single redundancy, involving all of the checks; this is therefore a \emph{global redundancy}, corresponding to the fact that any spin configuration contains an even number of domain walls. The situation is different in 2D. We still have global redundancies, defined by taking the product of all checks along either a row or a column of the 2D lattice. However, we now also have \emph{local redundancies}, corresponding to the four checks along the edges of a square plaquette (see Fig.~\ref{fig:IsingOnATorus}). Physically, this implies that domain walls in this case have to form \emph{closed loops} on the dual lattice. This fact is crucial to both of the differences mentioned above: it is responsible for the thermal stability of the symmetry broken order, as the finite line tension of the domain walls ensures that the different symmetry sectors are separated by diverging (free) energy barriers. It also means that the corresponding gauge theory exhibits a higher form symmetry (Fig.~\ref{fig:HigherForm}), which is crucial for understanding the appearance of the toric code phase. 

In this section, we explore how this distinction plays out in the context of generalized gauge dualities introduced in Sec.~\ref{subsec:KramersWannier}. In particular, we will make a distinction between two classes of cLDPC codes, depending on whether their domain wall excitations are point-like or extended objects and discuss how to make sense of this distinction in the general LDPC context. As the above discussion suggests, this distinction will be related to the structure of redundancies in the code, and in particular whether there are local redundancies present. In the LDPC contexts, we will define \emph{local redundancy} as one that involves a finite ($n$-independent) number of checks. We first consider codes with no such local redundancies, discuss the consequences for domain wall excitations and for the KW dualities defined above. We then go on to discuss codes with local redundancies. We will want to keep track of these explicitly, and include them in our description of the classical code, endowing it with additional structure, which we formulate in the language of chain complexes. Below, in Sec.~\ref{subsec:gauging}, we will see how this structure is related to the existence of a stable deconfined phase, associated to a quantum CSS code.

\subsection{Codes without local redundancies}\label{subsec:GlobalRed}

We first consider codes where all redundancies are global, i.e. involve some number of checks that diverges in the limit $n\to\infty$. We will first discuss the relationship between the structure of redundancies and the excitations of the model; in particular, we argue that in this case it is possible for a check to be excited with no other excitation nearby---a point-like domain wall. We then discuss the role of gauge fields and boundary conditions in this case. 

\subsubsection{Point-like excitations}

Let us begin with the most clear-cut case, when there are \emph{no} redundancies at all. As we already observed in Sec.~\ref{subsec:BackgroundGauge}, in this case there exist spin configurations (corresponding to the disorder operators $D_a$) that contain only a single excitation, corresponding the violation of a single one of the checks that define the code. Thus, in the absence of redundancies, domain walls are point-like in the sense that it is possible to excite them individually. We can also extend this to the case when code has a finite, $k^T = O(1)$ number of redundancies. We can take a linearly independent set of $m-k^T$ checks and use them to define disorder operators. By construction, these excite at most $1+k^T$ checks of the original code, thus creating a few, isolated excitations. 

When the number of redundancies grows with $n$, the above argument fails. A stronger claim can be made in the special case of translationally invariant codes defined on Euclidean lattices. In this case, one can show~\cite{HaahPersonal} that in the absence of local redundancies there exists a series of spin flips of unbounded support that only violate a finite number of checks. A characteristic example is given by the Newman-Moore model (Fig.~\ref{fig:ClassicalExamples}(c)), where the spin flips in question form a Sierpinski triangle whose sides are of length $2^p$ for integer $p$, creating a triplet of excitations at its corners; this is true despite the fact that the code exhibits $O(L)$ distinct redundancies. It is an interesting question what structure needs to be imposed on more general LDPC codes for this claim to remain true, for example whether a form of translation invariance (i.e., the existence of a symmetry that maps any any site $i$ to any other) would suffice. 

We note that the existence of point-like excitations does not mean that they can move around freely, without incurring extra energy costs. For example, in the Newman-Moore model, it is well known that in-between the aforementioned Sierpinski configurations, one encounters an energy barrier that grows logarithmically with the number of spins flipped~\cite{newman1999glassy,prem2017glassy}. The situation is even stronger for expander codes. For example, for the random expander codes discussed in Sec.~\ref{subsec:expander}, the number of violated checks grows linearly with the number of spins flipped, until one reaches a finite fraction $\gamma n$ of the entire system. Nevetheless, beyond this barrier, there exist configurations with $O(d)$ Hamming distance from the nearest codeword that only violate a single check, due to the fact that these codes can be made to have no redundancies at all (with high probability~\cite{ben2010locally}).

Before moving on, we briefly discuss an interesting example, originally from Ref. \onlinecite{siva2017topological} that further illustrates the intricate relationship between the structure of redundancies and excitations in the case when one relaxes the condition on translation invariance. This example involves an Ising model on an unusual 2D lattice, which looks like a regular square lattice when zoomed out, but with each edge composed of a long one dimensional chain, whose length $\ell$ itself grows as some positive power of the total system size $\ell \propto L^a$. There are two types of redundancies, ones along the plaquettes of the square lattice, and ones along its rows/columns. They are both global by our definition, but one is much smaller than the other, as $\ell \ll L$. This hierarchy of scales is reflected in the nature of domain walls: they appear point-like within each 1D segment, where they can move freely without creating any additional excitations, but they form closed loops once we go to scales larger than $\ell$. Nevertheless, as shown in Ref.~\onlinecite{siva2017topological}, this model does not order at any finite temperature. It also does not host a stable deconfined phase when gauged. In these senses, it is closer to the 1D than the 2D Ising model.

\subsubsection{Gauge fields and boundary conditions}\label{subsubsec:1DBC}

In Sec.~\ref{sec:Gauging}, we interpreted the gauge fields $\eta_r$ as describing holonomies around closed loops on the hypergraph description of a classical code. When all redundancies are global, these loops ``go around'' system; in this sense, we can think of the $\eta$'s as encoding ``boundary conditions'': periodic for $\eta_r = +1$ and anti-periodic when $\eta_r = -1$. Note that we can always choose a basis of redundancies, where the corresponding $\eta_r$ are supported on a single edge $a$. The sign of the coupling assigned to this edge can be toggled to change boundary conditions (see Fig.~\ref{fig:BackgroundGauge}).

This latter interpretation also suggests a different modification of the code. Rather than adding ancillas, another way to remove redundancies is by dropping some of the checks from the code. We can find a set of $k_T$ checks (the support of the $\eta_r$'s) such that dropping them removes all redundancies without changing any of the codewords. Continuing with the analogy, we can think of this as going from peropdic to open (free) boundary conditions. We will exploit this idea when we discuss SPT phases in Sec.~\ref{sec:SPT}.

\subsection{Codes with local redundancies}\label{subsec:LocalRed}

We now return to the case when the code $\mathcal{C}$ has local redundancies. We first discuss a geometric picture of what this means and then connect it to the Kramers-Wannier duality discussed above. 

\subsubsection{2-complex representation of local redundancies}

Let us consider a classical code that has local redundancies. Any local redundancy is associated to some finite subset of checks, similarly to how each check is associated to a finite subset of bits it is acting on. Indeed, similarly to bits and checks, we can form a $\mathbb{Z}_2$ linear space of local redundancies, where addition is defined by the symmetric difference of two subsets. In the spirit of LDPC, we will assume in the following that there exist a basis in this linear space such that each check of the code appears in finitely many of the basis elements and that we have identified such a basis set of local redundancies, $\ket{p}$, labeled by some set of integers $p = 1,\ldots,l$;  In analogy with the 2D Ising model, we will refer to elements of this basis set of ``plaquettes'' We will further assume that this basis set is complete in the sense that there are no local redundancies that are linearly independent of all $\ket{p}$\footnote{We allow the set to be overcomplete, so that the dimension of the linear space is smaller than $l$.}.

Let $\delta_2(p)$ denote the set of checks appearing in the local redundancy labeled by $p$, i.e. $R_p \equiv \prod_{a \in \delta_2(p)} C_a = + 1$. In analogy with the discussion in Sec.~\ref{sec:CodeReview}, $\delta_2$ defines a linear map between the linear spaces of local redundancies and that of the checks; it can be represented as a binary matrix, and by assumption it has a finite number of $1$'s in every row and every column. We will also change notation, and use $\delta_1$ (rather than $\delta$) to denote the linear map from checks to bits in this case. Thus, by a \emph{classical code with local redundancies} we will mean a set of three linear space, $(V_0,V_1,V_2)$, corresponding to bits, checks and local redundancies, respectively, along with two linear maps $\delta_{1},\delta_{2}$, with $\delta_q: V_q \to V_{q-1}$. Crucially, the fact that elements of $V_2$ are redundancies of the code implies the relation $\delta_2\delta_1 = 0$ (modulo 2). Thus, every classical code with local redundancies is associated to a $2$-level chain complex (see App.~\ref{app:ChainComplex} for a summary of chain complexes). This is the same mathematical object we encountered when we discussed quantum CSS codes in Sec.~\ref{subsec:CSS}, and as we shall see, this is no coincidence. 

The cycles of the $2$-complex correspond to the redundancies of the code, which naturally fall into two categories: the local redundancies, generated by the elements of $V_2$, correspond to trivial cycles, while the remaining non-trivial cycles are \emph{global redundancies}; the latter form equivalence classes, which are the homology classes of the chain complex. We can give a geometric interpretation to this distinction. Recall that in the hypergraph language, checks correspond to edges, and we interpreted redundancies as closed loops. The local redundancies labeled by $p$ are small finite loops, which was the reason for referring to them as `plaquettes'. In this sense, a code with local redundancies looks locally two-dimensional, while one without such redundancies is locally one-dimensional. The global redundancies, not generated by plaquettes, ``wrap around'' the system and we can interpret them as encoding periodic boundary conditions. 

The fact that the checks in the subset $\delta_2(p)$ form a redundancy means that they can only be violated in pairs, which we can visualize as an ``in-out'' condition, i.e. a domain wall that enters the plaquette $p$ must also leave it (see Fig.~\ref{fig:IsingOnATorus}), indicating that we should think of them as extended objects\footnote{For this to be the case, we should also require that the local redundancies are non-trivial, in the sense that every check appears in more than one basis redundancy.}. In the language of the chain complex, having to satisfy all local redundancies means that the domain walls correspond to cocycles, indicating that they form closed loops. If we also take the global redundancies into account, we further need to restrict to domain walls that are contractible, i.e. trivial cocycles. 

\subsubsection{Flat vs curved background gauge fields}

Let us now consider coupling to background gauge fields, as we did in Sec.~\ref{subsec:BackgroundGauge}. The key point is that there are now two classes of redundancies, global and local. Thus, we could perform the coupling to background fields in two ways: only removing the global ones, or removing both. In the former case, we can say that the background gauge fields are ``flat'', encoding only the (periodic or anti-periodic) boundary conditions, similarly to the case without local redundancies discussed above. Gauge fields that remove local redundancies, on the other hand, correspond to ``curvature'', or local defects. These become new local excitations when the gauge fields are made dynamical, endowing the gauge theory with qualitatively new features. 

To remove the global redundancies, we choose a set of representative (co)cycles from each (co)homology class of the chain complex, which we label by some integers $r$. They come in pairs: the cycle $\Gamma_r$ overlaps with the cocycle $\tilde{\Gamma}_r$ an odd number of times, and with all others an even number of times, such that $\braket{\Gamma_r|\tilde{\Gamma}_{r'}} = \delta_{rr'}$. This way, for each equivalence class of global redundancies, labeled by $r$, we associate a set of checks that constitute the cocycle $\tilde{\Gamma}_r$. We can now introduce a gauge field / ancilla $\eta_r$ and modify all checks $a \in \tilde{\Gamma}_r$ to $C_a \to C_a \eta_r$. 

In analogy with our discussion in Sec.~\ref{subsubsec:1DBC}, we can also consider open boundary conditions. We would do this by choosing $\tilde{\Gamma}_r$ to be a minimal representative within its cohomology class, and instead of introducing $\eta_r$, simply removing all checks belonging to $\tilde{\Gamma}_r$ from the code. This also removes a set of local redundancies that involve these checks. This gives rise to what, in the case of the 2D Ising model, are referred to as ``smooth'' boundary conditions. We can also define ``rough'' boundaries, by instead removing a non-trivial \emph{cycle} from the chain complex. 

We can also remove the local redundancies, labeled by the plaquettes $p$, by adding an ancilla variable $\eta_p$ that flips the sign of interactions along some ``seam'' of hyperedges, $\tilde\Gamma_p$ terminating at plaquette $p$, such that $\delta_2^T(\Gamma_p) = \{p\}$. This amounts to a ``curved'' background gauge field, and we can picture it as a ``flux defect'', located at the plaquette $p$, such that loops encircling the defect can now have a non-trivial holonomy, labeled by $\eta_p$. Note that if the set of plaquettes we have chosen are not all linearly independent, then there will be further constraints on the configurations of these flux defects (for example, in the 2D Ising model, they can only occur in pairs, while in 3D, they must form closed loops; see also the next subsection). 

It is also instructive to consider the role of flat and curved gauge fields in terms of domain wall configurations. In the absence of gauge fields, the domain walls, formed by violations of the checks $C_a$ have to form closed loops, that is cocycles of the chain complex. This is the crucial difference compared to the case with no local redundancies where, as we argued, domain walls can be thought of as point-like. 

In the original model, the presence of global redundancies in addition means that these cocycles also need to be trivial: they are coboundaries of some subset of sites in $V_0$. We can assign to these a disorder operator, $D_\Gamma = \prod_{i\in N} \sigma_i^x$ where $\Gamma = \delta_1^T(N)$ is a trivial cocycle. Under the Kramers-Wannier duality~\eqref{eq:KWmap}, these map onto a loop-operator $W_\Gamma = \prod_{a \in \Gamma} \tau_a^x$. 

When we remove all the global redundancies (but still keeping the local ones), the domain walls still need to form closed cocycles, but these need no longer be trivial: by opening/twisting the boundaries, we allowed for domain walls that wrap around the system in a non-trivial way. We can now assign a disorder operator for \emph{any} cocycle $\Gamma$. Finally, by removing all remaining redundancies, we can have domain walls that no longer form closed loops, but rather have endpoints at the locations of the plaquette defects with $\eta_p = -1$. 

A final comment is appropriate here. In this section, we used the chain complex representation of the code itself to define a notion of geometry, and we used it to motivate such notions as boundary conditions. Even when the code is defined on a finite-dimensional Euclidean lattice, this effective ``code geometry'' might be very different from the underlying geometry of the lattice. For example, while the codes defined in Fig.~\ref{fig:ClassicalExamples}(b,c) live on 2D lattices, they do not possess any local redundancies and therefore belong to the category discussed in Sec.~\ref{subsec:GlobalRed}, being effectively one-dimensional as codes. This is reflected in the definition of boundary conditions: for example, as we saw in Fig.~\ref{fig:BackgroundGauge} the 2D plaquette Ising model has a separate $\eta$ field for every row/column of the 2D lattice, rather than a single choice of periodic/anti-periodic boundary condition for \emph{all} rows and columns, as is the case in the Ising model where the chain complex coincides with the actual lattice geometry (see App.~\ref{app:Ising}). Physically, this reflects the restricted mobility of domain wall excitations in the case of the plaquette Ising model. 

\subsection{Higher dimensional chain complexes}\label{subsec:HigherForm}

In the previous section, we focused on cLDPC codes that appeared as two-dimensional chain complexes, with the $2$-dimensional object corresponding to the local redundancies. A natural generalization is to consider codes that correspond to the first to levels of a $D$-dimensional chain complex with $D > 2$. In this sense, one could speak of clsssical codes associated to a $D$-level chain complex (the $D$-dimensional Ising model being the obvious example). The third layer of such a chain complex would correspond to local \emph{meta-redundancies}, i.e. linear relations between the vectors $\ket{p}$ corresponding to the plaquettes / local redundancies (for example, the 3D Ising model has such a meta-redundancy for every elementary cube). In this case, domain walls form closed ``surfaces'', rather than loops etc. One can thus consider different families of cLDPC codes, depending on their local dimension as encoded in the structure of their chain complex representation. 

So far, we considered cases where the bits of the code correspond to the $0$-dimensional objects of a chain complex. We can further generalize to the case where the bits and checks correspond to the $r$ and $r+1$ levels of a chain complex, with $r \geq 0$. When $r > 0$, this means that the code has a small $O(1)$ code distance, with the logical operators corresponding to the objects that constitute the vector space $V_{r-1}$ in the chain complex\footnote{Note that we only consider chain complexes where the boundary maps $\delta$ are sparse (have finitely many non-zero elements in any row/column).}. Nevertheless, they can be physically interesting. An example is the three-dimensional \emph{classical} gauge theory~\cite{kogut1979introduction}, where bits (checks) correspond to edges (plaquettes) of a cubic lattice with the natural adjacency relations between them. Flipping the six spins adjacent to a site constitutes a logical operation. We could restrict ourselves to the symmetric subspace of all of these symmetries; in that case, we still have a non-zero ground state degeneracy corresponding to the \emph{global} logicals, not generated by the local ones. Geometrically, the local symmetry conditions means that the set of edges with $\sigma=-1$ must form closed loops; an example of a higher-form symmetry (see Fig.~\ref{fig:HigherForm}). A similar situation occurs naturally when we gauge classical codes with local redundancies, we we shall see below in Sec.~\ref{sec:Deconfined}.

Before concluding this section, let us make some general comments. In these discussion, we have assumed that the classical code arrives with a predefined set of local redundancies (meta-redundancies, etc.) that can be used to form a chain complex. An important question is whether one can construct these, given only the checks of the classical code and whether there is a canonical way of doing so at least in some cases. In general, one would probably want the support of each basis redundancy to be as small as possible. Assuming the code has additional structure, such as symmetries (e.g .translations), we could also impose that the set of local redundancies should respect these. Another approach is to construct the cLDPC code explicitly as part of a larger chain complex, which then automatically includes a set of local redundancies. One way of achieving it is by so-called \emph{product constructions}~\cite{tillich2013quantum,bravyi2014homological,breuckmann2021balanced}, which we will discuss in detail in future work~\cite{LDPCProduct}.

\section{Deconfined phases and quantum codes}\label{sec:Deconfined}

We can now return to the issue raised at the end of Sec.~\ref{sec:Gauging}: what is the phase diagram of the theory obtained from gauging a cLDPC code? In particular, when do we obtain a stable deconfined phase, similar to the one that appears in 2D $\mathbb{Z}_2$ gauge theory? 

Based on the discussion of Sec.~\ref{sec:Redundancies}, we expect that the answer should be related to the presence of local redundancies. Indeed, considering the gauge theories obtained in Sec.~\ref{subsec:gauging}, we can ask what are the possible gauge invariant terms that we can write down. In particular, is there an analogue of the plaquette term of the $\mathbb{Z}_2$ gauge theory, i.e. a gauge-invariant combination of $\tau^z$ operators? Such a combination would amount to a product of the minimal coupling terms, $C_a\tau_a^z$, where all the $\sigma^z$ cancel, i.e. a redundancy of $\mathcal{C}$. And since we want to maintain locality of the Hamiltonian, this would have to be a local redundancy. Thus, the allowed $\tau^z$ type terms in the gauge theory correspond exactly to local redundancies of the ungauged classical code. In this section we therefore focus on gauge theories obtained by gauging cLDPC codes with local redundancies and ask whether their phase diagram hosts a topologically ordered deconfined phase, analogous to the toric code phase in two dimensions.

\subsection{The deconfined phase}

Consider a cLDPC code with local redundancies, and assume that we have identified a particular basis set for them, as discussed in Sec.~\ref{subsec:LocalRed}, labeled by $p=1,\ldots,l$. To each of these, we now associate an operator in the gauge theory, $B_p = \prod_{a \in \delta_2(p)}\tau_a^z$. Note that these are related to the distinction between flat and curved gauge fields discussed above, i.e. the flat configurations are precisely those that satisfy $B_p = +1$, $\forall p$. 

Thus, when the classical code has local redundancies, we can extend the gauging procedure described in Sec.~\ref{subsec:gauging} by also adding the terms $B_p$ to the Hamiltonian, which have the effect of favoring flat gauge field configurations. Starting from the transverse field version of the code in Eq.~\eqref{eq:C_trans_field} this procedure gives the gauged Hamiltonian
\begin{equation}\label{eq:H_gauged}
    H = -J \sum_a C_a \tau_a^z - g \sum_i \sigma_i^x - K \sum_p B_p -\Gamma \sum_{a} \tau_a^x,
\end{equation}
along with the gauge constraints $G_i = \sigma_i^x A_i = + 1$. This Hamiltonian is the analogue of the usual Fradkin-Shenker Hamiltonian of a $\mathbb{Z}_2$ gauge theory, indeed it coincides with it when the original Hamiltonian is the two-dimensional Ising model (i.e., when the chain complex corresponds to a 2D lattice). 


What is the phase diagram of~\eqref{eq:H_gauged}? To gain some insight, let us go into the unitary gauge, $\sigma_i^z = +1$. This results in the following gauge-fixed Hamiltonian:
\begin{equation}\label{eq:H_gaugefixed}
    H_\text{gf} = -J \sum_a \tau_a^z - g\sum_i A_i -K \sum_p B_p -\Gamma \sum_a \tau_a^x.
\end{equation}
As a starting point, we consider the limit of a \emph{pure gauge theory}, $J=\Gamma=0$, so that only the $A_i$ and $B_p$ terms remain. By constructions, these all commute, thanks to the way they have been constructed out of a chain complex. Indeed, we recognize the pure gauge theory as the Hamiltonian of a \emph{quantum CSS code}, as in Eq.~\eqref{eq:H_quantum} (where we take $g=K=1$). Thus we find that upon gauging a classical code with local redundancies, the resulting gauge theory has a limit where it it described by a CSS code. This corresponding code is associated to the same 2-complex as the original classical code, as shown by comparing the first two rows in Fig.~\ref{fig:2complex}. 

The CSS code in general has multiple distinct ground states, whose number $k_\text{q}$ is determined by properties of the underlying chain complex, which it inherits from the classical code $\mathcal{C}$. Following the discussion in Sec.~\ref{subsec:CSS}, we can interpret this ground state degeneracy as arising from a spontaneous breaking of 1-form symmetries. In the present case, the $J=\Gamma=0$ limit exhibits two such symmetries, generated by the \emph{Wilson loops}, $W_\Gamma = \prod_{a \in \Gamma} \tau_a^z$, and \emph{`t Hooft loops}, $H_{\tilde \Gamma} = \prod_{a \in \tilde \Gamma} \tau_a^x$, where $\Gamma$ ($\tilde\Gamma$) is a cycle (cocycle) of the chain complex, satisfying $\delta_1(\Gamma)=0$ ($\delta_2^T(\tilde\Gamma)=0$). When these (co)cycles are non-contractible, the Wilson and `t Hooft loops correspond to the logical operators of the CSS code.

\begin{figure}[t!]
    \centering
    \includegraphics[trim={1cm 0 0 0},width = 0.7\linewidth]{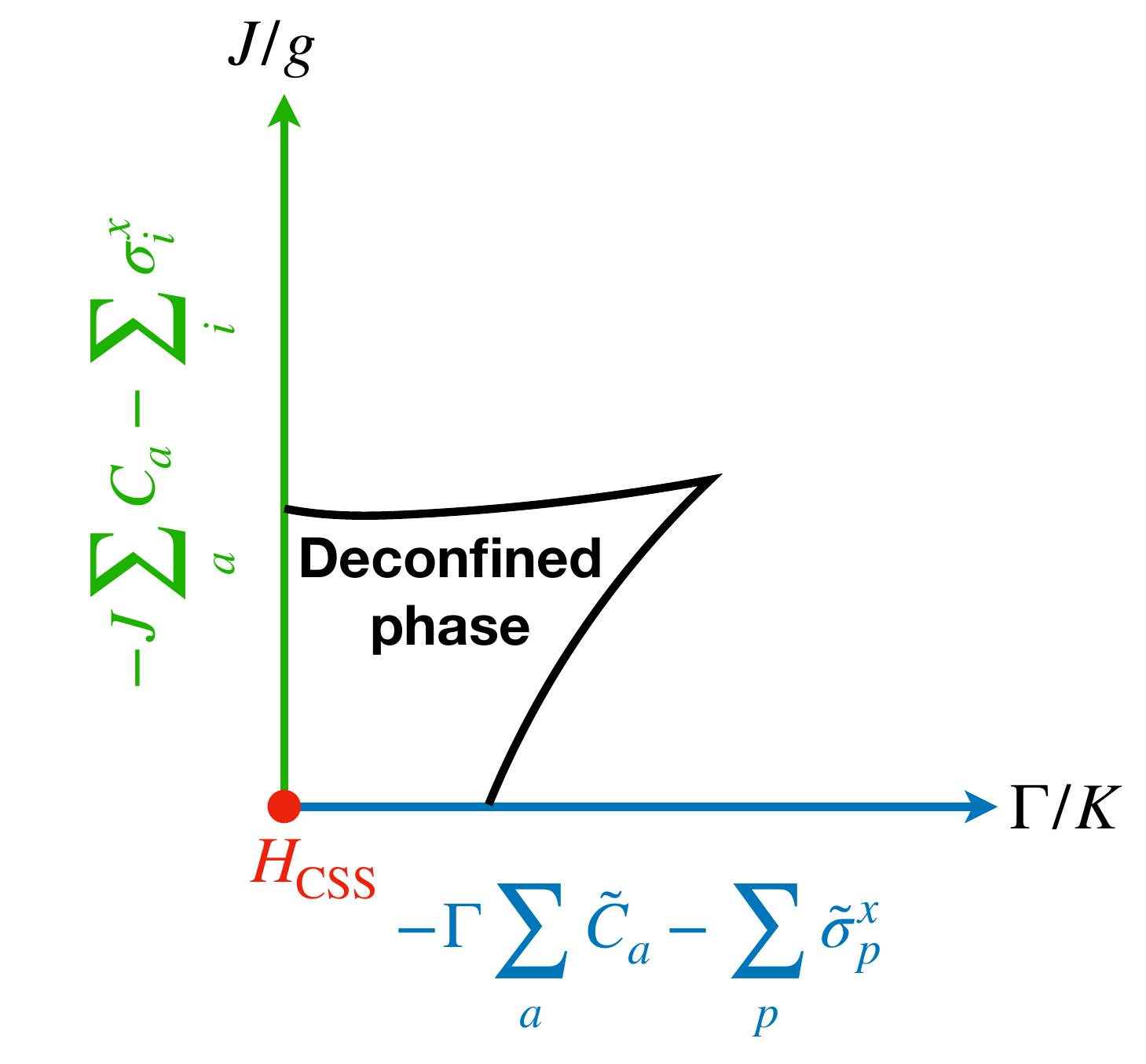}
    \caption{Schematic phase diagram of the gauge theory~\eqref{eq:H_gauged}. The two axes are dual to the two classical codes, $\mathcal{C}$ and $\tilde{\mathcal{C}}$, subject to an additional transverse field. The point where the two axes meet is described by a CSS stabilizer Hamiltonian, defining a quantum code.}
\label{fig:FradkinShenker}
\end{figure}

Considering Eq.~\eqref{eq:H_gaugefixed} away from the pure gauge theory limit, the CSS code Hamiltonian is perturbed by transverse and longitudinal fields. A natural question is whether there is a finite region of the phase diagram where the deconfined phase remains stable. On finite dimensional Euclidean lattices, there are rigorous results that establish the robustness of stabilizer Hamiltonians~\cite{bravyi2011short}. Generalizing these proofs to the non-Euclidean setting is, to the best of our knowledge, an open problem. However, a natural expectation is that the robustness of the CSS codes towards local errors should translate into a robustness against local perturbations of $H_\text{CSS}$; this results in the Fradkin-Shenker type phase diagram illustrated in Fig.~\ref{fig:FradkinShenker}. The robustness of the code itself is ensured as long as the quantum code distance $d_\text{q}$ grows at least logarithmically with the number of qubits~\cite{dumer2015thresholds}.

We can gain more insight into the phase diagram by considering the two axes, where either $\Gamma=0$ or $J=0$ but not both. Along these axes, one of the two higher form symmetries remains exact. Let us first consider the axis obtained by tuning $J$, while keeping $g=K=1$ and $\Gamma = 0$ fixed. Along this line, all terms continue to commute with $B_p$ so at low energies we can restrict our attention to the subspace with $B_p = +1$, $\forall p$. Within this subspace, Eq.~\eqref{eq:H_gaugefixed} reduces to $-J\sum_a\tau_a^z - g \sum_i A_i$, which we recognize to be the Hamiltonian of the dual code $\mathcal{C}^T$ with a transverse field. By the discusion of Sec.~\ref{sec:Gauging}, this is dual to a version of the original code $\mathcal{C}$ where we introduced gauge fields (the ancillas $\eta$) for global, but not for local redundancies. Thus, the stability of the deconfined phase along this axis is equivalent to the stability of the \emph{paramagnetic} phase of $\mathcal{C}$ in the presence of twisted boundary conditions. In Euclidean settings, the boundary conditions are usually unimportant in determining the critical field, but this might not be the case in other geometries~\cite{placke2023random}.  

It is worth briefly mentioning how the properties of the quantum code manifest in the classical model under KW duality. The `t Hooft loops map onto domain walls of the classical code, while the eigenvalues of non-contractible Wilson loops (the $Z$ logicals of the quantum code) correspond to the values of the $\eta$ fields that label the different ``boundary condition'' sectors. Twisting the boundaries, i.e. choosing $\eta=-1$, enforces a non-contractible domain wall. In the pure gauge theory, when $J=0$, there is no line tension associated to domain walls. Consequently, the different twisted boundary sectors are all degenerate, giving the (topological) degeneracy of the quantum code. In the opposite, ferromagnetic limit ($g=0$), domain walls cost energy and so the different sectors are split. 

Now let us consider the other axis of the phase diagram, defined by the condition $J=0$. This is dual to a \emph{different} classical code. In this case, we can take $A_i = + 1$ to get a low energy Hamiltonian $-K \sum_p B_p -\Gamma \sum_a \tau_a^x$. This is the classical code $\mathcal{C}^Z$ in a transverse field and it is the transpose of the classical code $\tilde{\mathcal{C}}$ defined by $\delta = \delta_2^T$. In terms of the 2-complex, this second code corresponds to assigning bits to the faces, while checks are still associated to the edges; vertices become local redundancies (see third row in Fig.~\ref{fig:2complex}). If we start with $\tilde{\mathcal{C}}$ and apply the same gauging procedure, we end up with the exact same gauge theory, except with the roles of $\tau_a^x$ and $\tau_a^z$ exchanged. We summarize the relationship between the three codes in Table~\ref{Table:Gauging}.

\begin{table*}[t!]
\centering
\begin{tabular}{|c| c c c|} 
 \hline
  Geometric object & $\mathcal{C}_\text{q}$ & $\mathcal{C}_\text{cl}$ & $\tilde{\mathcal{C}}_\text{cl}$ \\ [0.5ex] 
 \hline
 ``sites'' & $X$-checks & bits &local redundancies \\ 
 ``edges'' & qubits & checks & checks \\
 ``plaquettes'' & $Z$-checks & local redundancies & bits \\
 cycles ($\text{Ker}(\delta_1)$) & 't Hooft ($Z$) loops & redundancies & domain walls \\
 co-cycles ($\text{Ker}(\delta_2^\text{T})$) & Wilson ($X$) loops & domain walls & redundancies \\ [1ex]
 closed surfaces ($\text{Ker}(\delta_2)$) & $Z$ redundancies & meta-redundancies & logicals \\ [1ex] 
 closed co-surfaces ($\text{Ker}(\delta_1^\text{T})$) & $X$ redundancies & logicals & meta-redundancies \\ [1ex] 
 \hline
\end{tabular}
\caption{Relationship between properties of the three codes defined in Fig.~\ref{fig:2complex}.}
\label{Table:Gauging}
\end{table*}

\subsection{Quantum-classical dictionary}\label{subsec:Dictionary}

Summarizing our procedure so far, we have started with a classical code, $\mathcal{C}_\text{cl}$, defined in terms of its checks, which can be used as generators of terms that are symmetric under a set of spin-flip symmetries, given by the logical operators of the code. Furthermore, we have focused on codes which exhibit local redundancies between their checks, with some basis set of redundancies which are provided as part of the specification of the code. Out of this classical code, we have constructed a gauge theory, which contains a particular point where it is described bu a \emph{quantum} CSS code. The relationship between the original classical code and this CSS code is that they correspond to the same $2$-complex, just with a different interpretation of the three terms appearing in it, as shown in Fig.~\ref{fig:2complex}. 

Furthermore, there is another classical code, $\tilde{\mathcal{C}}_\text{cl}$, which is dual to $\mathcal{C}_\text{cl}$ in the sense that it is defined by inverting the roles of spins and local redundancies (i.e., taking the dual $2$-complex) also also indicated in Fig.~\ref{fig:2complex}. Gauging $\tilde{\mathcal{C}}_\text{cl}$ gives rise to the \emph{same} CSS code (up to a Hadamard basis transformation, $\tau_a^x \leftrightarrow \tau_a^z$). We can summarize the relationship between the three codes by considering the CSS code as a pair of classical codes, $(\mathcal{C}^X),\mathcal{C}^Z$, corresponding to the $X$ and $Z$ checks, respectively, so that we can write
\begin{align}
    \mathcal{C}^X = \mathcal{C}_\text{cl}^T, & &
    \mathcal{C}^Z = \tilde{\mathcal{C}}_\text{cl}^T.
\end{align}
Looking back at the discussion of Sec.~\ref{subsec:KramersWannier}, we recognize these as a pair of Kramers-Wannier dualities. Indeed, along the two axes of the phase diagram~\ref{fig:FradkinShenker} that meet at the CSS code point, we can use these Kramers-Wannier dualities to map the low-energy effective Hamiltonian of the gauge theory to the corresponding classical code subject to an additional transverse field. 

The two classical codes, $\mathcal{C}_\text{cl}$ and $\tilde{\mathcal{C}}_\text{cl}$, and the quantum CSS code $\mathcal{C}_\text{q}$ are all described by the same chain complex. The various geometric features of the chain complex have different interpretations in the three different codes, providing a dictionary between them, which we summarize in Table~\ref{Table:Gauging}. In particular, cycles of the chain complex correspond to Wilson loops of the quantum code $\mathcal{C}_\text{q}$, redundancies of $\mathcal{C}_\text{cl}$ and domain walls of $\tilde{\mathcal{C}}_\text{cl}$, while cocycles correspond to `t Hooft loops, domain walls and redundancies, respectively. Contractible (non-contractible) cycles correspond to local (global) redundancies of $\mathcal{C}$ and non-contractible cocycles define domain walls that do not appear in $\mathcal{C}_\text{cl}$ but only in its modification after global redundancies have been removed by adding background gauge fields; the situation is reversed when we consider $\tilde{\mathcal{C}}_\text{cl}$. In the quantum code, it is the non-contractible (co)cycles that give rise to non-trivial logical operators. 

We can also consider the closed surfaces and co-surfaces of the chain complex, which are elements of $\text{Ker}(\delta_2)$ and $\text{Ker}(\delta_1^T)$ respectively. The latter are the logical operators of the classical code $\mathcal{C}_\text{cl}$ while the former are logicals of $\tilde{\mathcal{C}}_\text{cl}$. In $\mathcal{C}_\text{q}$, the closed surfaces and co-surfaces correspond to redundancies between $Z$ and $X$ checks, respectively. This allows us to relate code rates of the three codes. In the absence of redundancies, the $k$ would simply be the difference between the number of spins, and the number of checks. We thus have
\begin{equation}
    k_\text{q} - k_\text{cl} - \tilde{k}_\text{cl} = m - n - \ell.
\end{equation}

\subsection{Local testability and good qLDPC codes}\label{subsec:LTC}

Having established the relationship between clasical codes, the gauge theories obtained from them, and quantum CSS codes, we will now comment on the recent constructions of \emph{good} qLDPC codes that have appeared in the literature~\cite{panteleev2022asymptotically,leverrier2022quantum,dinur2023good,lin2022good}, obtaining the optimal scaling $k_\text{q},d_\text{q} \propto n$. 

From the perspective developed here, a natural question to ask is what kind of classical codes give rise to such good qLDPC codes upon gauging? Our key observation is that in all known constructions, both ungauged classical codes, $\mathcal{C}_\text{cl}$ and $\tilde{\mathcal{C}}_\text{cl}$ are good cLDPC codes that have a particular property called \emph{local testability}~\cite{goldreich2005short}. This suggests that gauging locally testable codes (LTCs) is a promising way for obtaining interesting quantum codes and gauge theories in general. We will first define LTCs and then discuss some intuition on why they might be related to good codes, based on our preceding discussion of gauge dualities. 

The defining property of LTCs can be stated in terms of \emph{energy barriers}, which we define as follows (for more coding-based definitions, see \onlinecite{goldreich2005short,ben2010locally,dinur2022locally}). Let $\mathcal{X}$ denote the smallest logical operator of a classical code, supported on a set of spins $\Sigma$ of size $|\Sigma| = d$. We consider subsets $\Sigma' \subset \Sigma$ with sizes $|\Sigma'| = F \leq d/2$ and define 
\begin{equation}
E_\text{min}(F) \equiv \min_{\Sigma': |\Sigma'| = F}(|\delta^\text{T}(\Sigma')|),
\end{equation}
which is the minimal energy cost of flipping $F$ spins en route from the all ground state to the nearest codeword. In other words, this describes how the size of the domain walls created by flipping spins grow as we continuously interpolate between the two ground states. Locally testable codes are defined by an optimally fast growth of their energy barriers. More precisely, we have
\begin{equation}\label{eq:LTC_def}
    E_\text{min}(F) \geq \kappa F, \qquad 0 \leq F \leq d/2,
\end{equation}
for some $O(1)$ constant $\kappa$ (referred to as ``soundness'' in the coding literature). 

\begin{figure} 
    \centering
    \includegraphics[trim={1cm 0 0 0},width = 1.\linewidth]{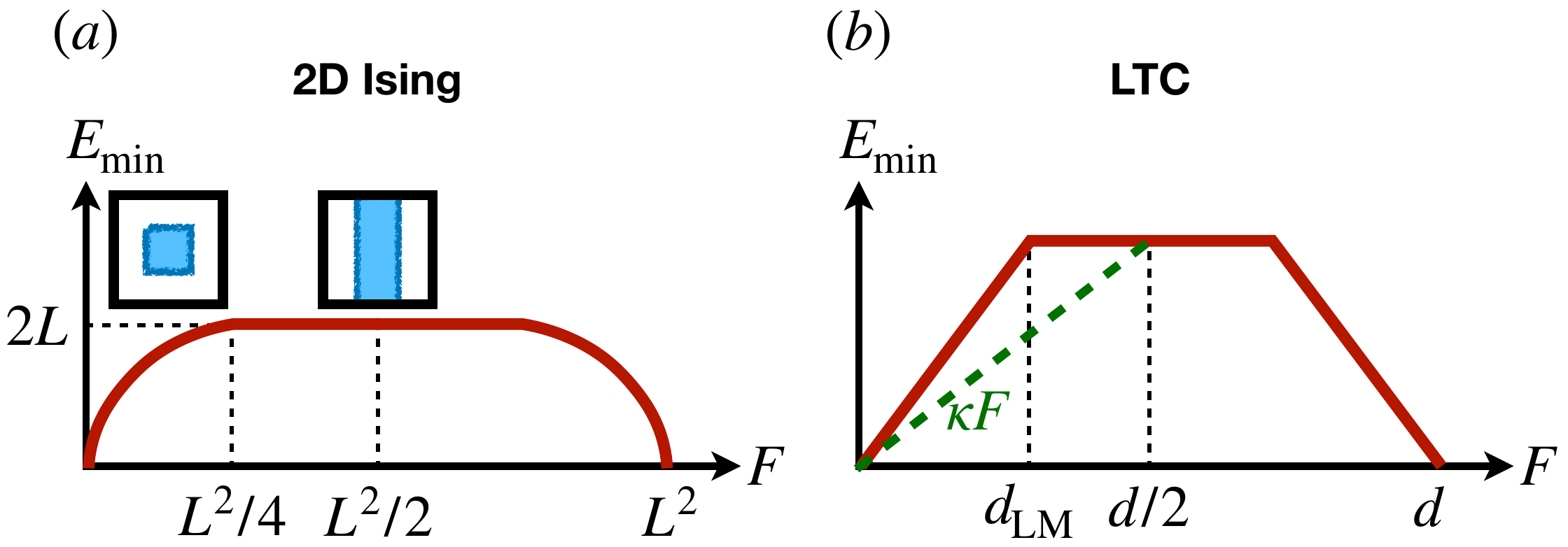}
    \caption{Scaling of energy barriers in various classical codes. (a) the 2D Ising model. Energy grows with surface area (square roof of volume in 2D) until it saturates at twice the size of the smallest non-contractible cocycle. (b) A possible scenario for the locally testable codes of Refs.~\cite{panteleev2022asymptotically,dinur2022locally,lin2022c,leverrier2022quantum}, with linear growth and saturation leading to an overall linear lower bound of Eq.~\eqref{eq:LTC_def}.}
    \label{fig:EnergyBarriers}
\end{figure}

From the perspective of coding theory, the interpretation of Eq.~\eqref{eq:LTC_def} is as follows. Given a bit-string (configuration of $\sigma_i$) we want to ``test'' whether it is a codeword or not, but randomly selecting a check, and asking whether it is satisfied or not. This test requires access to only the $O(1)$ number of bits in the support of the chosen check (hence \emph{local} testability). Eq.~\eqref{eq:LTC_def} then says that this local test succeeds (i.e., does not give false positives) with a probability that is linear in the Hamming distance between the bit-string and the nearest codeword. In other words, if a bit-string is a large, $O(n)$ distance from any codewords, we will be able to identify it with a non-vanishing probability. Note that this behavior is impossible in finite dimensions, where we can also create a large spin-flip configuration that only violates checks at its boundary and will be undetectable in its bulk, as in the 2D Ising example in Fig.~\ref{fig:EnergyBarriers}. 

The 1D and 2D Ising models again provide some useful intuition. In the former, $E_\text{min}$ is constant for $F > 1$, since we can move between the two ground states by nucleating a pair of domain walls and then moving them around the system at no additional energy cost. In 2D, on the other hand, we have $E_\text{min} \propto \sqrt{F}$ (up to $F \leq d/4$), reflecting the fact that a growing island of flipped spins has an extended domain wall at its boundary. Combined with our discussions in Sec.~\ref{sec:Redundancies}, this shows that the behavior of energy barriers is linked to the structure of redundancies. Indeed, the arguments presented in Sec.~\ref{subsec:GlobalRed} show that any locally testable code requires an $O(n)$ number of redundancies~\cite{ben2010locally}.

The existence of good cLDPC codes that are also locally testable has been an open question for a long time, until it has been answered affirmatively in a series of recent papers~\cite{dinur2022locally,panteleev2022asymptotically,lin2022c}. Interestingly, all of these classical codes are constructed explicitly in terms of a 2-term chain complex, with the third level corresponding to local redundancies, just like in the description we provided in Sec.~\ref{subsec:LocalRed}. Indeed, these LTCs were obtained from  very similar (or, in some cases, the same) constructions as the good qLDPC codes with related chain complexes~\cite{panteleev2022asymptotically,dinur2022locally,dinur2023good,lin2022c,lin2022good,leverrier2022quantum}. We can use this to show that all known examples of good qLDPC codes arise from gauging classical LTCs. We will provide a more detailed discussion of these examples and their concrete constructions in forthcoming work~\cite{LDPCProduct}.

A picture of why this might be true can be obtained from considering the quantum-classical dictionary summarized in Sec.~\ref{subsec:Dictionary}. The energy barrier scaling of $\mathcal{C}_\text{cl}$ and the $X$-distance of $\mathcal{C}_\text{q}$ both concern the behavior of cocycles in the underlying chain complex. The former is defined in terms of how the size of contractible cocycles grows with the area they enclose, while $d_X$ measures the size of the smallest non-contractible cycles. Indeed, in the existing proofs~\cite{panteleev2022asymptotically,leverrier2022quantum,dinur2023good,lin2022good} the two properties (local testability of $\mathcal{C}_\text{cl}$ and $d_X \propto n$ for $\mathcal{C}_\text{q}$) both follow from the same geometrical features of the chain complex. 

More concretely, in Refs. \onlinecite{panteleev2022asymptotically,lin2022good}, the authors consider what they call \emph{locally minimal cocycles}, meaning those whose length (i.e., the total number of edges participating in the cocycle) cannot be decreased by flipping a single site. In equations, a cocycle $\ket{c} = \sum_a \beta_a \ket{a}$ is locally minimal if
\begin{equation}
    |\ket{c}| \leq |\ket{c+\delta^T(i)}| \quad \forall i,
\end{equation}
where we used $|\ket{c}|$ to denote the length of a cocycle. One can then define the \emph{locally minimal distance} as the length of the smallest locally minimal cocycle:
\begin{equation}
    d_\text{LM} \equiv \text{min}_{\text{loc. min. cocycles } c} |\ket{c}|.
\end{equation}
A lower bound on $d_\text{LM}$ can then be used to bound \emph{both} the $X$-distance of the quantum code $\mathcal{C}_\text{q}$ and the soundness of $\mathcal{C}_\text{cl}$.

The first of these is easy to see: clearly, the smallest $X$-logical of $\mathcal{C}_\text{q}$ corresponds to a locally minimal cocycle, which implies that $d_X \geq d_\text{LM}$. On the other hand, one can argue~\cite{panteleev2022asymptotically} that $\mathcal{C}_\text{cl}$ satisfies the inequality Eq.~\eqref{eq:LTC_def} with $\kappa = 1$ whenever $E_\text{min}(F) < d_\text{LM}$. This can be seen as follows. Consider a minimal energy configuration for an $F$ that satisfies this condition. The corresponding domain wall configuration is a cocycle that cannot be locally minimal, therefore there exist some spin that one can flip to decrease the energy. This results in a new configuration with a smaller energy, which can also not be locally minimal by assumption, so we can flip a second spin to further lower the energy and so on. As the energy strictly decreases in each step, this process must terminate after at most $E_\text{min}(F)$ steps, by which point we must have reached a zero energy state, so we must have had $F \leq E_\text{min}(F)$ initially. While this is only true for $E_\text{min}(F) < d_\text{LM}$, which will be smaller than $d_\text{cl}/2$, if $d_\text{LM}$ itself is $O(n_\text{q})$, then we can find some $1 > \kappa > 0$ such that Eq.~\eqref{eq:LTC_def} holds (see Fig. \ref{fig:EnergyBarriers}(b)).

We can develop some further intuition about the relationship between $d_X$ and local testability by again considering the example of the 2D Ising model, depicted in Fig.~\ref{fig:EnergyBarriers}(a). After the initial $E_\text{min} \propto \sqrt{F}$ growth, the energy barrier saturates at $E_\text{min} = 2L$. This is due to a transition in the shape of the optimal energy configuration; while for $F < d/4$, the optimal configuration is a $\sqrt{F} \times \sqrt{F}$ square-shaped domain of down spins, when $F \geq d/4$ it becomes energetically favorable to instead form a ``strip'' of down spins, stretching along the entire system in one direction, which has energy $2L$ independent of the width of the strip. The energy of this strip is twice the code distance $d_X$ of the toric code, due to the fact that the boundary of the strip is composed of two non-contractible loops, thus connecting the quantum code distance to the behavior of classical energy barriers. More generally, a similar relationship would hold if it is possible to form a large island of flipped spins, of size $O(n_\text{cl})$, whose boundary consists of a finite number of non-contractible cocycles. This question is analogous to the one discussed in Sec.~\ref{subsec:LocalRed}, of whether one can create configurations of far separated point-like excitations (in the absence of local redundancies). Whether this condition is satisfied in the existing LTC constructions, is an interesting open question. 

As this discussion shows, the local testability of $\mathcal{C}$ is related to the $X$-distance of the quantum code. To obtain a genuinely good qLDPC code, we also need an optimal growth of the $Z$-distance $d_Z$, which is related to the scaling of energy barriers in the second classical code $\tilde{\mathcal{C}}_\text{cl}$. Indeed, in the construction of good qLDPCs, both $\mathcal{C}_\text{cl}$ and $\tilde{\mathcal{C}}_\text{cl}$ are LTCs. 

\section{Higgs-SPT phases}\label{sec:SPT}

In our discussion above, we combined the physical idea of gauging with the chain complex description of CSS codes. When applied to codes with extended domain walls, this gauging procedure connects two main types of phases of matter: those with spontaneously broken (discrete) symmetries and those with topological order (or, in a different parlance, with 0-form and higher form SSB). We will now illustrate the usefulness of this combination of ideas by showing how to derive from it a third kind of order: symmetry protected topological (SPT) phases. This follows on from earlier work~\cite{devakul2019fractal,verresen2022higgs}, which has identified such phases as Higgs phases of gauge theories. Here, we generalize this discussion to the large variety of generalized gauge theories we have been considering here.  

SPT phases are usually characterized by the fact that while they are trivial (have unique GS) on a closed manifold, they become non-trivial with open boundaries. Below, motivated by our previous discussion, we will use directly the hypergraph / chain complex representation as a way of defining the notion of a boundary and of changing boundary conditions. This provides a way of defining and probing SPT phases without reference to an underlying Euclidean lattice, opening up the possibility of studying them in unusual geometries, such as hyperbolic lattices or expander graphs. We view this perspective as ripe for further study. 

\subsection{Higgs-SPT phases from classical codes} 

Consider again the Fradkin-Shenker Hamiltonian~\eqref{eq:H_gauged}. In the cases of usual $D$-dimensional $\mathbb{Z}_2$ gauge theory (i.e., when the classical code $\mathcal{C}$ is the Ising model on a $D$-dimensional cubic lattice) it was pointed out in Ref. \onlinecite{verresen2022higgs} (and mentioned earlier in an appendix of Ref. \onlinecite{devakul2019fractal}) that the so-called \emph{Higgs phase} of the gauge theory can be understood as an SPT phase, protected by a combination of two $\mathbb{Z}_2$ symmetries. One of these is the original global $\mathbb{Z}_2$ symmetry of the Ising model. The nature of the other symmetry depends on the dimension: for $D=1$ it is another $0$-form symmetry, while for $D \geq 2$ it is a higher form symmetry. The SPT nature of these phases can be shown in a number of ways, e.g. by considering non-local string order parameters and by imposing open boundary conditions and studying the resulting edge modes. We will now describe how this discussion generalizes to the generalized gauge theories, constructed out of an arbitrary classical code $\mathcal{C}$ as in the previous section. 

The SPT phases are most easily understood in a modified setup, where we relax the Gauss law constraints\footnote{Nonetheless, note that it was argued in Ref. \onlinecite{verresen2022higgs} that the gauge theory exhibits many of the relevant SPT properties even in the case where the constraints are strictly enforced.}. Thus we now consider the model on the full Hilbert space $\mathcal{H}_0 \otimes \mathcal{H}_1$. To compensate, we add the constraints into the Hamiltonian as an additional term $-\lambda \sum_i G_i$. 

To uncover the SPT phase, we consider the limit $g = \Gamma = 0$ and $J=\lambda$. In this case, the system is described by the stabilizer Hamiltonian
\begin{equation}\label{eq:H_SPT}
    H_\text{SPT} = - \sum_a \tau_a^z \prod_{i \in \delta(a)} \sigma_i^z -\sum_i \sigma_i^x \prod_{a \in \delta^T(i)} \tau_a^x,
\end{equation}
where we have omitted the $B_p$ terms as they can be written as products of the $\tau_a^z C_a$ terms (however, they will be relevant when considering the model with open boundaries below). All terms in~\eqref{eq:H_SPT} commute, which follows from the fact $i \in \delta(a)$ iff $a \in \delta^T(i)$. They are also all independent from each other, with no redundancies between them and there is precisely one term for each qubit\footnote{This can be seen e.g. by noting that there is precisely one term that acts as $\tau_a^z$ on a given gauge qubit $a$, so there is no way to take products of multiple terms to yield an identity.}. Thus $H_\text{SPT}$ has a unique ground state. In fact, we can write it in a more familiar form by performing a Hadamard transform, $\tau_a^x \leftrightarrow \tau_a^z$ on all the gauge fields. The resulting Hamiltonian is nothing but the so-called \emph{cluster model} associated to the Tanner graph of the code $\mathcal{C}$, with its ground state the corresponding \emph{cluster state} (also known as a \emph{graph state}). 

By construction, $H_\text{SPT}$ is symmetric under all the logical operators of both $\mathcal{C}$ and $\mathcal{C}^T$, the former acting as $\mathcal{X}_\lambda \equiv \prod_{\Sigma_\lambda}\sigma_i^x$ and the latter as\footnote{In this section, we use a different notation from the one above, using $\mathcal{Z}_r$ to denote the symmetries of the $\tau$ variables, instead of $R_r$ as e.g. in Eq.~\eqref{eq:KWfull3}.} $\mathcal{Z}_r \equiv \prod_{R_r} \tau_a^z$. Since the ground state is unique, these symmetries are unbroken. Indeed, the system is trivial in the sense that it can be fully disentangled by a local unitary circuit of local \textit{CNOT} gates, defined as
\begin{equation}\label{eq:U_DW}
U_\text{DW} = \prod_i\prod_{a \in \delta_1^T(i)} (\text{CNOT})_{ia},  
\end{equation}
which turns it into $U_\text{DW}^\dagger H_\text{SPT}U_
\text{DW} = -\sum_a \tau_a^z - \sum_i \sigma_i^x$. Note, however, that this transformation does not locally respect the symmetries. 

We can interpret $H_\text{SPT}$ in terms of domain walls, as defined by either the classical code $\mathcal{C}$ or its transpose. In particular, consider, consider a spin-flip operator $X_N = \prod_{i \in N}\sigma_i^x$, creating a set of domain wall excitations at the (co)boundary $\delta^T(N)$. We can define a dressed version of this operator as $\tilde{X}_N \equiv X_N \prod_{a\in \delta^T(N)} \tau_a^x$. The dressing operator, $\prod_{a\in \delta^T(N)} \tau_a^x$ is charged under the $\mathcal{Z}_r$ symmetries, as the two operators anti-commute locally. As $\tilde{X}_N = \prod_{i\in N} G_i$, it takes eigenvalue $+1$ in the ground state of $H_\text{SPT}$, identifying it as a condensate of dressed domain walls, analogous to the `decorated domain wall' picture known for a variety of SPT phases~\cite{chen2014symmetry}. This picture is encompassed by a set of stabilizer relations obtained from requiring that the ground state takes eigenvalue $+1$ under all ther local terms in Eq.~\eqref{eq:H_SPT}:
First note that since the ground state takes eigenvalue $+1$ for all the terms in Eq.~\eqref{eq:H_SPT}, it satisfies the following set of relations:
\begin{align}\label{eq:SPT_stab}
    \tau_a^z \equiv C_a = \prod_{i \in \delta(a)} \sigma_i^z, & & \sigma_i^x \equiv A_i = \prod_{a \in \delta^T(i)} \tau_a^x. 
\end{align}
These tie the domain walls of the symmetries $\mathcal{X}_\lambda$ ($\mathcal{Z}_r$), measured by the eigenvalues of $C_a$ ($A_i$) to the spin flips of the other set of variables. The unitary $U_\text{DW}$ is exactly the operator that performs the appropriate dressing of the domain walls, mapping $\sigma_i^x \to \sigma_i^x A_i$ and $\tau_a^z \to \tau_a^z C_a$. Below, in Sec.~\ref{sec:OrderParams}, we will discuss how this can be used to construct non-local order parameters for the SPT phases.

\begin{figure}[t!]
    \centering
    \includegraphics[trim={1cm 0 0 0},width = 1.\linewidth]{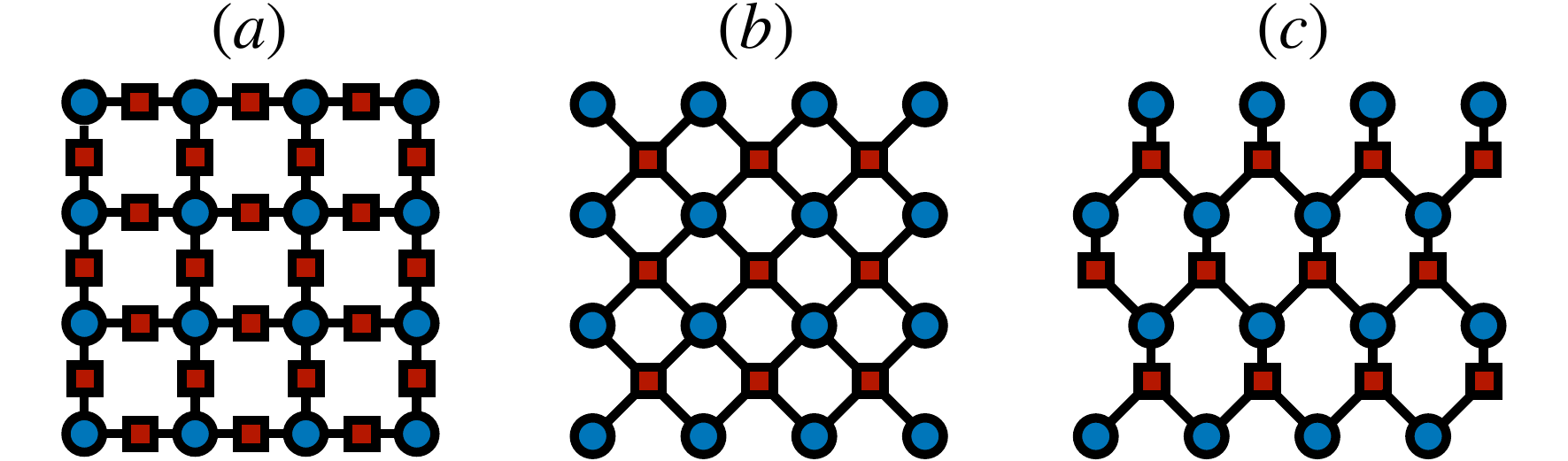}
    \caption{Examples of cluster state SPT phases obtained from the classical codes in Fig.~\ref{fig:ClassicalExamples}. Blue dots (red squares) denote $\sigma$ ($\tau$) degrees of freedom which constitute two sublattices of a bi-partite graph. The ground state of $H_\text{SPT}$ in Eq.~\eqref{eq:H_SPT} is a cluster / graph state on this graph. (a) has global 0-form and 1-form symmetries, (b) has line-like subsystem symmetries along rows and columns, (c) has fractal subsystem symmetries on Sierpinski triangles.}
    \label{fig:Higgs}
\end{figure}

The construction described here can be used to obtain many known examples of SPT phases. For Ising models, this was pointed out in Refs. \onlinecite{devakul2019fractal,verresen2022higgs}. If we take the 1D Ising model as our input classical code $\mathcal{C}$, we obtain the 1D cluster chain~\cite{raussendorf2001one}, which is a canonical examples of an SPT protected by a pair of $\mathbb{Z}_2$ global symmetries, which correspond to the logicals of $\mathcal{C}$ and $\mathcal{C}^T$. From the 2D Ising model, on the other hand, we obtain a cluster state on the so-called Lieb lattice, which exhibits SPT order protected by a combination of a 0-form and a 1-form $\mathbb{Z}_2$ symmetry~\cite{verresen2022higgs,yoshida2016topological}. From our perspective, this follows from the fact that the code $\mathcal{C}^T$, corresponding to the $X$-checks of the toric code, exhibits a 1-form symmetry due to the local redundancies of the 2D Ising model. This is a generic feature of the SPT Hamiltonians in our construction, whenever the corresponding gauge theory exhibits a deconfined phase corresponding to a CSS code with non-trivial distance scaling.

We can also apply the same procedure to other examples of classical codes mentioned in Sec.~\ref{sec:CodeReview} to recover known SPT phases. For example, using the 2D plaquette Ising model, we end up with the 2D cluster model on a (rotated) square lattice, which is the simplest example of an SPT with subsystem symmetries~\cite{you2018subsystem,raussendorf2019computationally}, once again originating from the logicals of the classical code and its transpose. Similarly, applying the procedure to the Newman-Moore model yields the cluster state on a hexagonal lattice, which is an SPT protected by fractal symmetries~\cite{devakul2019fractal}. Some of these examples are illustrated in Fig.~\ref{fig:Higgs}.

Yet another interesting model that can be understood this way is is the Raussendorf-Bravyi-Harrington (RBH) model~\cite{raussendorf2005long}, which is an SPT protected by a pair of $1$-form symmetries. In our language, it can be obtained from the classical code that corresponds to a classical gauge theory in $3D$, i.e. with bits assigned to edges of a cubic lattice, and $4$-spin checks acting on all the edges that form a face of a cube. This classical code is the classical Kramers-Wannider dual (in the sense of Sec.~\ref{sec:Gauging}) to the 3D Ising model whose local redundancies live on the faces. As such, it corresponds to a 3-dimensional chain complex and if we use the middle two layers of the complex in our construction, then the resulting SPT cluster-Hamiltonian has a pair of 1-form symmetries, in analogy of our discussion in Sec.~\ref{subsec:HigherForm}.

\subsection{Generalized Kennedy-Tasaki transformations}\label{subsec:KennedyTasaki}

\begin{figure} 
    \centering
    \includegraphics[trim={1cm 0 0 0},width = 0.75\linewidth]{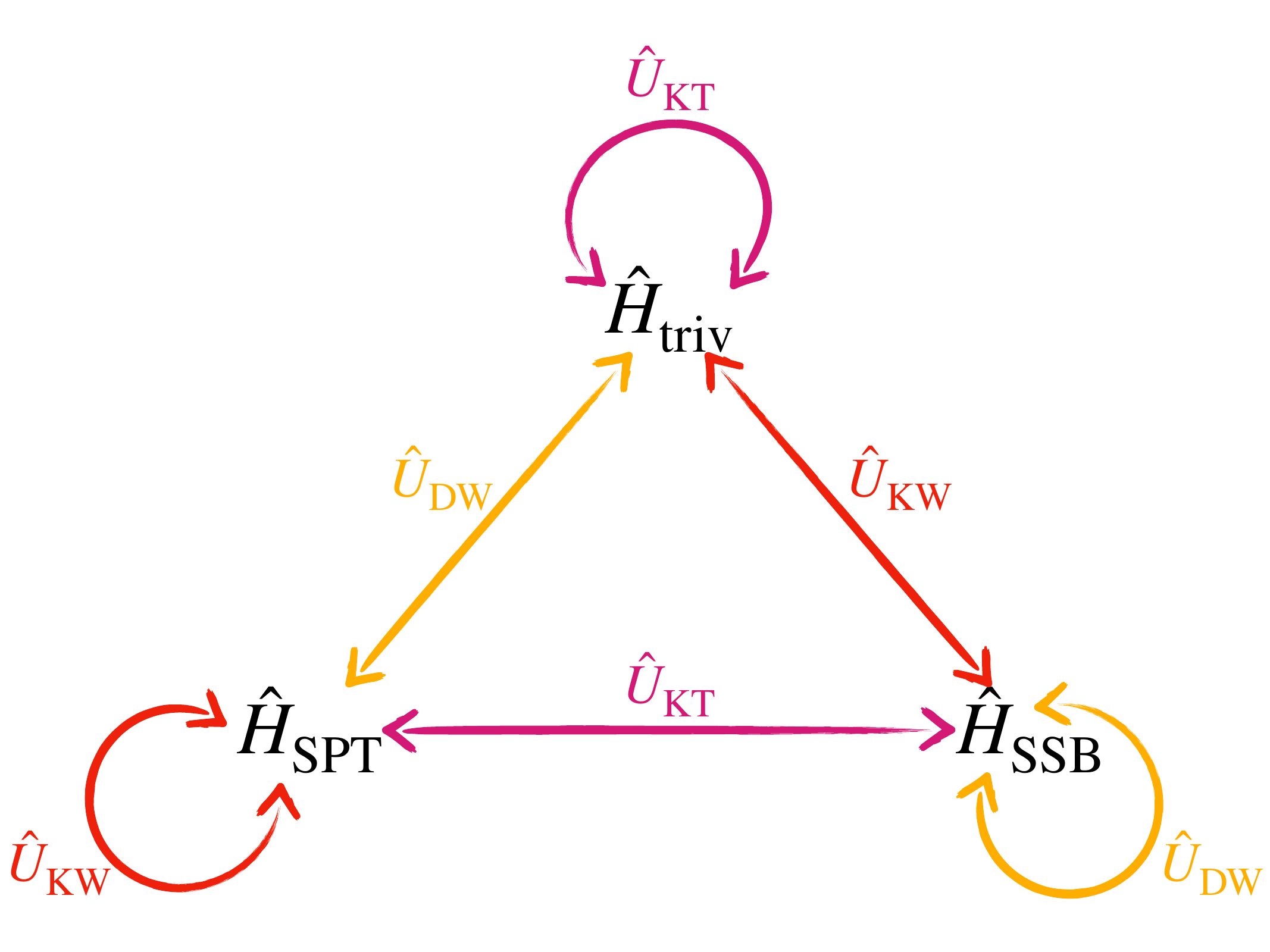}
    \caption{A set of dualities (Kramers-Wannier, Kennedy-Tasaki and domain wall dressing) that relate $H_\text{SPT}$ in Eq.~\eqref{eq:H_SPT} to (a) a trivial Hamiltonian with on-site terms alone and (b) a Hamiltonian that corresponds to two decoupled classical codes. 
    }
    \label{fig:KTduality}
\end{figure}

In this section we discuss how the SPT cluster Hamiltonians constructed in Eq.~\eqref{eq:H_SPT} fit into the picture of Kramers-Wannier dualities discussed in Sec.~\ref{sec:Gauging}. As noted, $H_\text{SPT}$ is symmetric under the combined symmetries of  the classical codes $\mathcal{C}$ and $\mathcal{C}^T$. This means that we can apply the Kramers-Wannier duality of Sec.~\ref{subsec:KramersWannier} to it; however, since it is symmetric between the two codes by construction, $H_\text{SPT}$ is in fact self-dual under this transformation. However, by combining $U_\text{KW}$ with the domain wall dressing operator $U_\text{DW}$, we can define a similar, but different non-local transformation that maps it onto two decoupled Hamiltonians, corresponding to $\mathcal{C}$ and $\mathcal{C}^T$. Following Ref. \onlinecite{li2023non}, we interpret this as a generalized version of the Kennedy-Tasaki (KT) transformation, originally used to identify the SPT order in the one-dimensional Haldane chain as a kind of ``hidden symmetry breaking''~\cite{kennedy1992hidden1,kennedy1992hidden2,pollmann2012symmetry,else2013hidden}. 

To define the KT transformation, let us again introduce ancillary variables $\eta$ and $\mu$ as we did in Sec.~\ref{subsec:BackgroundGauge}, to obtain an enlarged Hilbert space $\hat{\mathcal{H}}_0 \otimes \hat{\mathcal{H}}_1$. Let $\hat{\delta}$ denote the binary matrix that encodes the adjacency relations between the variables $(\sigma,\eta)$ on the one hand, and $(\tau,\mu)$ on the other. We can now define an enlarged version of the domain wall dressing operator, $\hat{U}_\text{DW}$, by replacing $\delta$ with $\hat{\delta}$ in the definition~\eqref{eq:U_DW}. This has the effect of mapping 
\begin{align}
    \hat{U}_\text{DW} \sigma_i^x \hat{U}_\text{DW}^\dagger = \sigma_i^x \hat{A}_i; & & \hat{U}_\text{DW} \sigma_i^z \hat{U}_\text{DW}^\dagger = \sigma_i^z; \\ 
    \hat{U}_\text{DW} \tau_a^z \hat{U}_\text{DW}^\dagger = \tau_a^z \hat{C}_a; & & \hat{U}_\text{DW} \tau_a^x \hat{U}_\text{DW}^\dagger = \tau_a^x,
\end{align}
where $\hat{C}_a$ ($\hat{A}_i$) are the modified checks of $\mathcal{C}$ ($\mathcal{C}^T$) obtained after coupling to the ancillas $\eta$ ($\mu$). 
This transformation maps the trivial Hamiltonian, $\hat{H}_\text{triv} = -\sum_i \sigma_i^x - \sum_a \tau_a^z$ onto the cluster Hamiltonian in the presence of background gauge fields, where one replaces $C_a$ with $\hat{C}_a$ and $A_i$ with $\hat{A}_i$.
 
With this in hand, we define the KT transformation on the extended space following Ref. \onlinecite{li2023non} as 
\begin{equation}
    \hat{U}_\text{KT} \equiv \hat{U}_\text{KW} \hat{U}_\text{DW} \hat{U}_\text{KW},
\end{equation}
with $\hat{U}_\text{KW}$ the Kramers-Wannier duality defined in Eqs.~\eqref{eq:KWfull1}-\eqref{eq:KWfull4}. $\hat{U}_\text{KT}$ has the effect of mapping $\hat{H}_\text{SPT}$ onto the sum of the decoupled classical code Hamiltonians of $\mathcal{C}$ and its transpose, both in the presence of background gauge fields:
\begin{equation}
    \hat{U}_\text{KT} \hat{H}_\text{SPT} \hat{U}_\text{KT}^\dagger = \hat{H}_\text{SSB} \equiv -\sum_a \hat{C}_a -\sum_i \hat{A}_i.
\end{equation}

The three transformations, $\hat{U}_\text{KW}$, $\hat{U}_\text{DW}$ and $\hat{U}_\text{KT}$ form a set of relationships between the trivial, SPT and SSB phases, which is summarized in Fig.~\ref{fig:KTduality} (see also Fig. 9 in Ref. \onlinecite{cao2023subsystem}). We could also restrict all of these transformations to the subspace where all the symmetries take eigenvalue $+1$ and fix $\eta_r^z = \mu_\lambda^x = +1$ to get mappings between symmetric subspaces of the three models. More generally, the transformations could be applied to any symmetric Hamiltonian of the combined $\sigma,\tau$ degrees of freedom. For example, transverse field terms remain unchanged under $U_\text{KT}$, so that e.g. a critical point between the SSB and trivial phases should map onto the critical point between the SPT and the trivial phase. We could also consider a Hamiltonian interpolating between $H_\text{SPT}$ and $H_\text{SSB}$, which would include a point that is self-dual under $U_\text{KT}$. Much like with the generalized Kramers-Wannier dualities, it will be interesting to explore the properties of these self-dual points and what constraints they put on possible critical points between different phases.

It is also worth briefly considering the interplay between the KT transformation and the configuration of background gauge fields $\eta^z$ and $\mu^x$. As we saw in Sec.~\ref{subsec:BackgroundGauge}, the KW duality exchanges the eigenvalues of the background fields with the symmetry eigenvalues of the dual theory, e.g. $\eta_r^z \to \mathcal{Z}_r = \prod_{a\in R_r}\tau_a^z$. The unitary $\hat{U}_\text{DW}$ then maps $\tau_a^z \to \tau_a^z \hat{C}_a$. Since $\prod_{a\in R_r} \hat{C}_a = \eta_r^z$ by definition, we overall have the following set of mappings
\begin{equation}
    \eta_r^z \xrightarrow{\text{KW}}
    \mathcal{Z}_r 
    \xrightarrow{\text{DW}}
    \mathcal{Z}_r \eta_r^z
    \xrightarrow{\text{KW}}
    \eta_r^z \mathcal{Z}_r.
\end{equation}
We thus see that the KT transformation mixes the background gauge field configurations of one set of variables with the symmetries of the other, a feature that was noted in the 1D case in Ref. \onlinecite{li2023non}. There is a similar set of transformation rules for the symmetry eigenvalues $\mathcal{X}_\lambda$ and the corresponding gauge dual gauge fields $\mu_\lambda^x$. 

\subsection{Edge modes}

A crucial feature of SPT phases is that they host non-trivial degrees of freedom at their boundary when defined with open boundary conditions. These edge modes are protected from being gapped out by symmetric perturbations on the boundary, as long as the bulk is in the SPT phase, a manifestation of so-called bulk-boundary correspondence. How might one go about defining edge modes in the generic case, where there is no underlying Euclidean lattice to speak of? Here, we import insights from our preceding discussion in Sec.~\ref{sec:Redundancies}, where we argued that one can define boundary conditions directly in terms of the classical code / chain complex itself. We discuss how to define a version of $H_\text{SPT}$ with open boundaries this way, and argue that the resulting Hamiltonian exhibits a number of edge modes localized at the boundary that we created. We will then discuss whether these edge modes can be gapped out by symmetric perturbations and relate this to properties of the underlying classical code. 

\subsubsection{Without local redundancies}

Consider the case when $\mathcal{C}$ has only global redundancies. We can follow the recipe outlined in Sec.~\ref{subsubsec:1DBC} to identify a set of labels $\mathcal{L} \equiv \{a_r\}_{r=1}^{k^T}$ that can be dropped to define open boundary conditions. The remaining chain complex has a set of ``boundary sites'' $\partial \Lambda$, defined as those sites $i$ that are connected to at least one of the edges $a$ in $\mathcal{L}$. We argue that the symmetries ``fractionalize'' onto operator acting on this boundary and lead to ground state degeneracies associatdd to operators localized at the boundary. 

To show this, we first want to appropriately modify the Hamiltonian $H_\text{SPT}$ to reflect the change of boundary conditions but without violating any of its symmetries. We can achieve this by dropping all terms from Eq.~\eqref{eq:H_SPT} which involve any $\tau_a^{x,z}$ with $a \in \mathcal{L}$ to get a new Hamiltonian $H_\text{SPT}^\text{OBC}$\footnote{Note that the resulting Hamiltonian is different from what we would get if we first dropped the checks $C_{{a_l}}$ from the classical code $\mathcal{C}$ and then constructed $H_\text{SPT}$ out of that modified code using Eq.~\eqref{eq:H_SPT}. The latter model would involve truncated versions of the $\sigma_i^z A_i$ terms for $i$ where $\delta^T(i)$ overlaps with $\mathcal{L}$. However, these terms would violate some of the $\mathcal{Z}_r$ symmetries.}. 

Consider now the effect of a symmetry operator $\mathcal{Z}_r$ on the ground state of $H_\text{SPT}^\text{OBC}$. Its support is now truncated from $R_r' \equiv R_r \setminus \mathcal{L}$. By construction, $R_r$ contains exactly one of the elements in $\mathcal{L}$, let us denote it by $a_r$. We thus have $\delta(R_r') = \delta(R_r) + \delta(a_r) = \delta(a_r) \subseteq \partial \Lambda$. Combining this with the stabilizer conditions from eq.~\eqref{eq:SPT_stab}, we find that in the ground state, $\braket{\mathcal{Z}_r}_\text{OBC} = \braket{\prod_{i\in\delta(a_r)} \sigma_i^z}$. Thus, the symmetry no longer acts trivially on the GS; instead it has some non-trivial action localized in the vicinity of the hyperedges $a_r$ that have been ``cut'' to impose OBC. Moreover, consider any $i \in \delta(a_r)$. Since $a_r \in \delta^T(i)$, the corresponding $\sigma_i^x A_i$ are absent from $H_\text{SPT}^\text{OBC}$ and therefore $[H_\text{SPT}^\text{OBC},\sigma_i^z] = 0$. $\mathcal{Z}_r$ is thus now a product of observables which all individually commute with the Hamiltonian. This is an example of symmetry fractionalization, characteristic of SPT phases.

An analogous fractionalization occurs for the other set of symmetries, $X_\lambda = \prod_{i \in {\Sigma_\lambda} }\sigma_i^x$; note that here the support does not need to be modified to accommodate OBC, since we only dropped $\tau$ degrees of freedom. Let $K_\lambda = \Sigma_\lambda \cap \partial\Lambda$ denote the set of boundary sites appearing in the support of $\mathcal{X}_\lambda$. Again using Eq.~\eqref{eq:SPT_stab} we have
\begin{align*}
\braket{\mathcal{X}_\lambda}_\text{OBC} &= \braket{\prod_{i \in K_\lambda} \sigma_i^x  \prod_{j \in \Sigma_\lambda \setminus K_\lambda}A_j}_\text{OBC} \\ &= \braket{\prod_{i \in K_\lambda} \sigma_i^x  \prod_{a \in \delta^T(K_\lambda)}\tau_a^x}_\text{OBC} = \braket{\prod_{i \in K_\lambda} G_i}_\text{OBC},
\end{align*}
where in the last equality we made use of the fact that $\delta^T(\Sigma_\lambda) = 0$. Since, by construction, this symmetry operator is only acting on the remaining degrees of freedom (after dropping the $\tau$ spins belonging to $\mathcal{L}$), we can replace $G_i$ in the expression above with a truncated version $G_i' \equiv \sigma_i^x \prod_{a \in \delta^T(i) \setminus \mathcal{L}} \tau_a^x$, such that $\braket{\mathcal{X}_\lambda}_\text{OBC} = \braket{\prod_{i \in K_\lambda} G_i'}_\text{OBC}$. By construction, $G_i'$ commutes with $\tau_a^z C_a$ for $a \notin \mathcal{L}$. It can anti-commute with terms centered around $a \in \mathcal{L}$; however, these are precisely the terms that were removed in taking OBC, so we have $[H_\text{SPT}^\text{OBC},G_i'] = 0$. Again, we see that the symmetry fractionalizes into multiple observables, each localized near the boundary, which individually commute with the Hamiltonian. Furthermore, since $G_i'$ involves $\sigma_i^x$ and no other matter qubit, we have that $G_i'\sigma_i^z = - \sigma_i^z G_i'$, i.e. the two types of edge operators anti-commute. Thus $H_\text{SPT}^\text{OBC}$ has a ground state degeneracy of $2^{|\partial\Lambda|}$.

This argument shows that the truncation of $H_\text{SPT}$ to open boundaries has a ground state degeneracy, which scales with the size of the boundary introduced. However, a crucial question is how much of this degeneracy can be lifted by additional symmetry-preserving perturbations on the boundary. One usually argues that in fact no such perturbations exist, and the individual edge modes remain well-defined, at least at low energies. This conclusion obtains when the individual edge modes are far separated from each other, so that they cannot be coupled by local perturbations. Whether this is the case is closely related to the question which we touched upon in Sec.~\ref{subsec:LocalRed}, of whether the bulk hosts point-like excitations that can be separated from each other: it is essentially these excitations that turn onto edge modes at the boundary. As discussed there, this is a general feature of translation invariant lattice models, and indeed, we have seen that the examples of this kind produce Hamiltonians that are known to realize non-trivial SPT phases. The situation on non-Euclidean geometries is less clear. As we discussed in Sec.~\ref{subsec:LocalRed}, in some cases one can argue that well-separated point-like excitations exist. More generally, understanding the conditions under which at least some of the edge modes are robust to symmetric perturbations is a problem that warrants further study. 

One way of approaching this problem is using the KT transformation defined in Sec.~\ref{subsec:KennedyTasaki}. By going to OBC and dropping the terms corresponding to $a \in \mathcal{L}$, we have removed the redundancies of the $\sigma$ variables, thus we can define the Kramers-Wannier transformation without the need to introduce $\eta$ variables or restrict to the symmetric subspace of the $\mathcal{Z}_r$ symmetries. Now consider the set $\mathcal{M}$ of sites $i$ that have a gauge field $\mu$ associated to them. If it is possible to choose these such that they are all on the boundary, $\mathcal{M} \subseteq \partial\Lambda$, then the corresponding $\sigma_i^x$ are missing from $H_\text{SPT}^\text{OBC}$. In this case, we can define a KT transformation that maps $H_\text{SPT}^\text{OBC}$ onto $H_\text{SSB}$ with the same choice of open boundary conditions. By construction, the latter has the same ground state degeneracy as it does with the original periodic boundaries, so that the GS degeneracy of the SPT is this case is equal to $k + k^T$. In this case, symmetric perturbations to $H_\text{SPT}^\text{OBC}$ map onto symmetric local perturbations to $H_\text{SSB}$, whose ground state degeneracy we expect to be robust due to its large code distance. However, it will not always be possible to choose $\mathcal{M}$ appropriately, e.g. if $k \gg k^T$. 

\subsubsection{With local redundancies}

Let us now discuss the case when the classical code has local redundancies and thus forms part of a $2$-complex. In this case, we will refer to the $Z$-type symmetries, $\mathcal{Z}_r$ of the SPT as ``magnetic symmetries'', in analogy of the case of usual $\mathbb{Z}_2$ gauge theory~\cite{verresen2022higgs} (which corresponds to the special case when we start from the Ising model as our classical code). Since we now have a $2$-complex, the change to open boundaries is achieved by removing checks that form either non-trivial cycles or co-cocycles; as discussed in Sec.~\ref{subsec:GlobalRed}, these two cases correspond to two different types of boundary conditions which can be called ``rough'' or ``smooth'' (one might picture the 2D square lattice for illustration). Here, we will focus on the former case, where the edge theory will take a simpler form, involving only gauge fields, while for smooth boundaries one generally needs to involve both matter and gauge fields in the edge theory~\cite{verresen2022higgs}. 

Let us choose a complete set of non-trivial cycles (i.e., one from each homology class) and denote by $\mathcal{M}$ the set of edges involved in them. Let furthermore $\mathcal{N}$ denote the set of vertices that are adjacent to some edge in $\mathcal{M}$. We change boundary conditions by dropping $\mathcal{N}$ and $\mathcal{M}$ from the chain complex, this arriving at a system with fewer matter and gauge qubits. There is now a set of remaining ``dangling'' edges $\mathcal{L}$, defined as those edges that are adjacent to some vertex in $\mathcal{N}$ but are not in the set $\mathcal{M}$ of edges we removed; these dangling edges will constitute the relevant edge degrees of freedom of the system. 

To define $H_\text{SPT}^\text{OBC}$, we drop not only the terms centered on elements of $\mathcal{M} \cup \mathcal{N}$ but also the edge terms centered on $\mathcal{L}$. This is necessary in order to preserve the symmetries. For example, consider some matter symmetry, truncated to the system with OBC: $\mathcal{X}_\lambda = \prod_{i\in \Sigma_\lambda \setminus \mathcal{N}} \sigma_i^x$. This commutes with the edge terms, $\tau_a^z C_a$, for bulk edges $a \notin \mathcal{L}$, but can fail to commute with them when $a \in \mathcal{L}$. The truncated magnetic symmetry, $\mathcal{Z}_r = \prod_{a\in R_r \setminus \mathcal{M}} \tau_a^z$ is maintained, since the only vertex terms ($\sigma_i^x A_i$) it might anti-commute with are those with $i \in \mathcal{N}$, which have already been dropped from the Hamiltonian. Since we now have local redundancies, $H_\text{SPT}$ also includes plaquette terms $B_p$. At the boundary, we need to truncate these appropriately, e.g. $B_p = \prod_{\delta_2(p) \setminus \mathcal{M}} \tau_a^z$. Since this only involves gauge qubits, it continues to commute with the matter symmetries. 

We can now ask how the symmetries fractionalize at the boundary. Since for all bulk vertices, $i \notin \mathcal{N}$, the vertex terms remain, we can use them to rewrite the matter symmetry as
\begin{equation}
    \mathcal{X}_\lambda = \prod_{\Sigma_\lambda \setminus \mathcal{N}} \sigma_i^x = \prod_{\delta_1^T(\Sigma_\lambda \setminus \mathcal{N})} \tau_a^x = \prod_{\delta_1^T(\Sigma_\lambda \cap \mathcal{N})} \tau_a^x, 
\end{equation}
where in the last step we used that $\delta_1^T(\Sigma_\lambda) = 0$ by virtue of it being a logical of the classical code. Since $\delta_1^T(\Sigma_\lambda \cap \mathcal{N}) \subseteq \mathcal{L}$, we find that the symmetry acts only on the dangling boundary edges. With regards to the magnetic symmetries, let $\bar{R}_r \equiv R_r \setminus \mathcal{M}$ be the support of the truncated symmetry operator. Then we can separate the symmetry operator into two pieces:
\begin{equation}\label{eq:Umag_OBC}
    \mathcal{Z}_r = \prod_{\bar{R}_r} \tau_a^z = \prod_{\bar{R}_r \setminus \mathcal{L}} \tau_a^z \prod_{\bar{R}_r \cap \mathcal{L}} \tau_a^z.
\end{equation}
The second product acts only on the boundary, since $\bar{R}_r \cap \mathcal{L} \subseteq \mathcal{L}$. The first product only involves edges in the bulk, for which the edge terms of the form $\tau_a^z C_a$ are present in the Hamiltonian. Thus, with regards to the ground state, we can rewrite this part as $\prod_{\delta_1(\bar{R}_r \setminus \mathcal{L})}\sigma_i^z$. Now, while the Gauss law is no longer enforced as a strict constraint, it is still effective at low energies, so we can use it to ``gauge fix'' the matter fields to $\sigma_i^z = +1$. This gets rid of the first product in~\eqref{eq:Umag_OBC} and leaves us with a symmetry acting solely on the danging edges $\mathcal{L}$.

Indeed, we can use the same gauge fixing idea to get rid of all bulk degrees of freedom and write down a theory for the edge alone. We use the Gauss law terms to enforce $\sigma_i^z = +1$ for all $i$. When this is done, the remaining edge terms become $\tau_a^z$ so that we have $\tau_a^z = +1$ for $a \notin \mathcal{L}$, leaving us with only the edge degrees of freedom. Having exhausted all the vertex and edge terms, we are left with the plaquette terms near the boundary, which define an effective classical code on the edge, with terms $\tilde{B}_p = \prod_{a \in \delta_2(p) \cap \mathcal{L}} \tau_a^z$. 

The effective boundary code with have some degeneracy $k_\text{bdry}$. The question remains how much of this degeneracy can be split by symmetry-preserving terms. The matter symmetries act as logicals of this code and restrict the kind of $\tau_a^z$ terms allowed (e.g. ruling out any longitudinal fields) and necessitating a degeneracy at the classical level. One could try to break these by adding off-diagonal perturbations, composed of products of $\tau_a^x$. These, however, are restricted by the magnetic symmetries. In particular, on any edge that is acted upon by some such symmetry, we cannot add a simple transverse field term to drive it into a trivial paramagnet. The precise nature of the resulting edge order depends on the details of the symmetries. In Euclidean models, the symmetry on the boundary acts on regions that are one dimension lower than their bulk counterparts. For example, in the case of the 2D ising model, the loop-like symmetries associated to Wilson loops terminate at point at the boundary; these therefore rule out \emph{any} $\tau_a^x$-type perturbation (see Ref. \onlinecite{verresen2022higgs} for details). In the RBH model, on the other hand, the bulk symmetries are associated to closed two-dimensional surfaces, which terminate on loops at the boundary. Such a symmetry is consistent with adding Toric-code type $X$-checks to the boundary, lifting the degeneracy at the boundary to be $k_\text{bdry} = 2$ [RBH]. A common feature of these models is that even when we allow for breaking of the higher form symmetry, and thus for transverse fields at the boundary, the edge redundancies remain up to some finite perturbation strength and are thus more stable than the edge modes associated to $0$-form symmetries. Once again, working out the nature of edge modes in non-Euclidean cases is an interesting problem for future work.

\section{Phases and order parameters}\label{sec:OrderParams}

In this paper we used cLDPC codes as a starting point, which we interpreted as spontaneous symmetry breaking phases, with the symmetries given by the logical operators of the code. From these, we constructed qLDPC codes that correspond to fixed points of the deconfined phase of a gauge theory. These exhibit a kind of topological order, or, equivalently, a spontaneously broken higher form symmetry, which can be interpreted in the chain complex language. Finally, in Sec.~\ref{sec:SPT} we used cLDPC codes to construct cluster states and we argued that they exhibit features of SPT phases, where the relevant symmetries are inherited from the classical code. From a physical perspective, a natural question is what happens when we perturbed away from these fixed-point stabilizer Hamiltonians. In this last section, we offer some remarks about the stability of these phases, and how one might go about diagnozing them. 

For phases that correspond to non-trivial stabilizer codes, we have already encountered this question in Sections \ref{sec:Gauging} and \ref{sec:Deconfined}, where cLDPC and qLDPC codes naturally appeared as special points in a larger phase diagram, subject to additional local fields. As mentioned there, we expect that the macroscopic code distance of the underlying code should translate into a stability of their Hamiltonian to these perturbations, as one needs to go to order $O(d)$ in perturbation theory to connect the different ground states (assuming symmetric perturbations in the cLDPC case). We thus expect a phase transition (or a series of phase transitions) separating the symmetry broken phases from the ``trivial paramagnet'' phase that occurs at large fields, with the symmetry broken phase still exhibiting the $2^k$-fold ground state degeneracy, up to a small $e^{-O(d)}$ energy splitting, in the vicinity of the stabilizer point. 

The situation with SPT phases is more subtle, since they have a unique ground state with ``periodic'' boundaries. One way of defining the SPT phase in this case would be in terms of robust degeneracies associated to edge modes.  As discussed in Sec.~\ref{sec:SPT}, this will likely depend on more detailed structure of the underlying classical code, i.e. whether domain walls (either point-like ones or non-contractible loops) can be separated from each other. Nevertheless, we expect that there will be examples even in the non-Euclidean context, where some robust edge degeneracy remains and can be used to define an SPT phase, which disappears only at a bulk phase transition. Another way of diagnozing the SPT phase, without the explicit need for open boundaries, might be in terms of its \emph{entanglement spectrum}~\cite{pollmann2010entanglement}.

The series of duality transformations we have described in this paper provide maps between these various phases (see Fig.~\ref{fig:KTduality} in particular). As such, they can be used to guide the exploration of these phase diagrams, relating the different phases and phase-transition to each other. In particular, they might be used to constrain critical points where they appear as symmetries.

Another question is how to diagnoze these phases, away from their solvable fixed point limits, without having to evaluate the ground state degeneracy explicitly. The constructions from codes can be used to motivate a set of order parameters, as we now discuss. However, as we will discuss, their expected behavior is a subtle question, at least on expander graphs. 

Let us first consider symmetry broken phases corresponding to cLDPC codes. As we noted before, the different codewords can be uniquely labeled by a set of operators $\mathcal{Z}_\lambda$, $\lambda=1,\ldots,k$, which we can choose to be single-site Pauli operators, $\mathcal{Z}_\lambda = \sigma_{i_\lambda}^z$. Therefore, we could use the expectation values of these as local order parameters. The $2^k$ ground states, which we expect to exist in the fully symmetry broken phase, could be labeled by $\braket{\sigma_{i_\lambda}^z}$. 

A downside of the local order parameter is that it requires one to be in a symmetry broken ground state, rather than their symmetric superposition. A more agnostic diagnostic is to look at long-range correlations. In particular, consider the operator $O_M \equiv \prod_{a \in M} C_a = \prod_{i \in \delta(M)} \sigma_i^z$; by definition, this takes the value $+1$ in any of the codewords. Away from the cLDPC point, but still within the SSB phase, we expect that it should exhibit a \emph{perimeter law}, $\braket{O_M} = e^{-O(|\delta(M)|)}$, as opposed an \emph{area law}, $\braket{O_M} = e^{-O(|M|)}$ in the symmetry unbroken phase. However, this distinction will only make sense for certain classical codes, where one can achieve a separation of scales between the size of $M$ and the size of its boundary. This might not be the case e.g. if the Tanner graph of $\delta$ is sufficiently quickly expanding, as discussed in Sec.~\ref{sec:CodeReview}\footnote{Note that the relevant question here is expansion from checks to bits, rather than the other way around, which is important for establishing a good code distance. For example, for the Ising model on an expander graph, a simple two-point function $\braket{\sigma_i^z\sigma_j^z}$ would still work as an order parameter.}.

We can relate these two diagnostics to each other by a kind of Peierls' argument~\cite{peierls1936ising}. In this case, we can consider $\{i_\lambda\}$ as located at the boundary of the system. We could imagine introducing a set of strong magnetic field on these boundary sites that pin them to $\sigma_{i_\lambda}^z = \mu_\lambda = \pm 1$, effectively removing them from the code, where the sign $\mu_\lambda$ is some value we can fix arbitrarily. One can then ask whether different pinning patterns (i.e., different values of $\mu_\lambda$) have different local expectation values far from the ``boundary''. As the pinning is designed to pick out a unique ground state, it removes all logical operators; this means that for any $i$, there exist some set of checks $N_i$, and some set of boundary sites $K_i$ such that $\sigma_i^z = \prod_{\lambda \in K_i} \mu_\lambda \prod_{a \in N_i} C_a$ where $K_i \cup \{i\} = \delta(N_i)$; therefore, we have that $\braket{O_{N_i}} = \braket{\sigma_i^z} \prod_{\lambda \in K_i} \mu_\lambda$, relating the local expectation value $\braket{\sigma_i^z}$ to the correlator $\braket{O_{N_i}}$. 

The disorder operators, introduced in Sec.~\ref{sec:Gauging}, provide a complementary viewpoint on the two phases. As these measure violations of the checks $C_a$, they have expectation value $0$ in any of the ground states of the cLDPC code. On the other hand, in ther trivial phase, $H_\text{triv} = -\sum_a \sigma_i^x$, they have expectation value $1$. Thus, in a code without redundancies, we should be able to use the expectation value $\braket{D_a}$ as an order parameter that is dual to $\braket{\sigma_i^z}$, taking finite values in the trivial phase but not in the symmetry broken one. The analogue of $O_M$ would then be the multi-point function of disorder operators, i.e. a spin flip $P_N = \prod_{i\in N} \sigma_i^x = \prod_{a \in \delta^T(i)} D_a$. The two viewpoints are connected by the KW duality: as we saw, the disorder operator $D_a$ maps onto $\tau_a^x$, which serves as a local order parameter for the transpose code $H(\mathcal{C}^T)$. Similarly, $P_N$ maps onto $\prod_{a \in \delta^T(N)} \tau_a^x = \prod_{i \in N} A_i$. Therefore, the perimeter vs area law of the disorder parameter $P_N$ is equivalent to the area vs perimeter law of the order parameter in the dual theory. This also provides diagnostics for the topologically ordered deconfined phase of Sec.~\ref{sec:Deconfined}, which can be obtained by applying the KW duality to order/disorder parameters of either of the two classical codes $\mathcal{C}_\text{cl}$ and $\tilde{\mathcal{C}}_\text{cl}$. 

Moving on the SPT phases, we can use the decorated domain wall picture to motivate non-local order parameters for these, which can be obtained from the aforementioned order parameters via the duality transformations in Fig.~\ref{fig:KTduality}. For example, while the naive disorder operator $\braket{P_N}$ vanishes in the cluster state (despite it being fully symmetric), we can conjugate it with $U_\text{DW}$ to obtain a dressed disorder operator $\tilde{P}_N = \prod_{i \in N} \sigma_i^x \prod_{a \in \delta^T(N)} \tau_a^x$. Noting that this is a product of Gauss law terms, $\tilde{P}_N = \prod_{i \in N} G_i$, we can see that this now takes expectation value $\braket{\tilde{P}_N} = 1$ in the cluster state, identifying it as a condensate of dressed domain walls. Note that the dressing at the boundary, $\prod_{\delta^T(N)}\tau_a^x$, is exactly the operator charged under the ``magnetic'' symmetries, $\mathcal{Z}_r$. This feature, i.e. the boundary dressing having non-trivial symmetry charge, is often taken as a defining feature of SPTs~\cite{chen2014symmetry}. Thus, $\tilde{P}_N$ provides a non-local order parameter for the SPT phase. Obviously, there is another dual order parameter of the form $\tilde{O}_{M} = \prod_{a\in M}\tau_a^z \prod_{i \in \delta(M)}\sigma_i^z$. Away from its stabilizer fixed point, one would again expect that the SPT phase is characterized by a perimeter, as opposed to area law decay of these observables.

Here, we proposed the perimeter vs area law scaling of certain observables as a possible diagnostic of various phases. This is well defined in any finite dimensional Euclidean space, and might also be well-defined in some other graphs, but not always. For example, as mentioned above, the distinction between perimeter and area might become ill-defined for Hamiltonians defined on expander graphs. For example, in the expander codes discussed in Sec.~\ref{sec:CodeReview}, we have that whenever $|N| \leq \gamma n$ (with $\gamma > 0$ some constant), the boundary has a size $|\partial^T(N)| \propto |N|$, so that the disorder parameter $\braket{P_N}$ would be expected to have the same scaling in both the SSB and the trivial phases. This is an issue in the usual order of limits where we take the system of the system to infinite first and then consider operators of increasing size. In some cases, we can get around this by relaxing the definition of an order parameter and considering operators whose size scales with $n$ itself (but still below $d/2$): in this case, one might be able to find $N$ where $|N|$ and $|\delta(N)|$ scale differently with $n$, such that they can be used as a diagnostic. Note, however, that even this would fail for the locally testable codes discussed in Sec.~\ref{subsec:LTC}, where $|N|$ and $|\delta^T(N)|$ are always proportional. However, one should be able to use either order or disorder parameters as long as $\mathcal{C}$ and $\mathcal{C}^T$ are not both locally testable, which might be impossible. 

A possible alternative characterization of a phase, which could get around the issue of perimeter vs area law, is to probe the symmetry broken phase directly by using background gauge fields to induce domain walls and look at the energy difference between the resulting configurations and the original ground states. For the completely trivial paramagnetic Hamiltonian, these background fields have no effect on the ground state energy---indeed, this is precisely how the ground state degeneracy of the cLDPC shows up on the other side of the KW duality. In stabilizer cLDPC limit, on the other hand, domain walls have a finite line tension, so introducing them comes at an energy cost. We expect that these different scaling will survive away from the fixed point limits as well. 

\section{Conclusions and outlook}\label{sec:Conclusions}

In this paper, we discussed error correcting codes from the perspective of defining quantum phases of matter, motivated by recent advances that resulted in the construction of so-called good quantum LDPC codes, which require qubits to be arranged on expander graphs that go beyond the possibilities of finite-dimensional lattices. Here, we initiated a discussion of such codes from a physical perspective, relying on the ideas of gauging and gauge dualities~\cite{wegner1971duality,levin2012braiding,vijay2016fracton,williamson2016fractal,shirley2019foliated,kubica2018ungauging}, and combining them with notions from the theory of error correcting codes, in particularly the language of chain complexes. 

In particular, we have discussed how to interpret generic classical LDPC codes as physical systems with a potentially large number of symmetries corresponding to logical operators of the code. This gives rise to the idea of gauging these symmetries and we described how to use the structure of the code to define background gauge fields that achieve this. We discussed how this gauging procedure gives rise to a general set of Kramers-Wannier dualities where the background gauge fields on one side of the duality become symmetries on the other. Furthermore, we have highlighted how the structure of local redundancies can be used to distinguish two kinds of codes, depending on whether their excitations are point-like or extended. We showed how the latter give rise to quantum CSS codes, that appear as a special point within the larger phase diagram of the corresponding gauge theory and discussed how good qLDPC codes arise from gauging locally testable classical codes. While here we only outlined briefly the kind of constructions that lead to both classical LTCs and good qLDPC codes, in upcoming work~\cite{LDPCProduct} we will discuss them in more detail and explore them also from the perspective of constructing new phases of matter from existing ones more generally.

In the last part of the paper, we used the gauge theory interpretation to construct a large class of SPT Hamiltonians from arbitrary classical codes, which correspond to graph states on the associated Tanner graph. Building on our discussion of Kramers-Wannier dualities, we described a set of generalized Kennedy-Tasaki transformations that map these Hamiltonians back to classical codes. We utilized the relationship between codes and geometries to discuss non-local order parameters and edge modes in these Hamiltonians in a unified language that covers many interesting examples and can be applied directly to non-Euclidean geometries. 

Our work raises a number of interesting follow-up questions. Some of these concern the relationship between redundancies and domain wall excitations discussed in Sec.~\ref{sec:Redundancies}. There, we advocated for a view that in codes without local redundancies, domain walls can be thought of as point-like excitations. However, while we described some cases where this claim can be made more precise, understanding when exactly this can be done remains an open problem. In particular, it would be useful to know what structure of redundancies is necessary to ensure that it is possible to create a finite, $O(1)$ number of excitations all separated from from each other. As discussed in Sec.~\ref{subsec:LTC}, a similar issue arises in trying to relate local testability of classical codes to the quantum code distance of their gauge duals. One possibility for making progress on these issues might be to focus on codes that exhibit an analogue of translation invariance, i.e. a symmetry that can map any spin $i$ to any other. In general, we expect that many of the connections we discussed, such as boundary conditions and edge modes, can be put on a more solid ground in this case.

Another obvious avenue for future work, which we briefly discussed in Sec.~\ref{sec:OrderParams}, is to understand which features of the stabilizer Hamiltonians constructed from codes are robust to small (symmetry-preserving) perturbations, which can be used to define phases of matter in the non-Euclidean context. In particular, we conjectured that the finite decoding threshold of LDPC codes with macroscopic code distance should also imply a robustness to perturbations of the Hamiltonian; substantiating this claim, or finding counter-examples to it, is an important open problem. 

A further interesting direction is to understand the nature of phase transitions in non-Euclidean geometries. The various dualities we developed here can be used to relate different critical points to each other, e.g. the critical points of the deconfined phase along one of the axes of the Fradkin-Shenker phase diagram to the critical point of the ungauged classical code in the presence of a transverse field. On the other hand, cLDPC codes without local redundancies can be self-dual under Kramers-Wannier. This raises the possibility of self-dual critical points, where the KW duality transformation becomes a ``non-invertible symmetry''~\cite{mcgreevy2023generalized,shao2023s}. It will be interesting to compute the fusion relations of these symmetries, along the lines of Refs. \cite{li2023non,cao2023subsystem}, and understand how they might be used to constrain possible critical points. 

Another set of open questions concerns the generalization of our gauge theory construction from $\mathbb{Z}_2$ to more general gauge groups. In one direction, combining the ideas presented here with Kitaev's quantum double construction~\cite{kitaev2003fault} should allow for gauge theories on generic graphs with discrete gauge groups, which can have interesting features, especially if the underling symmetry is non-Abelian. Generalizing the understanding to continuous gauge groups, especially $U(1)$ to this setting, is also an interesting direction. 

In this paper, we applied to gauging procedure to the simplest possible symmetric Hamiltonian, given by the appropriate classical code and a transverse field. This gave rise to the analogue of the usual Fradkin-Shenker Hamiltonian, whose deconfined phase is naturally described by the CSS Hamiltonian obtained from the corresponding chain complex. However, one advantage of our general view of gauging is that it can be applied to a host of other symmetric Hamiltonians, that can potentially be used to build other, distinct topologically ordered phases, similarly to how ``twisted'' versions of the usual $\mathbb{Z}_2$ gauge theory~\cite{dijkgraaf1990topological} arise from gauging SPT Hamiltonians~\cite{levin2012braiding}. Finally, the 1D Ising model can be solved via the Jordan-Wigner transformation which is closely related to Kramers-Wannier duality~\cite{radicevic2018spin}. Generalizations to higher dimensions have been developed recently~\cite{chen2018exact,tantivasadakarn2020jordan} and also rely on introducing gauge fields; understanding whether similar spin-to-fermion mappings can be obtained from the general gauge dualities we developed here is an interesting open question as well. 

\begin{acknowledgments}
We thank Jeongwan Haah, Curt von Keyserlingk, Nathan Seiberg, Steven Shenker, Sagar Vijay, Dominic Williamson and especially Anirudh Krishna for enlightening discussions. TR thanks Ruben Verresen, Ryan Thorngren and Ashvin Vishwanath for previous collaboration related to the contents of Sec.~\ref{sec:SPT}.
T.R. is supported in part by the Stanford Q-Farm Bloch Postdoctoral Fellowship in Quantum Science and Engineering. 
V.K. acknowledges support from the US Department of Energy, Office of Science, Basic Energy Sciences, under Early Career Award No. DE-SC0021111, the Alfred P. Sloan Foundation through a Sloan Research Fellowship and the Packard Foundation through a Packard Fellowship in Science and Engineering.
\end{acknowledgments}

\appendix

\section{Chain complexes and $\mathbb{Z}_2$ homology}\label{app:ChainComplex}

Here we review some properties of $\mathbb{Z}_2$ chain complexes, which play a prominent role throughout the paper. See Ref. \onlinecite{breuckmann2018phd} for more details.

A $D$-level $\mathbb{Z}_2$ chain complex is a series of vector spaces, $V_r$, $r=0,\ldots,D$ over binary variables $\mathbb{Z}_2$, along with \emph{boundary maps} $\delta_r: V_r \to V_{r-1}$. We can think of a $\mathbb{Z}_2$ vector space as formed by all subsets of some base set $X_r$. Elements of $X_r$ form a standard basis of $V_r$ and each vector in $V_r$ corresponds to a subset of $X_r$ defined by those elements that appear with coefficient $1$. Addition of two vectors corresponds to taking the symmetric difference of the two subsets. In this paper, when we talk about chain complexes, we imagine that they arise in this way, which means that their vector spaces come equipped with some fixed basis. We refer to elements of $X_r$ as $r$-dimensional cells (sites, edges, plaquettes, etc.). If $c_r$ is an $r$-cell, we denote the corresponding basis vector in $V_r$ by $\ket{c_r}$. Elements of $V_r$ are referred to as $r$-\emph{chains}.

The map, $\delta_r$ maps a subset of $r$-cells onto their ``boundary''. The defining property of a chain complex is that \emph{the boundary of a boundary is zero}, meaning that $\delta_r\delta_{r+1} = 0$, where multiplication is over $\mathbb{Z}_2$ (i.e., modulo 2). We can define $\delta_r$ by its action on the basis vectors; this defines an adjacency relation between elements of $X_r$ and $X_{r-1}$ and we say that $c_{r-1} \in X_{r-1}$ is incident on $c_r \in X_R$ if $\ket{c_{r-1}}$ appears in $\delta_r\ket{c_r}$ (in other words, if $\braket{c_{r-1}|\delta_r|c_r}$ = 1). This gives rise to the representation of the chain complex as a $D$-partite graph, where we draw a node for each cell, which can be grouped into $D$ groups depending on which $X_r$ they belong to, and we draw an edge between $c_r$ and $c_{r-1}$ if the latter is incident on the former. The resulting graph is called the \emph{Hasse diagram} of the chain complex and we arrange the levels horizontally, starting from $X_D$ on the left, as in Fig.~\ref{fig:2complex}. The Tanner graph introduced in the main text is a special case of this, for $D=1$.

Given a chain complex, we can define $Z_R \equiv \text{Ker}(\delta_r)$, the set of \emph{closed} $r$-\emph{cycles}. We can also define $B_r \equiv \text{Im}(\delta_{r+1})$, which is the set of $r$-chains that arise as the boundary of $(r+1)$-chain. Due to the defining property of chain complexes, $B_r \subseteq Z_r$. Vectors in their difference, $Z_r \setminus B_r$, are \emph{non-trivial} (also called non-contractible, or \emph{essential}) cycles. We can define an equivalence relation, where two vectors in $V_r$ are equivalent if their difference is in $B_r$; the equivalence classes then form the $r$-th \emph{homology group} $H_r \equiv Z_r / B_r$. For $r=D$, one defines $H_D = Z_D$ (which is formally equivalent to taking $\delta_{D+1}$ to be the zero map). 

The transposes of $\delta_r$ define \emph{coboundary maps}, which define cocycles $Z^r \equiv \text{Ker}(\delta_{r-1}^T)$, which include coboundaries of objects one dimension lower, i.e. $B^r \equiv \text{Im}(\delta_r^T)$. Non-trivial cocycles belong to $Z^r \setminus B^r$ and they form cohomology classes that constitute the cohomology group $H^r \equiv Z^r / B^r$. We also define $H^0 \equiv \text{Ker}(\delta_1^T)$. Given a chain complex $\mathcal{C}$, we can define its \emph{dual}, $\tilde{\mathcal{C}}$, which consists of vector spaces $\tilde{V}_r \equiv V_{D-r}$ (whose elements are called \emph{cochains}) and boundary maps $\tilde{\delta}_r \equiv \delta_{D-r+1}^T$. This amounts to reflecting the Hasse diagram around its horizontal axis. The homology (cohomology) classes of $\tilde{\mathcal{C}}$ are the cohomology (homology) classes of $\mathcal{C}$. 

As the kernel of a matrix is the orthogonal complement of the image of its transpose, we have that 
\begin{align}\label{eq:cycles}
  B^r = Z_r^\perp, & & Z^r = B_r^\perp
\end{align}
In words: set of cocycles are exactly those $r$-chains that are orthogonal to any boundary. By the equivalence between $V_r$ and subsets of $X_r$, $\braket{c_r|c'_r} = 0$ iff $c_r,c'_r$ overlap on an even number of $r$-cells. Thus any boundary overlaps with any cocycle on an even number of places, and similarly for cycles and coboundaries. A consequence of this is that if we take an $r$-cycle $c_r$ and a $(D-r)$-cocycle $\tilde{c}_{D-r}$, their inner product $\braket{\tilde{c}_{D-r}|c_r}$ is a function only of their (co)homology classes. We can always choose a basis of (co)homology classes in such a way that for a given homology class there is precisely one cohomology class where this inner product is $1$.


\section{Kramers-Wannier duality and gauging for the Ising model}\label{app:Ising}

Here, we give a review of the gauging procedure for the Ising model in one and two dimensions, following our discussions in Sec.~\ref{sec:Gauging} and~\ref{sec:Redundancies}.

\subsection{One dimension}

Consider the transverse field Ising model of spin-$1/2$ particles arranged on a periodic chain of sites labeled by $i=1,\ldots,L$:
\begin{equation}\label{eq:1D_TFIM}
    H_\text{1D TFIM} = - J \sum_{i} \sigma_{i}^z \sigma_{i+1}^z - g \sum_i \sigma_i^x.
\end{equation}
In the limit $g=0$, the model is classical and describes the fixed point limit of a ferromagnet, which spontaneously breaks the spin-flip symmetry $\prod_i \sigma_i^x$, with two perfectly degenerate ground states given by the ``all up'' and ``all down'' configurations. In the coding language, this is the classical repetition code and it has a single (global) redundancy, given by the product of all of its checks. The transverse field $g$ drives a transition to a paramagnet, which happens at $g=1$. 

The Kramers-Wannier dual variables are assigned to the bonds of the 1D chain, and we denote them by $\tau_{i+1/2}$. Following Eq.~\eqref{eq:KWmap}, we define the KW duality as
\begin{align}\label{eq:1DKW}
    \tau_{i+1/2}^z \equiv \sigma_{i}^z \sigma_{i+1}^z; & & \tau_{i-1/2}^x \tau_{i+1/2}^x \equiv \sigma_{i}^x.
\end{align}
Physically, these new variables correspond to domain wall configurations in the Ising model. In terms of these, the Hamiltonian~\eqref{eq:1D_TFIM} becomes $-J \sum_i \tau_i^z - g \sum_i \tau_{i-1/2}^x \tau_{i+1/2}^x$, which is the same as Eq.~\eqref{eq:1D_TFIM}, up to a Hadamard transform $\tau_{i+1/2}^z \leftrightarrow \tau_{i+1/2}^x$ , but with the roles of the two terms interchanged. Therefore the Kramers-Wannier transformation maps the two phases of the model onto each other. 

\begin{figure} 
    \centering
    \includegraphics[trim={1cm 0 0 0},width = 0.4\linewidth]{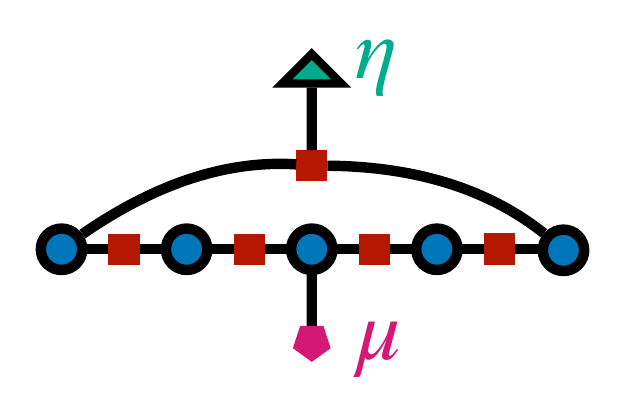}
    \caption{Coupling the 1D Ising model to background gauge fields. $\sigma$ spins (blue circles) live on sites and $\tau$ spins (red squares) on bonds. There is a single background field $\eta$ which toggles between periodic and anti-periodic boundary conditions by switching the sign of the check that crosses the boundary. There is also a gauge field $\mu$ for the dual $\tau$ variables.}
    \label{fig:1DIsing}
\end{figure}

The duality transformation~\eqref{eq:1DKW} applies within the symmetric subspaces,  $\prod_{i} \sigma_i^x = \prod_{i}\tau_{i+1/2}^z = 1\!\! 1$; thus the mapping is only defined between the symmetric subspaces, satisfying $\prod_i \sigma_i^x = 1$ and $\prod_i \tau_{i+1/2}\tau^z = 1$, respectively. We can extend it to the full space by adding ancilla variables, $\eta$ and $\mu$, following the prescription introduced in Sec.~\ref{subsec:BackgroundGauge}. For the former, we choose the check $C_L = \sigma_L^z \sigma_1^z$ and modify it to $\hat{C}_L = \eta^z \sigma_L^z \sigma_1^z$ as shown in Fig.~\ref{fig:1DIsing}. When $\eta^z = -1$, any configuration of $\sigma$ spins contains an odd number of domain walls (violated checks), amounting to anti-periodic boundary conditions. On the other side of the duality, We can similarly introduce an ancilla $\mu$, by modifying the check near the boundary as $\tau_{L-1/2}^x\tau_{L+1/2}^x \to \mu^x \tau_{L-1/2}^x\tau_{L+1/2}^x$ (see Fig.~\ref{fig:1DIsing}). In terms of the $\sigma$ variables, this is a ``pinning field'' that can be used to choose one or ther other of the two ground states. 

We can also follow the gauging prescription of Sec.~\ref{subsec:gauging} and consider a $\mathbb{Z}_2$ gauge theory, minimally coupled to the Ising ``matter'' degrees of freedom. Starting from Eq.~\eqref{eq:1D_TFIM}, we introduce ``gauge fields'' $\tau_{i+1/2}^{x,z}$ on the bonds and modify the Hamiltonian to
\begin{equation}\label{eq:1D_LGT}
    H_{\text{1D } \mathbb{Z}_2 \text{ LGT}} = - J \sum_{i} \sigma_{i}^z \tau_{i+1/2}^z\sigma_{i+1}^z - g \sum_i \sigma_i^x - \Gamma \sum_i \tau_{i+1/2}^x,
\end{equation}
where the last term adds dynamics to the gauge fields. Eq.~\eqref{eq:1D_LGT} is symmetric under a \emph{local} spin-flip $\sigma_i^x$, provided that it accompanied by a rearrangement of the gauge fields by $\tau_{i-1/2}^x \tau_{i+1/2}^x$, encoded in the Gauss law constraint
\begin{equation}
    \tau_{i-1/2}^x \sigma_{i}^x \tau_{i+1/2}^x = + 1 \quad\quad \forall i.
\end{equation}
This defines the physical Hilbert space $\mathcal{H}_\text{phys}$ of the theory. One can use this condition to ``gauge fix'' $\sigma_i^z = + 1$ on all sites. Using this, and taking $\Gamma = 0$, turns~\eqref{eq:1D_LGT} into $-J \sum_i \tau_i^z - g \sum_i \tau_{i-1/2}^x \tau_{i+1/2}^x$, where the $\tau$ variables are now unconstrained, reproducing the KW duality in the presence of background gauge fields. The periodic vs anti-periodic boundary conditions are now encoded in $\prod_{i} \tau_{i+1/2}^{z} = \pm 1$  

Finally, we can construct another Hamiltonian out of~\eqref{eq:1D_LGT}, by taking $g=0$ and relaxing the Gauss law from a strict constraint on the Hilbert space to an energetic term in the Hamiltonian, to get
\begin{equation}
    H_\text{1D Cluster} = - J \sum_{i} \sigma_{i}^z \tau_{i+1/2}^z\sigma_{i+1}^z - \lambda \sum_{i} \tau_{i-1/2}^x \sigma_{i}^x \tau_{i+1/2}^x.
\end{equation}
This is the Hamiltonian of the so-called cluster chain~\cite{raussendorf2001one,verresen2022higgs} (a more familiar form is obtained after the Hadamard transform $\tau^x \leftrightarrow \tau^z$), which is a paradigmatic model of an SPT protected by the two $\mathbb{Z}_2$ symmetries $\prod_i \sigma_i^x$ and $\prod_i \tau_{i+1/2}^z$. In particular, when open boundary conditions are imposed on it, it hosts a spin-$1/2$ edge mode at each boundary, which cannot be gapped out by any weak symmetric perturbation. 

\subsection{Two dimensions}

We can consider the two-dimensional version of the TFIM
\begin{equation}\label{eq:2D_TFIM}
    H_\text{2D TFIM} = - J \sum_{\langle i,j\rangle} \sigma_{i}^z \sigma_{j}^z - g \sum_i \sigma_i^x,
\end{equation}
with $i$ labeling sites on a square lattice with periodic boundary conditions and $\langle i, j\rangle$ describing nearest neighbor pairs. We can interpret the interaction terms as checks $C_{ij}$ of a classical code. As $g/J$ is tuned, Eq.~\eqref{eq:2D_TFIM} again interpolates between a ferromagnetic and a paramagnetic fixed point. 

We can again use the KW duality and introduce ``domain wall'' variables as 
\begin{align}\label{eq:2DKW}
    \tau_{e}^z \equiv \sigma_{i}^z \sigma_{j}^z; & & A_i \equiv \sigma_{i}^x,
\end{align}
where $e = (i,j)$ denotes an edge and $A_i = \prod_{e \in \foo } \tau_{e}^x$ is a product over a star shape defined by the four bonds that meet at site $i$ (see also the lower panel on Fig.~\ref{fig:ClassicalExamples}(a)). Unlike the 1D case, this maps Eq.~\eqref{eq:2D_TFIM} to a different, inequivalent Hamiltonian, $-J \sum_{\langle i,j\rangle} \tau_{\langle i,j\rangle}^z -g \sum_i A_i$. As in 1D, the definitions~\eqref{eq:2DKW} impose constraints, both local and global. Locally $B_p = \prod_{e \in \msquare} \tau_e^z = 1\!\!1$, where the product involves four edges around a square plaquette labaled by $p$. Globally, we have that the product of $\tau_e^z$ along any row or column also gives the identity. In the original model, these conditions stem from the fact that domain walls must for closed loops (on the dual lattice) while for the dual model, they can be seen as delineating a symmetric subspace for a \emph{1-form} $\mathbb{Z}_2$ symmetry~\cite{mcgreevy2023generalized}. 

\begin{figure} 
    \centering
    \includegraphics[trim={1cm 0 0 0},width = 0.65\linewidth]{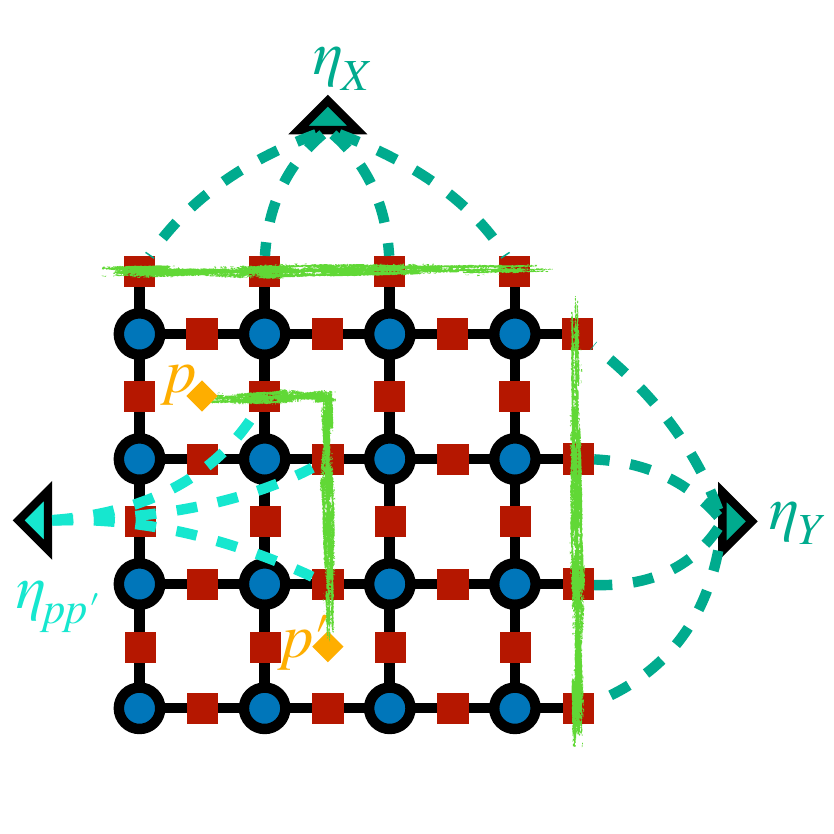}
    \caption{Coupling the 2D Ising model to background gauge fields. Two of the $\eta$ variables correspond to twisting the boundary conditions in either the $X$ or $Y$ directions, which is achieved by flipping the signs of all checks along a non-contractible loop on the dual lattice. The remaining $\eta$ variables correspond to open lines on the dual lattice, which can be used to remove the redundancies associates to the plaquettes at their endpoints. The support of one such ``seam'', connecing plaquettes $p,p'$ is shown. The checks along it are all connected to the ancilla variable $\eta_{pp'}$.}
    \label{fig:2DIsing}
\end{figure}

We can again introduce background gauge fields using the definitions of Sec.~\ref{subsec:BackgroundGauge}. As discussed in Sec.~\ref{subsec:LocalRed}, there are now two distinct types of redundancies: local ones, corresponding to the plaquettes, and global ones, corresponding to rows and columns of the 2D lattice. Let us consider the latter kind first: these can be removed by adding two ancilla variables, $\eta_X$ and $\eta_Y$. $\eta_X$ couples to all the checks in a row of the dual lattice, as shown in Fig.~\ref{fig:2DIsing}. As a result, the product of checks for any column (labeled by $j$) now becomes $\prod_i C_{ij} = \eta_X$. Similarly, $\eta_Y$ toggles checks along a column of the dual lattice and removes all the redundancies associated to rows: $\prod_j C_{ij} = \eta_Y$. 

As for the local redundancies, they are not all linearly independent, so they can only be removed in pairs. For example, we can choose a pair of plaquettes $p,p'$ and a path on the dual lattice connecting them. We can then change all checks along this path to $C_{ij} \to \eta_{pp'} \hat{C}_{ij}$. If we then take the product of checks along a loop that encircles one but not both of the two plaquettes, we get $\prod C_{ij} = \eta_{pp'}$. 

Alternatively, we can introduce $\tau_e$ variables at every edge, and modify $\sigma_i^z \sigma_j^z \to \sigma_i^z \tau_e^z \sigma_j^z$ to ensure symmetry under local spin flips. There are now two types of gauge invariant terms we can add for the gauge fields: $\tau_e^x$ and $B_p$. Combining these, we get the celebrated Fradkin-Shenker Hamiltonian~\cite{fradkin1979phase}
\begin{align}\label{eq:2D_LGT}
    H_{\text{2D } \mathbb{Z}_2 \text{ LGT}} = &- J \sum_{e=(i,j)} \sigma_{i}^z \tau_{e}^z\sigma_{j}^z - g \sum_i \sigma_i^x \nonumber \\ &-K \sum_p B_p - \Gamma \sum_e \tau_{e}^x,
\end{align}
along with a Gauss law constraint $\sigma_i^x A_i = +1$ $\forall i$. We can again use this to gauge fix and get
\begin{equation}\label{eq:TC_in_field}
    H = -J \sum_e \tau_e^z - g \sum_i A_i - K \sum_p B_p -\Gamma \sum_e \tau_e^x.
\end{equation}
Along the $\Gamma = 0$ axis, all terms commute with $B_p$ so at low energies we can take $B_p =+1$ and recover the dual Hamiltonian in the domain wall variables, but now without the global constraints on $\tau_e^z$. At the point $\Gamma = J = 0$, we obtain the toric code Hamiltonian, which exhibits topological order~\cite{kitaev2003fault}.

We can also lift the Gauss constraint to be energetic, adding it as a term $-\lambda \sum_i \sigma_i^x A_i$ to the Hamiltonian. Then, going to the extreme ``Higgs'' limit of $J \gg g,\Gamma$, we can describe the system with the Hamiltonian
\begin{equation}
    H_\text{2D Cluster} = - J \sum_{e=(i,j)} \sigma_{i}^z \tau_{e}^z\sigma_{j}^z - \lambda \sum_{i} \sigma_i^x \prod_{e \in \foo} \tau_e^x.
\end{equation}
As pointed out in \onlinecite{verresen2022higgs}, this is nothing but the cluster Hamiltonian on the so-called Lieb lattice, which has been shown previously~\cite{yoshida2016topological} to be an SPT protected by a combination of a 0-form and a 1-form $\mathbb{Z}_2$ symmetry. In particular, when put on a manifold with a boundary, the degrees of freedom at the edge form a 1D ferromagnetic with a 2-fold degeneracy~\cite{verresen2022higgs}.

\section{Gauge theories for subsystem codes}\label{app:subsystem}

In the main text we focused on stabilizer codes, wherein all checks defining the code commute. A more general type of code is given by \emph{subsystem codes}, defined by non-commuting checks. Here we give a brief description of how the gauge theory perspective we developed could be modified to accommodate these. We still focus on CSS codes, such that all checks are either purely $X$ or purely $Z$ type. 

A CSS subsystem code is again defined by two sets of checks, $A_i = \prod_{a\in \delta^T(i)} \tau_a^x$ and $B_p = \prod_{a \in \tilde{\delta}(p)} \tau_a^z$, which we defined through the maps $\delta$ and $\tilde{\delta}$. As the checks are now not required to commute, there is no restriction on these maps. The check define a set of \emph{stabilizers}. An $X$-type stabilizer is a product of $A_i$ that commute with all $B_p$ and vice versa. The code subspace is defined by the shared eigenspaces (with eigenvalue $+1$) of all the stabilizers, modded out by the action of checks. 

\begin{figure} 
    \centering
    \includegraphics[trim={1cm 0 0 0},width = 1.\linewidth]{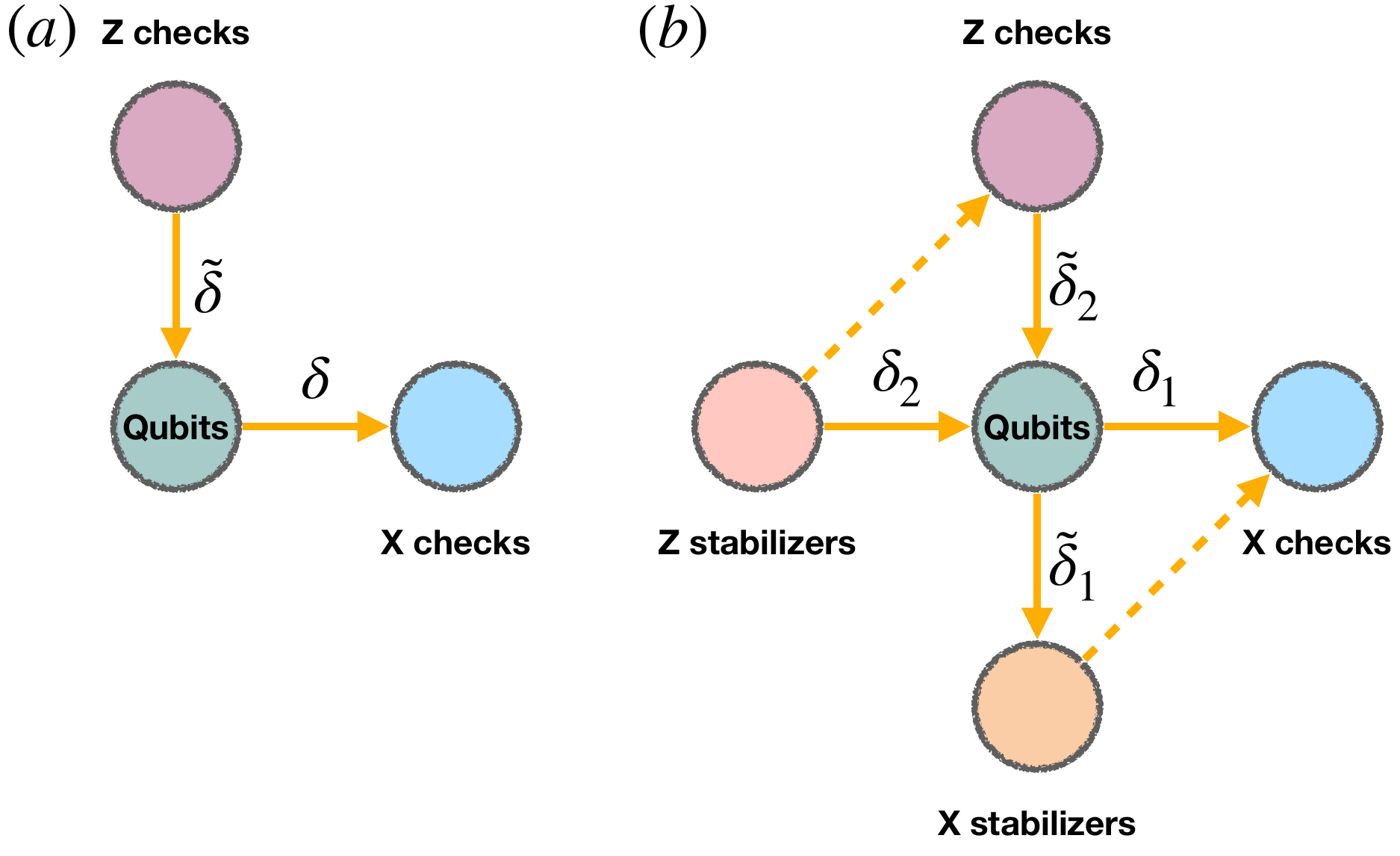}
    \caption{Chain complex representation of subsystem codes (a) without and (b) with local stabilizers.}
    \label{fig:SubsystemCodes}
\end{figure}

In a stabilizer code, like the ones discussed in the main text, there is no distinction between checks and stabilizers, which automatically implies that the stabilizers are local. For subsystem codes, however, we need to distinguish two cases, depending on whether there exist a local generating set for the stabilizers or not\footnote{In principle of course we also have mixed cases, e.g. where only one type of stabilizer has local generators, but we ignore this here.}. In the former case, we can include the stabilizers to form $2$-term chain complexes, formed by the triple of checks, qubits and stabilizers. However, unlike the case in the main text, there are now \emph{two} distinct chain complexes, one involving $X$ checks and $Z$ stabilizers and another involving $Z$ checks and $X$ stabilizers. These two chain complexes are intertwined by the fact that the same set of qubits forms the $1$-dimensional term in both; see Fig.~\ref{fig:SubsystemCodes}(b). Note that the $X$ and $Z$ stabilizers are themselves products of the checks of the same type, indicated by the dashed arrows in Fig.~\ref{fig:SubsystemCodes}(b). Consequently, the two types of stabilizers commute with each other, forming a stabilizer code, with the chain complex associated with the maps $\delta_2$ and $\tilde\delta_1$ (obviously, $\delta_2\tilde\delta_1 = 0$). When the stabilizers are non-local, on the other hand, we have only a pair of $1$-term chain complexes (Fig.~\ref{fig:SubsystemCodes}(a)).  

We can associate a Hamiltonian to the subsystem code, although due to the failure of checks to commute, there is no longer a strict relationship between it and the corresponding code ([Yaodong]). For example, in the case with local stabilizers, we could write
\begin{align}\label{eq:SubsysHam}
    H_\text{CSS} = -&J_x \sum_i \prod_{a \in \delta_1^T(i)} \tau_a^x - J_z \sum_p \prod_{a \in \tilde\delta_2(p)} \tau_a^z \nonumber \\ 
    - &K_z \sum_{q} \prod_{a\in \delta_2(q)} \tau_a^z - K_x \sum_{j} \prod_{a\in \tilde\delta_1^T(j)} \tau_a^x,
\end{align}
where the first two terms correspond to checks and the last two to stabilizers; when there no local stabilizers, the latter ones would be omitted. 

We can obtain this Hamiltonian from two different gauge theories coupled together by a strong local coupling, one corresponding to the classical code defined by $\delta_1$ and another to $\tilde\delta_2^T$. In particular, consider the following two ``classical code in a transverse field'' type Hamiltonians:
\begin{align}
    H = -J \sum_{a} \prod_{i \in \delta_1(a)} \sigma_i^z - g \sum_i \sigma_i^x, \\
    \tilde{H} = -\tilde{J} \sum_{a} \prod_{p \in \tilde\delta_2^T(a)} \tilde{\sigma}_p^z - \tilde{g} \sum_p \tilde{\sigma}_p^x.
\end{align}
Applying the gauging procedure to $H$ (and gauge fixing to get rid of matter fields) gives rise to 
\begin{align*}
    H_\text{gauged} = -J \sum_a \tau_a^z - g \sum_i \prod_{a \in \delta_1^T(i)} \tau_a^x \\ - K \sum_{q} \prod_{a \in \delta_2(q)} \tau_a^z - \Gamma \sum_{a} \tau_a^x, 
\end{align*}
where the third term arises from local redundancies of the classical code (whose presence vs absence is equivlent ot the question of whether the CSS code has local $Z$ stabilizers or not). We see that in the limit $J=\Gamma=0$ we recover the first and third term in Eq.~\eqref{eq:SubsysHam}, with $g \to J_x$ and $K \to K_z$. 

Similarly, we can gauge $\tilde{H}$ to get
\begin{align*}
    \tilde{H}_\text{gauged} = -\tilde{J} \sum_a \tilde{\tau}_a^z - \tilde{g} \sum_p \prod_{a \in \tilde\delta_2(p)} \tilde{\tau}_a^x \\ - \tilde{K} \sum_{j} \prod_{a \in \tilde\delta_1(j)} \tilde{\tau}_a^z - \tilde{\Gamma} \sum_{a} \tilde{\tau}_a^x, 
\end{align*}
in which we recognize the ramaining terms of~\eqref{eq:SubsysHam}, along with some on-site fields. 

Combining these, a strategy for obtaining $H_\text{CSS}$ within a larger gauge theory phase diagram is as follows. We start with $H + \tilde{H}$ as our ungauged Hamiltonian. We gauge the combined symmetries of both models, to get two decoupled gauge theories, for $\tau_a$ and $\tilde{\tau}_a$ (labeled by the same set of labels $a$). We can then couple these via a term\footnote{In the full theory, before gauge fixing, we should replace $\tau_a^z$ with $\tau_a^z C_a$ in this expression, where $C_a$ are the checks that appear in $H$, and similarly $\tilde{\tau}_a^z \to \tilde{\tau}_a^z \tilde{C}_a$.} $-\lambda \sum_a (\tau_a^x \tilde{\tau}_a^z + \tau_a^z \tilde{\tau}_a^x)$. In the limit of $\lambda \to \infty$ we project down to a subspace $\tilde{\tau}_a^x = \tau_a^z$ and $\tilde{\tau}_a^z = \tau_a^x$, so, after switching off the local fields ($J=\tilde J = \Gamma = \tilde\Gamma = 0$) we recover~\eqref{eq:SubsysHam}. One can then study the phase diagram of this gauge theory both by switching on the local fields, and making the coupling $\lambda$ finite. 

If the ungauged classical codes have no redundancies, then we can also write the coupling between them, before going to th gauged variables. It corresponds to $-\lambda(D_a \tilde{C}_a + C_a \tilde{D}_a)$, where $C_a$ ($\tilde{C}_a$) are the checks and $D_a$ ($\tilde{D}_a$) are the disorder operators of the codes corresponding to $H$ ($\tilde{H}$). Note, however, that this coupling is highly non-local, due to the fact that it explicitly involved disorder operators. 

\bibliography{Part1.bbl}

\end{document}